\documentclass[aps,amsmath,amssymb,showkeys]{revtex4}
\usepackage{graphicx,color}
\usepackage{times}
\usepackage{xcolor}
\usepackage[%
  colorlinks=true,
  urlcolor=blue,
  linkcolor=red,
  citecolor=blue
]{hyperref}
\usepackage{orcidlink}
\usepackage{mathrsfs}
\usepackage{amsmath}
\usepackage{accents}
\usepackage{mathtools}
\usepackage{amssymb}
\usepackage{graphicx}
\usepackage{epsfig}
\usepackage{dcolumn}
\usepackage{bm}
\usepackage{tikz}
\usetikzlibrary{arrows, decorations.markings, calc, fadings, decorations.pathreplacing, patterns, decorations.pathmorphing, positioning, angles, quotes}
\newcommand{\orcid}[1]{\href{https://orcid.org/#1}{\includegraphics[width=8pt]{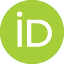}}}

\begin{document}
\title{Gravitational Lensing Signatures of a Black Hole in Mimetic Euler-Heisenberg theory}

\author{Nashiba Parbin$^{\orcidlink{0000-0002-3570-0117}}$}
\email[Email: ]{nashibaparbin91@gmail.com}

\affiliation{Pacif Institute of Cosmology and Selfology (PICS), Sagara, Sambalpur, 768224, Odisha, India}

\begin{abstract}
The present study delves into the gravitational lensing phenomena in the weak field limit of a charged black hole in the mimetic Euler-Heisenberg theory. The deflection angle of light is computed for the specified asymptotically flat spacetime assuming that the observer and the light source are both located at finite distances from the lens using the Gauss-Bonnet theorem. Additionally, the Einstein ring radius is analytically derived as a direct implementation of the deflection angle. The relevant influence of the Euler-Heisenberg parameter and the charge parameter on the Einstein ring radius is briefly discussed. We also investigate the observational implications of lensing in the weak gravitational limit of the black hole. The magnification and distortion parameter are also analyzed in this manuscript. Furthermore, the time delay of light due to gravtitational lensing is also briefly discussed. Our study reveals the significant influence of the model parameters of the specified black hole on the gravitational lensing signatures.

\end{abstract}

\keywords{Modified gravity; Deflection angle; Einstein ring; Magnification}

\maketitle

\section{Introduction}
\label{sec.1}

The key idea of gravitational lensing has emerged from the interpretation of gravitational deflection of light in the framework of geometrical optics of a lens \cite{schneider, trimble, renn, valls}. This concept in fact predates Einstein's groundbreaking theory of general relativity (GR)  \cite{einstein}. Pioneering scientists including Newton, Michell, Cavendish, Laplace, Soldner, and Einstein himself speculated that massive objects could bend light passing near the gravitational field of such objects \cite{schneider, schneider1}. Once GR was established in $1915$, Einstein calculated that a light ray grazing the Sun's surface tangentially would experience a total deflection twice the value predicted by Newtonian physics. This monumental prediction was experimentally verified by observations carried out during the total solar eclipse of $1919$ \cite{eddington}, providing compelling empirical support for GR. It was in $1979$, when the first instance of gravitationally lensed object, the double quasar Q$0957 + 561$A,B was identified \cite{young}. In the modern observational cosmology, gravitational lensing proved to be a crucial tool for exploring the exoplanets and for measuring the distribution of dark matter across a wide range of masses in galaxies. This has further enabled significant elevations in our understanding of galactic dynamics and cosmic structures. 

Past its publication, GR has gained utmost importance in describing many gravitational phenomena such as black holes and gravitational waves. However, studies of the accelerated expansion of the universe \cite{reiss, perlmutter} as well as the observed rotational dynamics of the galaxies \cite{rubin, young1, harko, gergely, parbin} have revealed that GR is not a theory entirely without drawbacks. GR needs some exotic stuff called dark energy and dark matter \cite{bertone, swart, tim, frenk, strigari, will2014} in order to overcome its shortcomings. This leads us to a need for gravitational theories beyond GR. Presently,  a number of modified theories have been proposed to resolve unexplained issues of the universe \cite{Clifton, wagoner, ronald, capozziello, clifton, oikonomou, odintsov, sergei, gogoi1, nashiba}. 

Black hole physics has become one of the most vibrant and rapidly advancing frontiers of modern physics. This dynamic evolution is driven by an influx of high-precision observational data and diverse theoretical frameworks. The optical properties of black holes have captured intense interest of the scientific community, fueled largely by the Event Horizon Telescope's (EHT) landmark capture of the first black hole image \cite{akiyama1, akiyama2}. This historic milestone has opened up new observational pathways for the study of black holes. This image has reaffirmed the robustness of Einstein's GR but also simultaneously has raised concerns about its limits in extreme surroundings, thereby encouraging researchers to probe deeper into physics beyond standard GR. The gravitational field features of a black hole, such as its gravitational lensing and shadow can be established from the null geodescis around the black hole. Gravitational lensing phenomena around black holes have been probed in a plethora of scenarios such as the Schwarzschild black hole \cite{virbhadra}, AdS/dS black hole \cite{zhao,rindler}, naked singularity and horizonless ultracompact objects \cite{ellis, shaikh}, etc. Several studies have also delved in examining the gravitational lensing effects around black hole spacetimes in theories beyond GR \cite{nashiba1, jamil, carvalho, nashiba2, kumar, ghosh, asada, takizawa, pantig1, pantig2, ovgun1, haximjan, filho, jamil1, gayatri}. Exploring the beautiful phenomena of gravitational lensing not only yields the informaton about black holes but also helps us understand and corroborate theories beyond GR in the weak as well as strong field regimes of the black hole. Gravitational lensing in the weak field limit of a black hole has been the aim of numerous studies \cite{izzet, pingsu, ovgun3, anisur, nashiba3}. In 2008, Gibbons and Werner \cite{gibbons} reformed the typical perspective by introducing a new geometrical approach to deduce the deflection angle in the weak field limit using the Gauss-Bonnet theorem (GBT) \cite{carmo, bonnet} for the static and asymptotically flat spacetimes. In 2016, Ishihara $\textit{et al.}$ further reconstructed this method for source and observer at finite distances employed to both static as well as stationary black hole solutions \cite{ishihara}. For asymptotically non-flat black hole solutions, study of gravitational lensing features can be found in the Refs.\,\cite{asada, takizawa, carvalho, nashiba3}. Another interesting result of gravitational lensing is the Einstein ring, which appears when the light source, the observer and the lens remain perfectly positioned. In order to deduce the Einstein ring radius of the black hole, gravitational deflection angle of the light beam is computed. When light passes through the gravitational field of the black hole, it undergoes a large deviation and even result in several complete loops around the black hole before reaching the receiver. Consequently, an infinite set of relativistic image can be observed on each side of the black hole. Virbhadra and Ellis \cite{virbhadra} introduced a new lens equation which made the study of observables such as magnification, distortion of images convenient. 

Following the aforementioned elements, we were motivated to implement the GBT to explore the gravitational lensing features in the weak field limit of a new black hole solution in mimetic Euler-Heisenberg (MEH) theory \cite{nashed}. This theory is an advanced framework in theoretical physics that combines mimetic gravity \cite{chamseddine1, chamseddine2} with the non-linear quantum electrodynamics of the Euler-Heisenberg (EH) theory \cite{heisenberg} to study modified spacetimes and black hole solutions. In this work, we assume a finite distance between the lens and the source as well as the receiver and derive the deflection angle of light. We then derive the expression for the Einstein ring radius and the magnification and distortion parameters from the lens equation introduced by Virbhadra. Furthermore, we briefly discuss the time delay of light due lensing by the black hole. The effects of the EH parameter and the charge parameter of the black hole on each of these gravitational lensing features in the weak field limit are examined.

Our paper is organized as follows. In Sec.\ \ref{sec.2}, we briefly review the 
field equations related to the MEH theory. In Sec.\ \ref{sec.3}, we derive the deflection angle in the spacetime of a charged black hole in this theory applying Ishihara $\textit{et al.}$ approach for asymptotically flat spacetime. The Einstein ring radius is computed and the magnification and distortion of images due to lensing by the black hole are studied in Sec.\ \ref{sec.4}. In Sec.\ \ref{sec.5}, we have studied the time delay of light as a consequence of the deflection of light by the black hole. Finally, in Sec.\ \ref{sec.6}, we summarize and conclude the results of our work. Throughout our work, we have imposed the sign convention ($-, +, +, +$) and $G=c=\hslash=1$.

\section{MEH gravitational theory and charged black hole solution} 
\label{sec.2}

A. H. Chamseddine and V. Mukhanov presented a new formulation of Einstein's GR known as the mimetic theory where they covariantly isolate the conformal degree of freedom \cite{chamseddine1, chamseddine2}. They defined the metric $g_{\mu\nu}$ in terms of the auxiliary metric $\tilde{g}_{\mu\nu}$ and a scalar field $\phi$ as
\begin{equation}
g_{\mu\nu} = (\tilde{g}^{\alpha\beta}\partial_\alpha\phi \partial_\beta\phi)\tilde{g}_{\mu\nu}
\label{eqn.1}
\end{equation} 
This scalar field ($\phi$) captures the conformal mode of gravity, ensuring that the theory remains explicitly invariant under Weyl transformations of the auxiliary metric. While the field equations of this theory closely mirror standard Einstein equations, they introduce a distinct longitudinal gravitational mode that stays dynamical despite the absence of normal matter. Their study demonstrates that this additional degree of freedom can effectively mimic cold dark matter and they termed it as ``mimetic dark matter".

The action for MEH theory can be expressed as \cite{nashed}
\begin{equation}
S=\frac{1}{16\pi} \int d^4 x \sqrt{-g(\tilde{g}_{\mu\nu}, \phi)} R(\tilde{g}_{\mu\nu}, \phi) - \int d^4 x \sqrt{-g(\tilde{g}_{\mu\nu}, \phi)} \mathscr{L}\mathcal{(F, G)} ,
\label{eqn.2}
\end{equation}
where, $g(\tilde{g}_{\mu\nu}, \phi)$ is the determinant of the metric tensor with scalar field $\phi$ and $R$ is the Ricci scalar.

$\mathscr{L}\mathcal{(F, G)}$ denotes the Lagrangian of the non-linear electromagnetic field. The Lagrangian is a function of the invariants, $\mathcal{F} = \frac{1}{4}\mathcal{F}_{\alpha\beta}\mathcal{F}^{\alpha\beta}$ and $\mathcal{G} = \frac{1}{4}\mathcal{F}_{\alpha\beta}\mathcal{^*F}^{\alpha\beta}$. Here, $\mathcal{F}_{\alpha\beta}$ is the strength tensor of the electromagnetic field and its dual is defined as $\mathcal{^*F}^{\alpha\beta} = \frac{\sqrt{-g(\tilde{g}_{\mu\nu}, \phi)}}{2} \epsilon^{\alpha\beta\gamma\delta}\mathcal{F}_{\gamma\delta}$, where the completely skew-symmetric tensor $\epsilon^{\alpha\beta\gamma\delta}$ meets the condition $\epsilon^{\alpha\beta\gamma\delta}\epsilon_{\alpha\beta\gamma\delta} = -4!$.

For the scenario of EH theory, the Lagrangian of the non-linear electromagnetic field can be expressed as \cite{carroll}
\begin{equation}
\mathscr{L}\mathcal{(F, G)} = -\mathcal{F} + \frac{\mu}{2}\mathcal{F}^2 + \frac{7\mu}{8}\mathcal{G}^2 ,
\label{eqn.3}
\end{equation}
where, $\mu = \frac{8\alpha^2}{45M^2}$ is the EH parameter, $\alpha$ is the fine structure constant and $M$ is the mass of electron. 

It is clear from Eq.(\ref{eqn.3}) that the standard Maxwell electrodynamics can be retrieved for $\mu = 0$ in which case $\mathscr{L}\mathcal{(F)} = - \mathcal{F}$. In the framework of non-linear electromagnetism, the field equations can be formulated using one of two distinct geometric frames. The first is the classical frame, which constructs the Lagrangian density directly from the standard electromagnetic field strength tensor $\mathcal{F}_{\alpha\beta}$. The other is the $\mathcal{C}$ frame with the tensor $\mathcal{C}^{\alpha\beta}$ as the primary field interpreted as
\begin{equation}
\mathcal{C}^{\alpha\beta} = -\left(\mathscr{L}_\mathcal{F} \mathcal{F}^{\alpha\beta} + \mathcal{^*F}^{\alpha\beta} \mathscr{L}_\mathcal{G}\right),
\label{eqn.4}
\end{equation}
where, $\mathscr{L}_\mathcal{F}$ is the derivative of $\mathscr{L}$ w.r.t. $\mathcal{F}$ and $\mathscr{L}_\mathcal{G}$ is the derivative w.r.t. $\mathcal{G}$. In the MEH theory, $\mathcal{C}^{\alpha\beta}$ can be expressed as \cite{nashed}
\begin{equation}
\mathcal{C}^{\alpha\beta} = (1-\mu\mathcal{F})\mathcal{F}^{\alpha\beta} - \frac{7\mu}{4}\mathcal{^*F}^{\alpha\beta}\mathcal{G}.
\label{eqn.5}
\end{equation}

The two invariants inherent to the $\mathcal{C}$ frame are defined as
\begin{equation}
\mathcal{C} = -\frac{1}{4}\mathcal{C}_{\alpha\beta}\mathcal{F}^{\alpha\beta} , \hspace{10mm} \mathcal{O} = -\frac{1}{4}\mathcal{C}_{\alpha\beta}\mathcal{^*C}^{\alpha\beta} ,
\label{eqn.6}
\end{equation}
where, $\mathcal{^*C}^{\alpha\beta} = \frac{1}{2\sqrt{-g(\tilde{g}_{\mu\nu}, \phi)}} \epsilon_{\alpha\beta\gamma\delta}\mathcal{C}^{\gamma\delta}$.

Based on the relations mentioned above, the MEH field equations emerge as \cite{nashed}
\begin{equation}
I_\alpha{}^\beta \equiv G_\alpha{}^\beta - {T^{\mathrm{EH}}}_\alpha{}^\beta - \tilde{T}_\alpha{}^\beta = 0 , \hspace{10mm} \nabla_\alpha \mathcal{C}^{\alpha\beta} = 0 ,
\label{eqn.7}
\end{equation}
where $G_\alpha{}^\beta$ is the Einstein tensor, $\tilde{T}_\alpha{}^\beta$ is the energy-momentum tensor of the scalar field in mimetic gravity and ${T^{\mathrm{EH}}}_\alpha{}^\beta$ is the energy-momentum tensor of the MEH theory in the $\mathcal{C}$ frame which reads
\begin{equation}
{T^{\mathrm{EH}}}_\alpha{}^\beta = \frac{1}{4\pi}\left([1-\mu\mathcal{C}]\mathcal{C}^\mu{}_\alpha \mathcal{C}^\beta{}_\mu + \delta_\alpha{}^\beta \left[\mathcal{C} - \frac{3\mu}{2}\mathcal{C}^2 - \frac{7\mu}{8}\mathcal{O}^2 \right]\right).
\label{eqn.8}
\end{equation}

The contribution of the mimetic field to the field equations can be expressed as
\begin{equation}
\tilde{T}_{\alpha\beta} = - (G - T^{\mathrm{EH}}) \partial_\alpha\phi \partial_\beta\phi ,
\label{eqn.9}
\end{equation}
where, $G$ is the trace of the Einstein tensor. Notably, the energy-momentum tensors are conserved, i.e. $\nabla_\mu {T^{\mathrm{EH}}}_{\mu\nu} = 0 = \nabla_\mu \tilde{T}_{\mu\nu}$ with $\nabla^\mu$ as the covariant derivative.

In the present work, we consider charged black holes in the MEH theory and the metric ansatz for such a black hole spacetime has the following form \cite{nashed}
\begin{equation}
ds^2 = A(r)dt^2 - B(r) dr^2 - r^2 (d\theta^2 +r^2 \sin^2\theta d\phi^2) ,
\label{eqn.10}
\end{equation}
and the metric co-efficients are expressed as
\begin{equation}
A(r) = \frac{1}{B(r)} = 1 - \frac{2M}{r} + \frac{q^2}{r^2} - \frac{q^4 \mu}{10r^6},
\label{eqn.11}
\end{equation}
where, $M$ is the mass, $q$ is the charge parameter and $\mu$ is the EH parameter of the aforementioned black hole.

\section{Deflection Angle of light for static spherically symmetric spacetime by Gauss-Bonnet theorem}
\label{sec.3}

In astrophysical observations, electromagnetic radiation propagating across cosmological distances does not travel along strictly Euclidean straight lines. Instead, photons emitted from distant luminous sources - such as quasars, active galactic nuclei, and primordial galaxies - are continuously deflected and distorted as they traverse the gravitational fields of massive celestial structures, which include supermassive black holes residing at galactic cores, galaxy clusters, and massive dark matter halos. This captivating phenomenon, widely known as gravitational lensing, serves as one of the most powerful probes of the cosmos, offering a direct window into the distribution of mass across the universe. A key parameter which quantifies the bending of photon beams around massive objects is the deflection angle of light. 

In this section, the deflection angle shall be derived within the weak field limit of a black hole~(\ref{eqn.10}) in the MEH gravitational theory. In this limit, the point of closest approach of light propagating from a distant source tends to lie far away from the black hole lens. We aim to analyze the effect of the charge parameter and the EH parameter on the bending angle of light as it approaches the weak gravitational field of the black hole. In this respect, the Gibbons-Werner (GW) method is utilized, which is a geometric approach based on the application of the GBT on optical geometry. Ishihara et al. extended this approach considering finite distance to a gravitational lens object from the light source and the receiver. Figure\ \ref{fig.1} shows a source of light ($S$), a receiver on earth ($R$) located at a finite distance from a black hole which acts as lens ($L$). At the equatorial plane ($\theta=\pi/2$), the deflection angle can be written as \cite{ishihara, ono}
\begin{equation}
\hat{\alpha} = \mathrm{\bf{\Psi}}_R - \mathrm{\bf{\Psi}}_S + \mathrm{\bf{\Phi}}_{RS},
\label{eqn.12}
\end{equation}
where, where ${\bf{\Psi}}_R$ and ${\bf{\Psi}}_S$ denote the angles of light measured
in relation to the lens at the positions of $R$ and $S$ respectively. ${\bf{\Phi}}_{RS}$ is the separation angle between $R$ and $S$ expressed as $\mathrm{\bf{\Phi}}_{RS} = \Phi_R - \Phi_S$ with $\Phi_R$ and $\Phi_S$ as the longitudes of $R$ and $S$ respectively.

\begin{center}
\begin{figure}[htbp]
\begin{tikzpicture}[scale=1.8]
    \def\R{2}
    
    \draw[dashed, blue, thick] (0,0) -- (140:1);
    \draw[line width=1pt, blue] (140:1) -- (140:2);
    
    \draw[dashed, blue, thick] (0,0) -- (0:1);
    \draw[line width=1pt, blue] (0:1) -- (0:2);
    
    
    \draw[thick, red] (0:\R) arc[start angle=0, end angle=140, radius=\R];
    
    \draw[blue, thick] (1,0) to [bend right=18] (-0.74,0.6);
    
    \draw (1.2,0) arc[start angle=0, end angle=142, radius=0.2];
    \node at (11:1.3) {$\bf{\Psi_S}$};
    
    \draw (-0.5,0.6) arc[start angle=0, end angle=142, radius=0.2];
    \node at (-60:-1.1) {$\bf{\pi - \Psi_R}$};
    
    \draw (1.8,0) -- (1.8,0.2) -- (1.98,0.2);    
    
    \draw (-1.4,1.17) -- (-1.25,1.30) -- (-1.40,1.44);    
    
    \fill (0,0) circle (2.8pt) node[below left] {$\bf{Black ~hole(L)}$};
    \node[right] at (0:\R) {$\infty$};
    \node[below left] at (140:\R) {$\infty$};
    \node[above right] at (80:\R) {$Q_r (r_C \rightarrow \infty)$};
    
    \draw[fill=red] (-0.74,0.6) circle (1.2pt) node[below left] at (140:1) {$\bf{Receiver(R)}$};
    \draw[fill=red] (1,0) circle (1.2pt) node[below right] at (0:1) {$\bf{Source(S)}$};
    
\end{tikzpicture}
\caption{Schematic illustration of gravitational deflection of light for light source and receiver at finite distance from a black hole. The quadrilateral $\stackrel{\infty}{R}\!\square\!\stackrel{\infty}{S}$ is embedded in a curved space.}
\label{fig.1}
\end{figure}
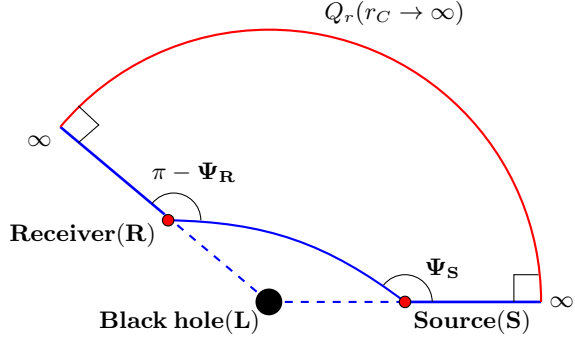
\end{center}

Figure~\ref{fig.1} portrays a quadrilateral $\stackrel{\infty}{R}\!\square\!\!\stackrel{\infty}{S}$ enclosed in a 3D Riemannian space $^{(3)}\! \mathcal{M}$, which consists of light rays emanating from $S$ to $R$ denoted by a spatial curve, two outgoing radial lines from $S$ and $R$, and a circular arc fragment $Q_r$ having the coordinate 
radius $r_C$ ($r_C \rightarrow \infty$). Putting the GBT to work in the domain $\stackrel{\infty}{R}\!\square\!\!\stackrel{\infty}{S}$, the deflection angle~(\ref{eqn.12}) can be recast as a function of the Gaussian curvature $\mathcal{K}$ of the surface of propagation of light and the geodesic curvature $k_g$ of the light curves \cite{gibbons},
\begin{equation}
\hat{\alpha} = - \int\!\! \int_{\stackrel{\infty}{R}\square\stackrel{\infty}{S}} \mathcal{K}\, dS + \int_S^R k_g\, dl,
\label{eqn.13}
\end{equation}
where $dS$ is the infinitesimal area element of the surface and $dl$ is the infinitesimal 
line element of the arc. 

For an asymptotically flat, non-rotating black hole, Ishihara et al. utilized the GBT to derive the weak deflection angle as
\begin{equation}
\hat{\alpha} = - \int\!\! \int_{\stackrel{\infty}{R}\square\stackrel{\infty}{S}} \mathcal{K}\, dS,
\label{eqn.14}
\end{equation}

As a consequence of the condition for null geodesics, $ds^2 = 0$ followed by light rays, the black hole metric~(\ref{eqn.10}) can be rewritten as
\begin{equation}
dt = \pm\, \sqrt{\zeta_{ij} dx^i dx^j},
\label{eqn.15}
\end{equation}
with $\zeta_{ij}$ as the optical metric which for the spacetime under study in this paper is given by 
\begin{align}
\zeta_{ij} dx^i dx^j & = \frac{dr^2}{\left(1 - \frac{2M}{r} + \frac{q^2}{r^2} - \frac{q^4 \mu}{10 r^6}\right)^2} + 
\frac{r^2 d\theta^2}{\left(1 - \frac{2M}{r} + \frac{q^2}{r^2} - \frac{q^4 \mu}{10 r^6}\right)} 
+ \frac{r^2 \sin^2 \theta d\phi^2}{\left(1 - \frac{2M}{r} + \frac{q^2}{r^2} - \frac{q^4 \mu}{10 r^6} \right)}.
\label{eqn.16}
\end{align}
With this optical metric the Gaussian curvature of propagating light is 
defined as \cite{gibbons}
\begin{equation}
\mathcal{K} = \frac{^{(3)\!} R_{r\phi r\phi}}{\zeta} = \frac{1}{\sqrt{\zeta}} \left[\frac{\partial}{\partial \phi} \left(\frac{\sqrt{\zeta}}{\zeta_{rr}}\ ^{(3)\!}\zeta_{rr}^\phi\right) - \frac{\partial}{\partial \phi} \left(\frac{\sqrt{\zeta}}{\zeta_{rr}} \ ^{(3)\!}\zeta_{r\phi}^\phi\right) \right],
\label{eqn.17}
\end{equation}
where $\zeta \equiv \det(\zeta_{ij})$. For the charged black hole in MEH gravitational theory, the Gaussian curvature $\mathcal{K}$ is computed as
\begin{equation}
\mathcal{K} \simeq -\frac{2M}{r^3} + \frac{3(M^2 + q^2)}{r^4} - \frac{6Mq^2}{r^5} + \frac{2q^2}{r^6} - \frac{21q^4 \mu}{10r^6}.
\label{eqn.18}
\end{equation}
In deriving the Gaussian curvature, the weak-field approximation has been taken into account. Hence, we have calculated only the leading order contributing terms. Using Eq.~(\ref{eqn.18}), the surface integral of $\mathcal{K}$ over the closed quadrilateral can be deduced as
\begin{equation}
- \int \!\! \int_{\stackrel{\infty}{R}\square\stackrel{\infty}{S}} \mathcal{K}\, dS = \int_{\phi_S}^{\phi_R}\!\! \int_{\infty}^{r_0} \mathcal{K} \sqrt{\zeta}\,dr\, d\phi,
\label{eqn.19}
\end{equation}
where, $r_0$ is the closest distance to the black hole. In order to make the above relation finite, we introduce $r$ in terms of $u$ as $r \equiv \frac{1}{u}$. Next, we need to study the photon orbit equation in the equatorial plane for the metric~(\ref{eqn.10}) in order to derive the integrals in Eq~(\ref{eqn.19}). To this end, the derivation of the reduced geodesic equation proves to be of paramount significance. Considering the equatorial plane ($\theta = \pi/2$), the condition for massless particles moving along null geodesics reduces Eq.~(\ref{eqn.10}) to
\begin{equation}
0 = A(r) \left(\frac{dt}{d\lambda}\right)^2 - B(r) \left(\frac{dr}{d\lambda}\right)^2 - r^2 \left(\frac{d\phi}{d\lambda}\right)^2
\label{eqn.20}
\end{equation}
where, $\lambda$ is an affine parameter along the photon orbit. It should be noted that there are two constants of motion along the orbit in the equatorial plane introduced from the symmetry and Killing vectors of the spacetime,
\begin{equation}
J \equiv r^2 \frac{d\phi}{d\lambda} \hspace{5mm}; \hspace{5mm} E \equiv A(r) \frac{dt}{d\lambda}
\label{eqn.21}
\end{equation}
where $J$ is the conserved angular momentum and $E$ is the conserved energy along a photon orbit. From these two constants, we obtain the impact parameter as $\xi = \frac{J}{E}$. Therefore, using Eq.~(\ref{eqn.21}) in Eq.~(\ref{eqn.20}) the reduced geodesic equation can be obtained as
\begin{equation}
\left(\frac{dr}{d\lambda}\right)^2 + A(r)\frac{J^2}{r^2} = E^2.
\label{eqn.22}
\end{equation}

The light orbit equation can thus be written as
\begin{equation}
\left(\frac{dr}{d\phi}\right)^2 = -r^2 A(r) + \frac{r^4}{\xi},
\label{eqn.23}
\end{equation}
and as mentioned earlier, we have chosen $r \equiv \frac{1}{u}$. Hence, the above expression can be recast as
\begin{equation}
\left(\frac{du}{d\phi}\right)^2 = F(u),
\label{eqn.24}
\end{equation}
with $F(u) \equiv -u^2 A(u) + \frac{1}{\xi}$, where $\xi$ is the impact parameter. Within the weak field limit, the iterative solution of Eq.~(\ref{eqn.24}) reads
\begin{equation}
u \simeq \frac{\sin\phi}{\xi} + \frac{M(1 + \cos^2\phi)}{\xi^2}.
\label{eqn.25}
\end{equation}
Hence, Eq.~(\ref{eqn.19}) can be recast as
\begin{equation}
- \int\!\! \int_{\stackrel{\infty}{R}\square\stackrel{\infty}{S}} \mathcal{K} dS = \int_{\phi_s}^{\phi_R} \!\!
\int_0^u - \,\frac{\mathcal{K}\sqrt{\zeta}}{u^2}\, du\, d\phi
\label{eqn.26}
\end{equation}
which for the metric~(\ref{eqn.10}) is computed to be
\begin{equation}
\begin{split}
- \int\!\! \int_{\stackrel{\infty}{R}\square\stackrel{\infty}{S}} \mathcal{K} dS 
& = \left(\frac{2M}{\xi} - \frac{21Mq^2}{4\xi^3} + \frac{115Mq^4}{32\xi^5} + \frac{255Mq^4\mu}{128\xi^7}\right) \left(\sqrt{1-\xi^2 u_R^2} + \sqrt{1-\xi^2 u_S^2} \right) \\
& + \left(-\frac{Mq^2}{12\xi^3} - \frac{25Mq^4}{64\xi^5} - \frac{199Mq^4\mu}{640\xi^7}\right) \left[(1-\xi^2 u_R^2)^{3/2} + (1-\xi^2 u_S^2)^{3/2}\right] + \left(-\frac{Mq^4}{320\xi^5} + \frac{87Mq^4\mu}{3200\xi^7}\right) \\
& \times \left[(1-\xi^2 u_R^2)^{5/2} + (1-\xi^2 u_S^2)^{5/2}\right] + \frac{81Mq^4}{31360\xi^7} \left[(1-\xi^2 u_R^2)^{7/2} + (1-\xi^2 u_S^2)^{7/2}\right] \\
& + \left(\frac{15M^2}{4\xi^2} - \frac{3q^2}{4\xi^2} - \frac{279M^2 q^2}{32\xi^4} + \frac{15q^4}{64\xi^4} + \frac{2565M^2 q^4}{256\xi^6} + \frac{7q^4 \mu}{64\xi^6} + \frac{2889M^2 q^4 \mu}{512\xi^8}\right) \\
& \times \left[\pi - \arcsin(\xi u_R) - \arcsin(\xi u_S)\right].
\label{eqn.27}
\end{split}
\end{equation}

In the expression deduced above for the surface integral of $\mathcal{K}$ over the closed quadrilateral, $u_R$ and $u_S$ denote the reciprocal of the distance of the receiver and the light source respectively from the black hole. Also, we have used $\cos \phi_R = - \sqrt{1 - \xi^2 u_R^2}$ and $\cos \phi_S = \sqrt{1 - \xi^2 u_S^2}$ \cite{ono}. Now, in the asymptotically far distance limit $u_R \rightarrow 0$ and $u_S \rightarrow 0$, Eq.~(\ref{eqn.27}) reduces to
\begin{equation}
\begin{split}
- \int\!\! \int_{\stackrel{\infty}{R}\square\stackrel{\infty}{S}} \mathcal{K} dS 
& = \frac{4M}{\xi} - \frac{32Mq^2}{3\xi^3} + \frac{32Mq^4}{5\xi^5} + \frac{5467Mq^4\mu}{1600\xi^7} + \frac{81Mq^4}{15680\xi^7} \\
& + \left[\frac{15M^2}{4\xi^2} - \frac{3q^2}{4\xi^2} - \frac{279M^2 q^2}{32\xi^4} + \frac{15q^4}{64\xi^4} + \frac{2565M^2 q^4}{256\xi^6} + \frac{7q^4 \mu}{64\xi^6} + \frac{2889M^2 q^4 \mu}{512\xi^8}\right]\pi.
\label{eqn.28}
\end{split}
\end{equation}
Thus, the angle of deflection of light by a charged black hole in MEH gravitatioanl theory can be arrived at by substituting the above expression in Eq.~(\ref{eqn.14}) as
\begin{equation}
\begin{split}
\hat{\alpha} 
& = \frac{4M}{\xi} - \frac{32Mq^2}{3\xi^3} + \frac{32Mq^4}{5\xi^5} + \frac{5467Mq^4\mu}{1600\xi^7} + \frac{81Mq^4}{15680\xi^7} \\
& + \left[\frac{15M^2}{4\xi^2} - \frac{3q^2}{4\xi^2} - \frac{279M^2 q^2}{32\xi^4} + \frac{15q^4}{64\xi^4} + \frac{2565M^2 q^4}{256\xi^6} + \frac{7q^4 \mu}{64\xi^6} + \frac{2889M^2 q^4 \mu}{512\xi^8}\right]\pi.
\label{eqn.29}
\end{split}
\end{equation}

From this relation, it is clear that if we put $q = \mu = 0$, we arrive at the Schwarzschild spacetime which reads
\begin{equation}
\hat{\alpha} \simeq \frac{4M}{\xi} + \frac{15\pi M^2}{4\xi^2}.
\label{eqn.30}
\end{equation}

\begin{figure}[h!]
\includegraphics[scale=0.33]{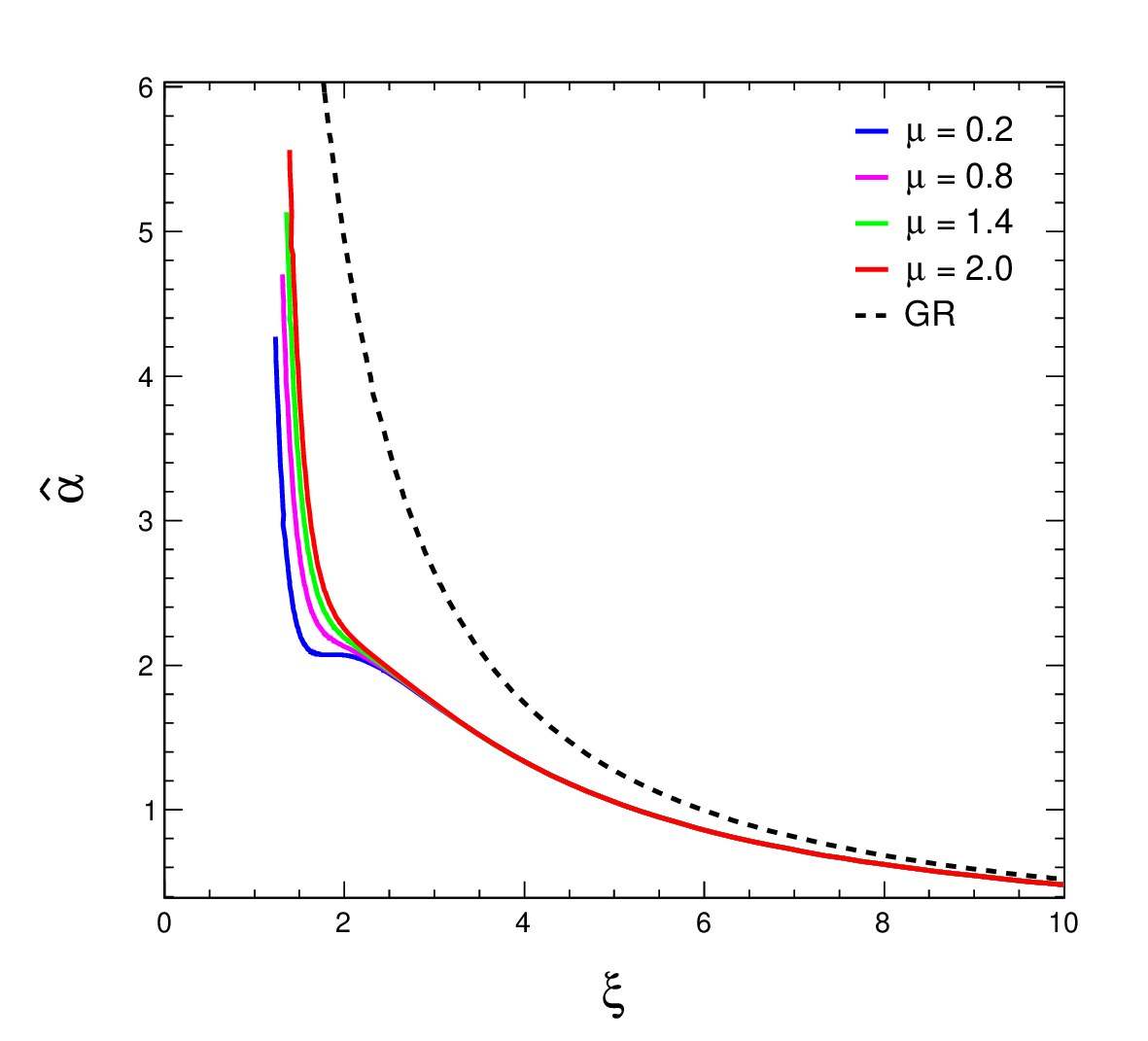}\hspace{2mm}
\includegraphics[scale=0.33]{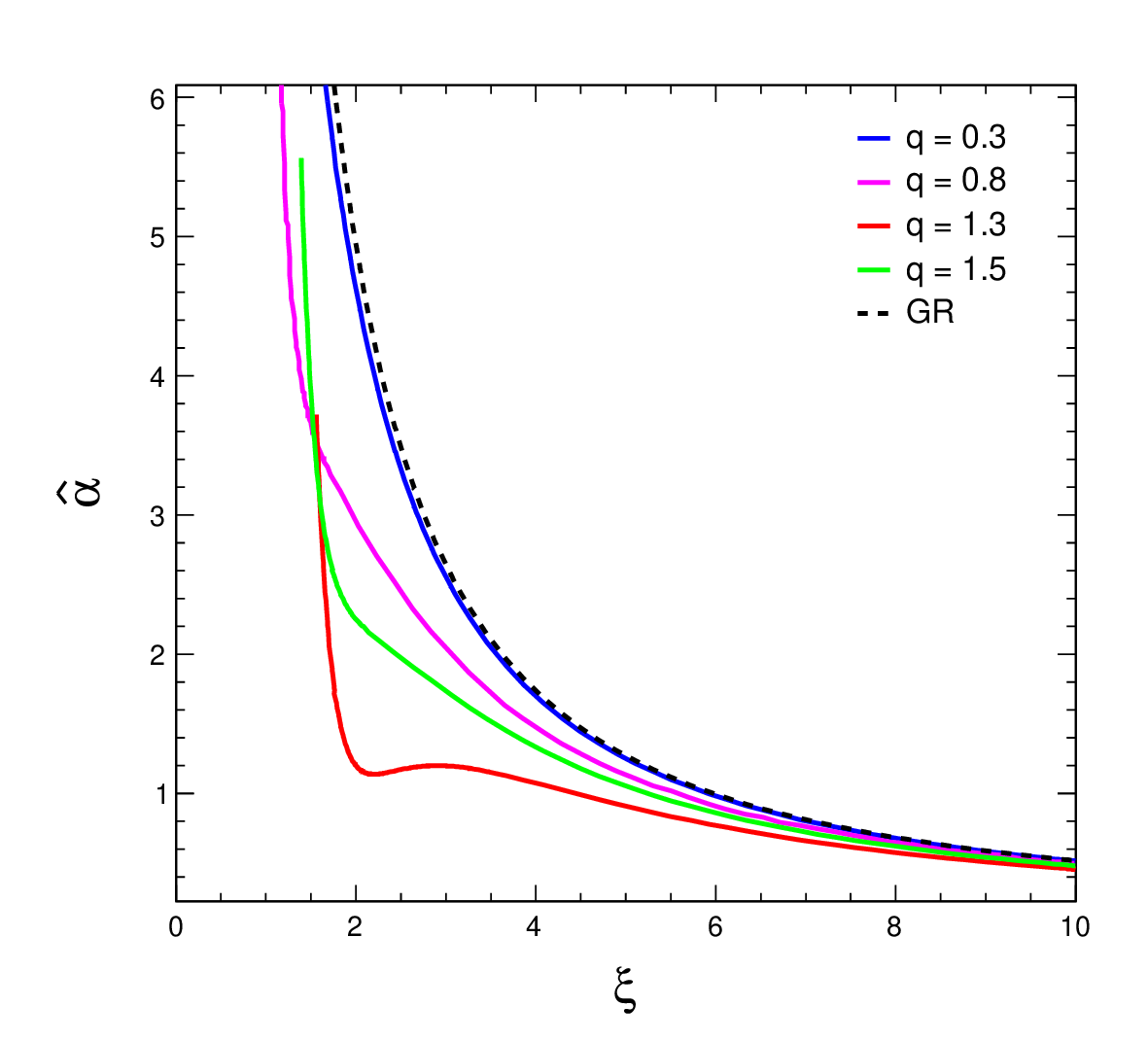}
\caption{Deflection angle as a function of impact parameter for a black hole in MEH theory for different values of the EH parameter, $\mu$ (in the first panel) and the charge parameter, $q$ (in the second panel).}
 \label{fig.2}
\end{figure}

In Figure~\ref{fig.2}, we portray the deflection angle formed as a consequence of the gravitational bending of light in the weak field of the black hole in MEH theory. The deflection angle is examined as a function of the impact parameter and the effects of the EH parameter $\mu$ and the charge parameter $q$ on the deflection angle are analyzed. It is evident from the two graphical representations that for each plot the computed results are compared with that of a black hole in GR theory. It can be inferred from the figures that the deflection angle decreases with increase in the impact parameter. Infact, it is evident that the deflection angle pattern is quite similar to that in GR except for a slight dip in the deflection angle for small impact parameter values. The first panel is portrayed for different values of $\mu$ while the second panel is for different values of $q$. It is seen that for higher values of $\mu$ this small dip tends to disappear and the pattern becomes analogous to that in GR as $\mu$ increases. On the contrary, higher values of $q$ increases the dip in the deflection angle and for very small values of $q$, the pattern of deflection angle almost overlaps with that in GR. Thus, it can be said that by increasing the EH parameter and decreasing the charge parameter, it is possible to recover the Schwazschild behaviour for the black hole in MEH theory.

\section{Einstein Ring, Magnification and Distortion}
\label{sec.4}

Gravitational lensing produces a fascinating optical phenomenon known as the Einstein ring, whose morphology and behavior depend strongly on the source-lens alignment as well as the surrounding dark matter distribution. A complete Einstein ring is observed when a background light source is positioned perfectly behind the lens. In this study, we investigate the influence of the charge parameter as well as the EH parameter on the resulting Einstein ring. To establish our geometric framework, let us define $D_{RS}$ as the distance between the light source and the receiver, $D_{LS}$ as the distance between the lens and the light source, and $D_{LR}$ as the distance between the receiver and the lens. With these notations the lens equation can be expressed as \cite{bozza}
\begin{equation}
D_{RS}\tan\beta = \frac{D_{LR}\sin\theta - D_{LS}\sin(\hat{\alpha} - \theta)}{\cos(\hat{\alpha} - \theta)}.
\label{eqn.31}
\end{equation}
In this context, $\beta$ represents the angular position of the source, and $\theta$ represents the angular position of the lensed image of the source as detected by the 
receiver. Furthermore, in the regime of weak gravitational lensing, if the source and observer are at infinite distances from one another, the above lens equation reduces to \cite{papnoi},
\begin{equation}
\beta = \theta - \frac{D_{LS}}{D_{RS}}\hat{\alpha}.
\label{eqn.32}
\end{equation}
An Einstein ring is developed under the condition $\beta = 0$ which indicates that the light source, the receiver and the lens are perfectly aligned. Thus the above relation~(\ref{eqn.32}) results in
\begin{equation}
\Theta_\mathrm{E} = \frac{D_{LS}}{D_{RS}}\hat{\alpha}.
\label{eqn.33}
\end{equation}
Substituting Eq.~(\ref{eqn.29}) into the above equation, we arrive at the angular radius of the Einstein ring as
\begin{equation}
\begin{split}
\Theta_\mathrm{E} &= \frac{D_{LS}}{D_{RS}}\bigg[\frac{4M}{\xi} - \frac{32Mq^2}{3\xi^3} + \frac{32Mq^4}{5\xi^5} + \frac{5467Mq^4\mu}{1600\xi^7} + \frac{81Mq^4}{15680\xi^7} \\
& + \left(\frac{15M^2}{4\xi^2} - \frac{3q^2}{4\xi^2} - \frac{279M^2 q^2}{32\xi^4} + \frac{15q^4}{64\xi^4} + \frac{2565M^2 q^4}{256\xi^6} + \frac{7q^4 \mu}{64\xi^6} + \frac{2889M^2 q^4 \mu}{512\xi^8}\right)\pi \bigg].
\label{eqn.34}
\end{split}
\end{equation}

Again, in the weak field gravitational lensing, we observe a small Einstein ring. Hence, the impact parameter satisfies the relation, $\xi = D_{LR}\sin\Theta_\mathrm{E} \approx D_{LR} \Theta_\mathrm{E}$. Hence, we obtain the final expression of the angular radius of the Einstein ring which reads
\begin{equation}
\begin{split}
\Theta_\mathrm{E} &= \sqrt{\frac{D_{LS}}{D_{RS} D_{LR}}}\bigg[4M - \frac{32Mq^2}{3\xi^2} + \frac{32Mq^4}{5\xi^4} + \frac{5467Mq^4\mu}{1600\xi^6} + \frac{81Mq^4}{15680\xi^6} \\
& + \left(\frac{15M^2}{4\xi} - \frac{3q^2}{4\xi} - \frac{279M^2 q^2}{32\xi^3} + \frac{15q^4}{64\xi^3} + \frac{2565M^2 q^4}{256\xi^5} + \frac{7q^4 \mu}{64\xi^5} + \frac{2889M^2 q^4 \mu}{512\xi^7}\right)\pi \bigg]^{1/2}.
\label{eqn.35}
\end{split}
\end{equation}

\begin{figure}[h!]
\includegraphics[scale=0.29]{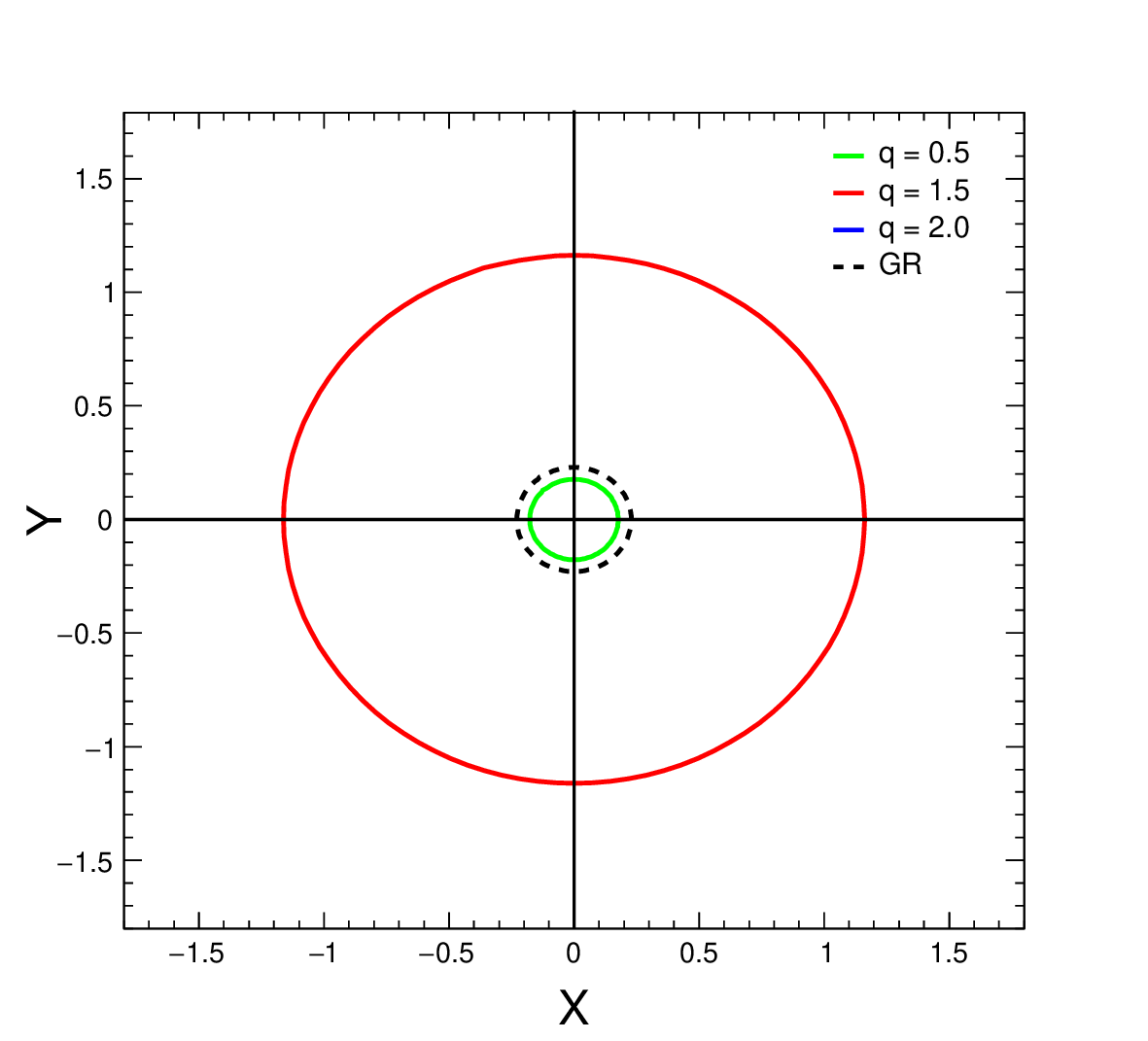}
\includegraphics[scale=0.29]{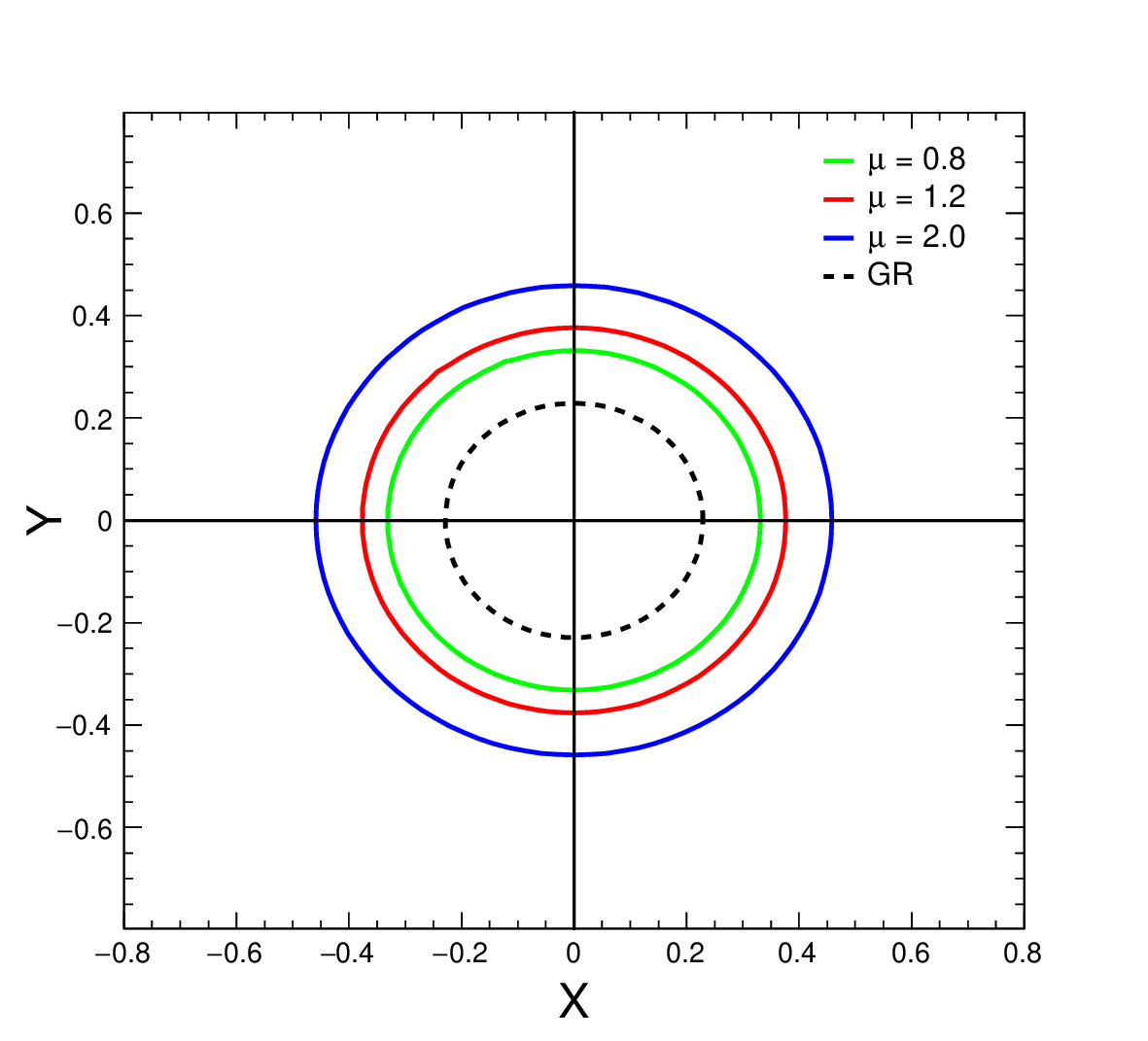}
\includegraphics[scale=0.29]{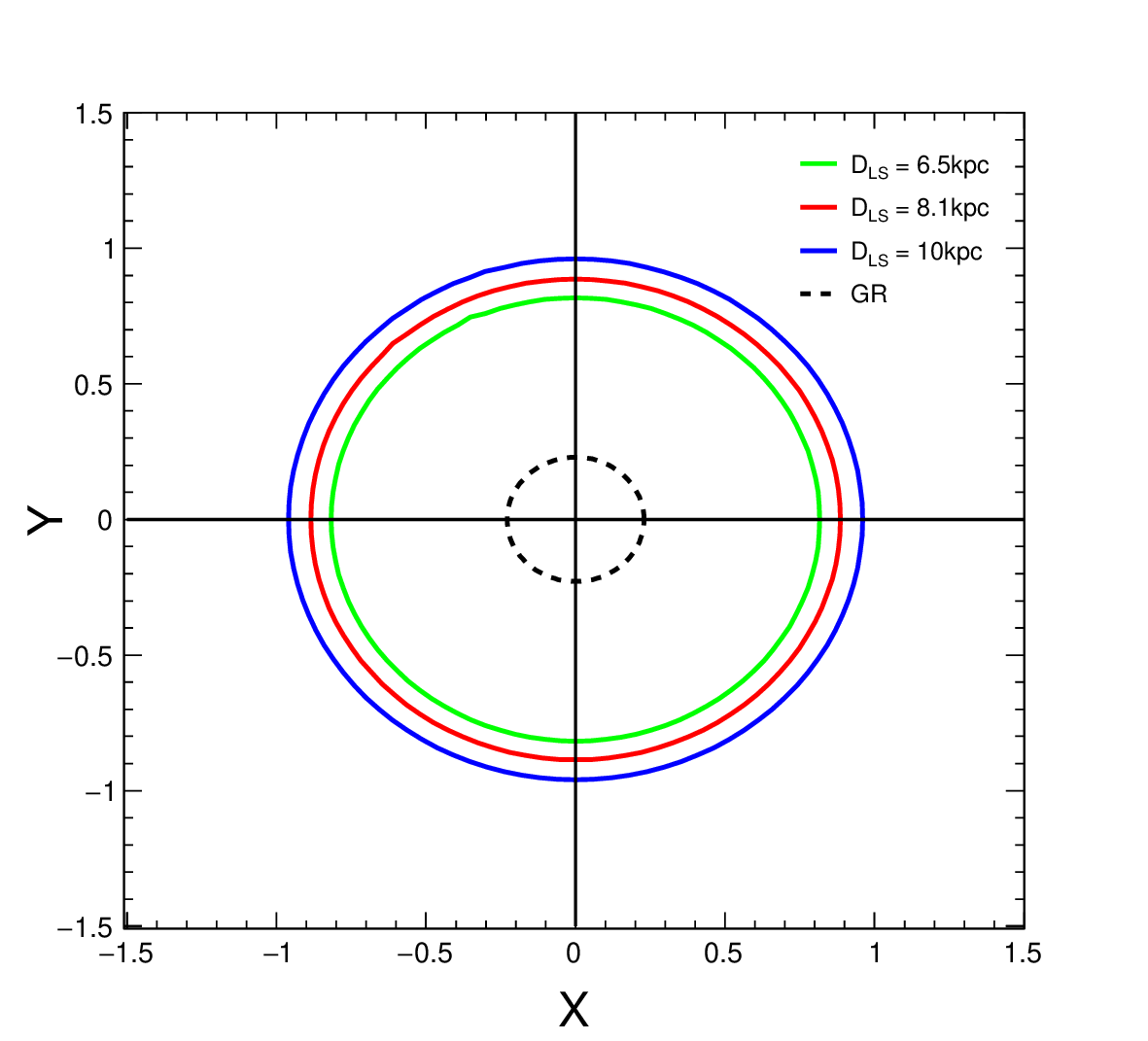}
\caption{The Einstein rings for different values of the charge parameter $q$ (first panel), the EH parameter $\mu$ (second panel) and the lens to source distance $D_{LS}$ (third panel) for a black hole in MEH theory.}
 \label{fig.3}
\end{figure}

Figure~\ref{fig.3} depicts the Einstein rings resulting from the lensing in the weak field limit of a black hole in MEH theory. First panel indicates the Einstein ring radius for different values of the charge parameter $q$. The second panel shows the Einstein rings for different values of the EH parameter $\mu$ and the third panel is for different lens to source distances $D_{LS}$. In each figure the black dashed rings correspond to the Einstein rings formed in GR. It is evident that the radii increases as the values of $q$, $\mu$ and $D_{LS}$ increases. Infact, for $q = 2$, the Einstein ring is much bigger and including it in the figure makes it difficult to depict the Einstein ring for other values of $q$. Hence, we have zoomed into the figure to represent the Einstein ring for $q = 0.5$ and $1.5$ and that for GR.

Next we investigate the image magnification which acts as another vital source of information. Ref.~\cite{virbhadra1} defines image magnification as the ratio of the flux of the lensed image to the ratio of the flux of the unlensed source. However, following Liouville's theorem which dictates that surface brightness remains conserved during deflection of light, the magnification factor simplifies to the ratio between the solid angles of the image and the source. Hence, the total magnification of an image due to lensing in a spherically symmetric spacetime can be written as
\begin{equation}
\zeta = \zeta_t \zeta_r,
\label{eqn.36}
\end{equation}
where, $\zeta_t$ and $\zeta_r$ are the tangential and radial magnifications respectively and are expressed as
\begin{equation}
\zeta_t = \left(\frac{\sin\beta}{\sin\theta}\right)^{-1} \hspace{5mm} ; \hspace{5mm} \zeta_r = \left(\frac{d\beta}{d\theta}\right)^{-1}.
\label{eqn.37}
\end{equation}

\begin{figure}[h!]
\includegraphics[scale=0.29]{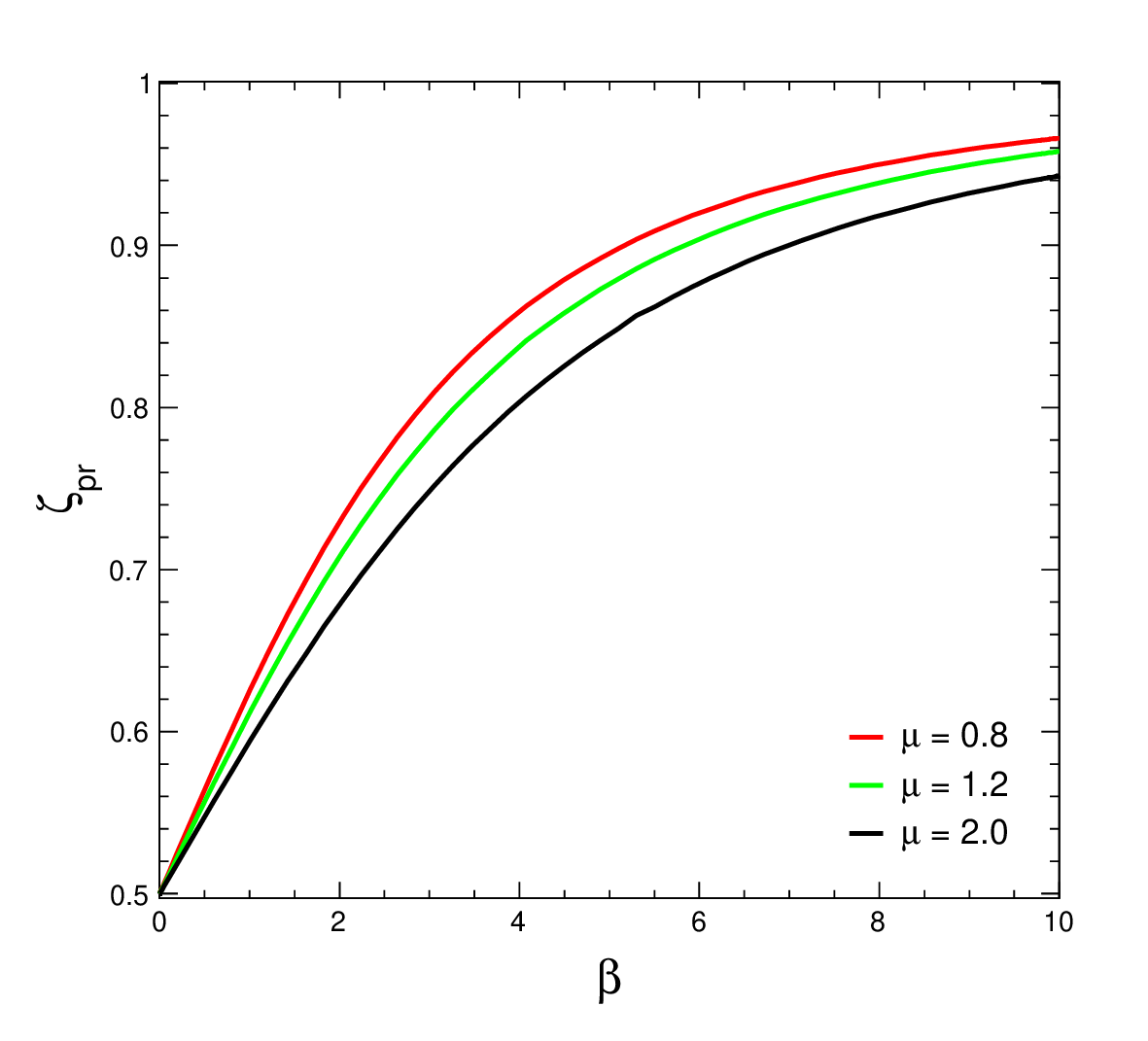}
\includegraphics[scale=0.29]{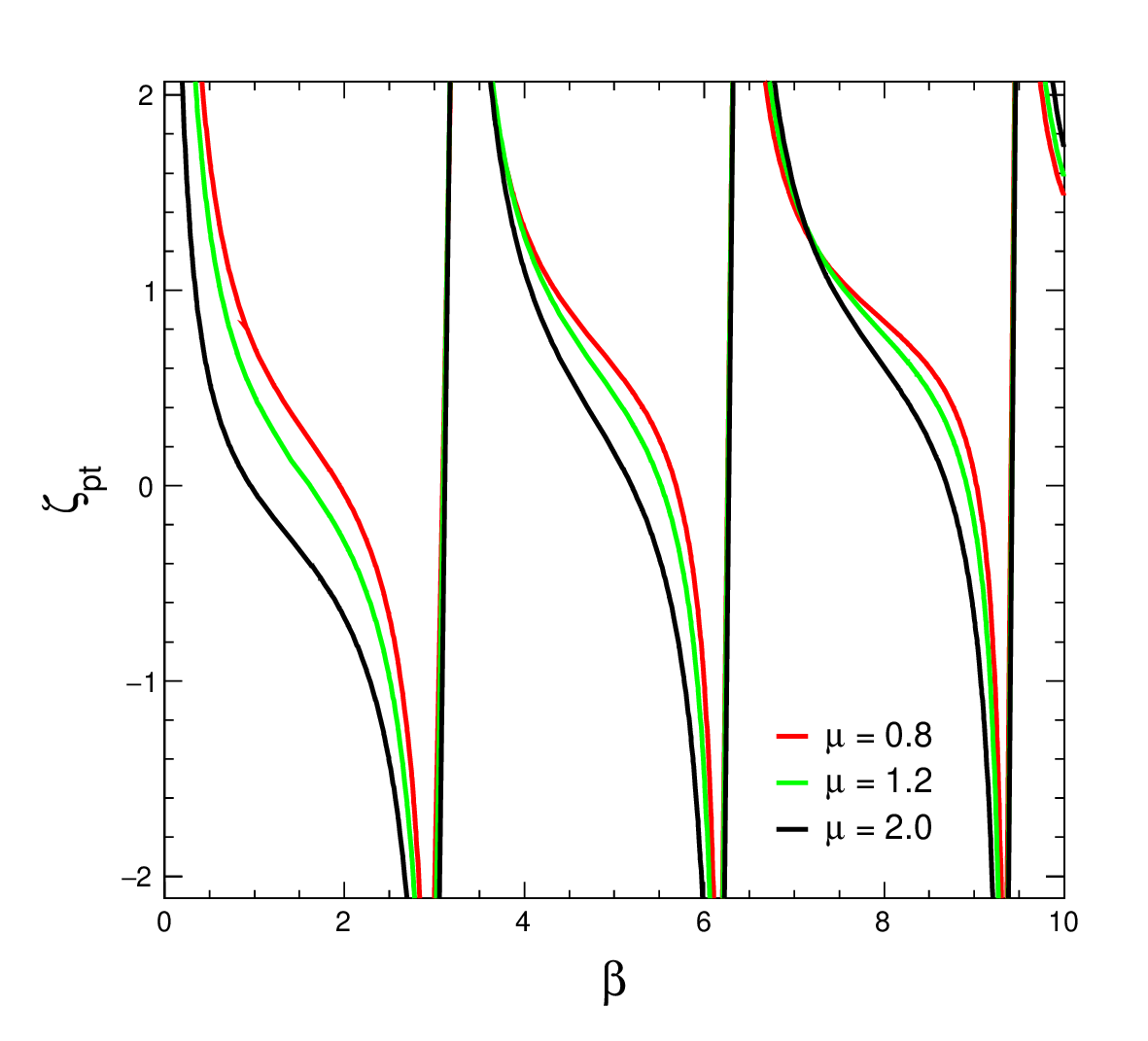}
\includegraphics[scale=0.29]{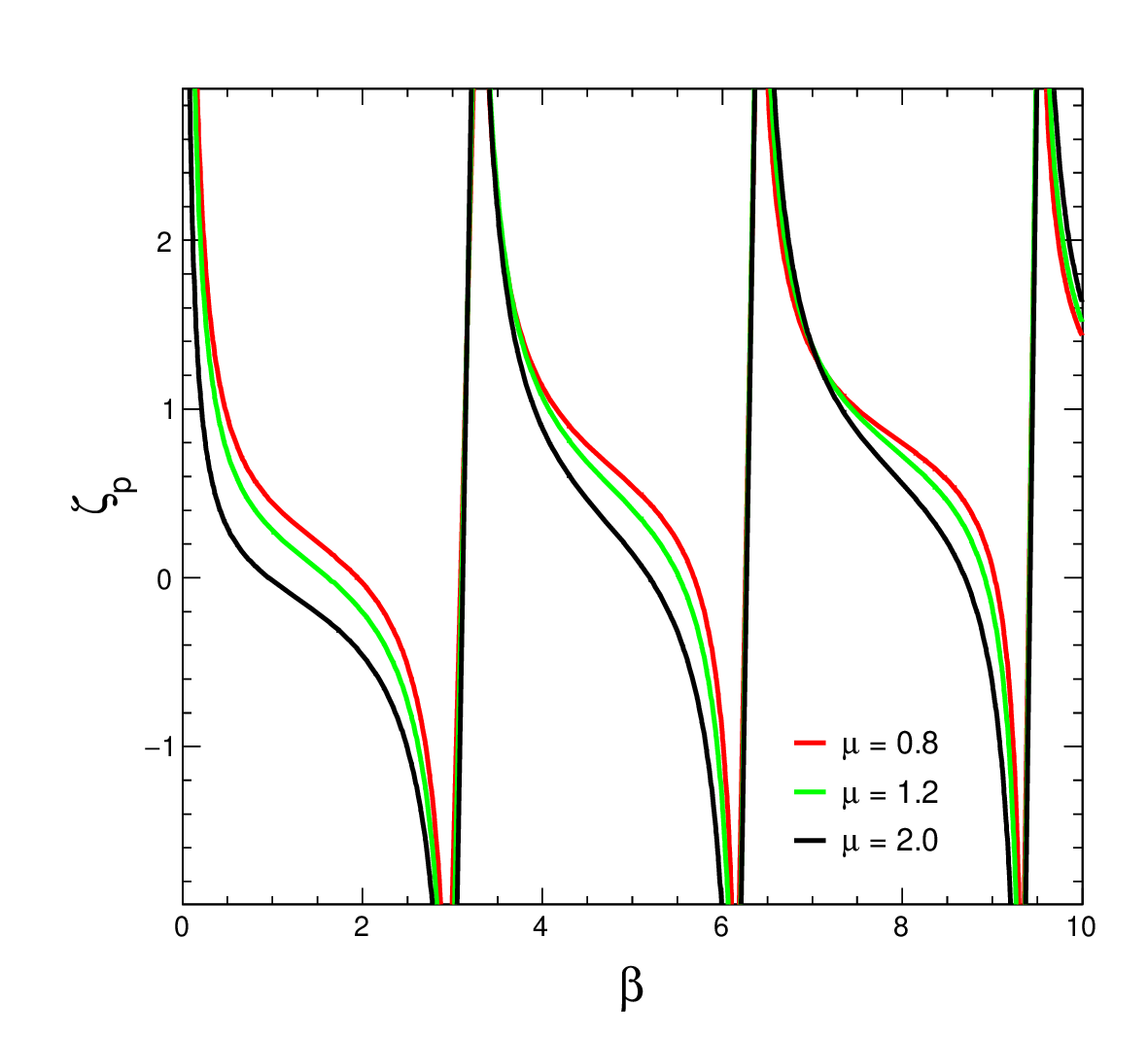}\\
\includegraphics[scale=0.29]{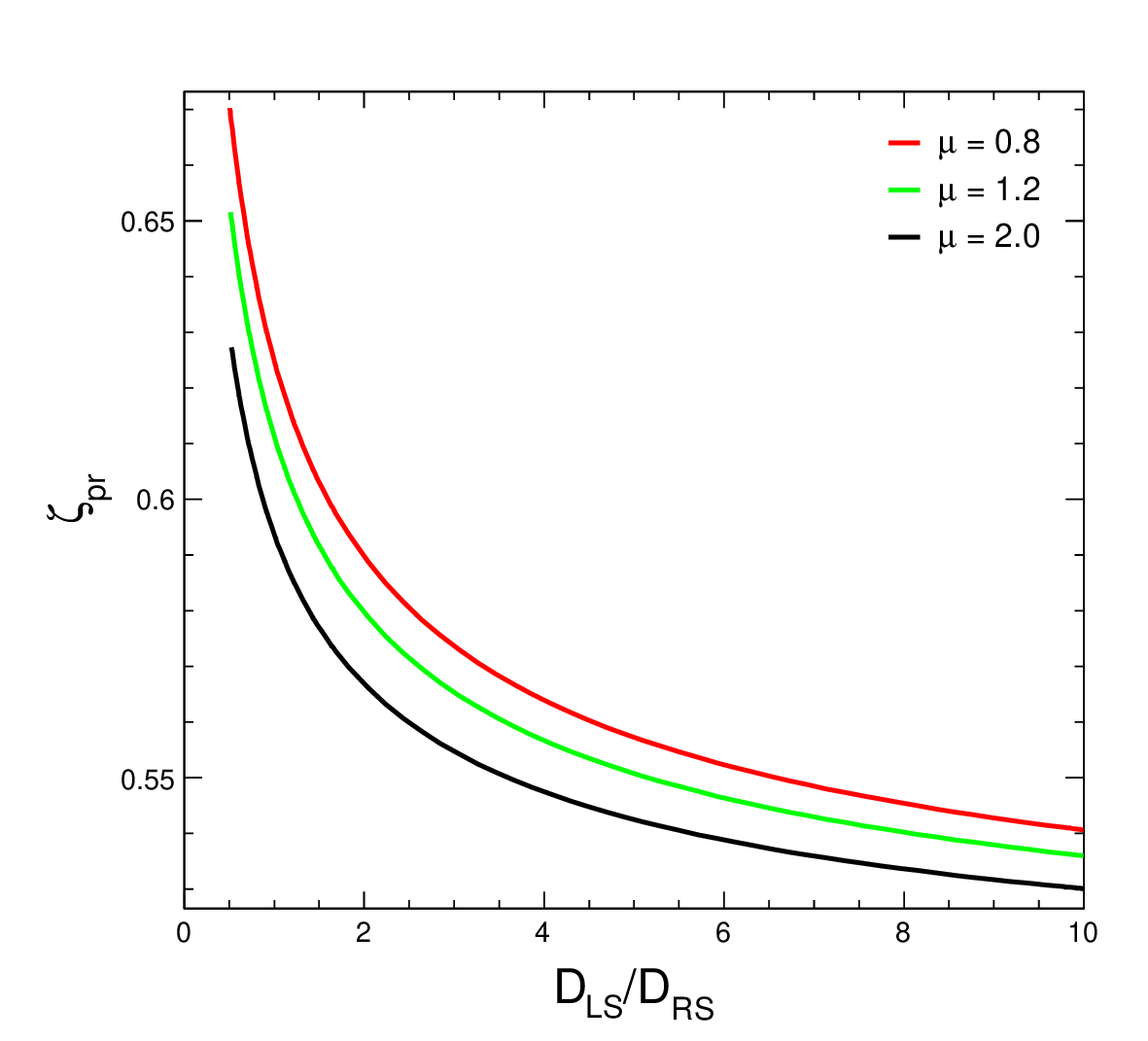}
\includegraphics[scale=0.29]{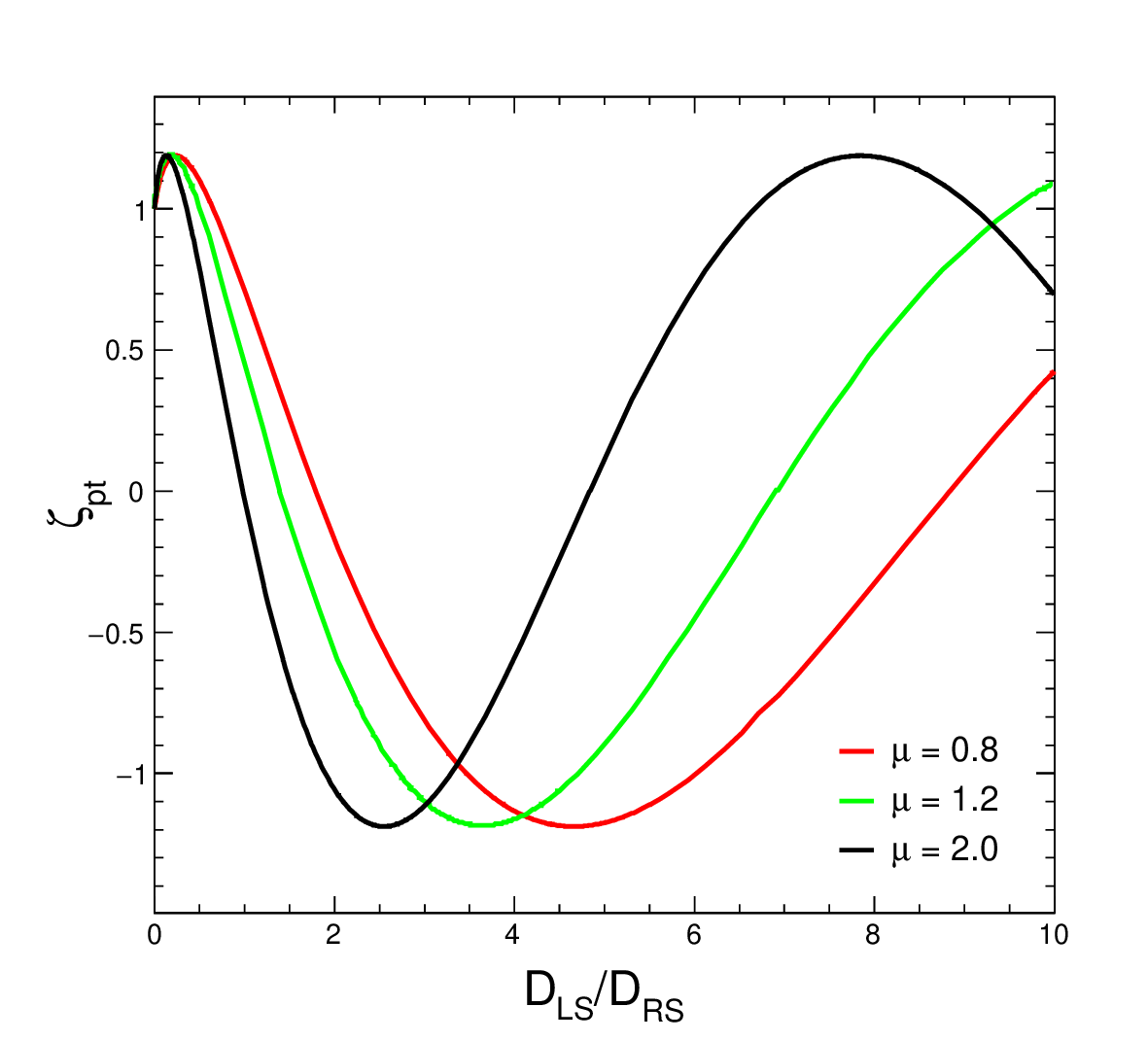}
\includegraphics[scale=0.29]{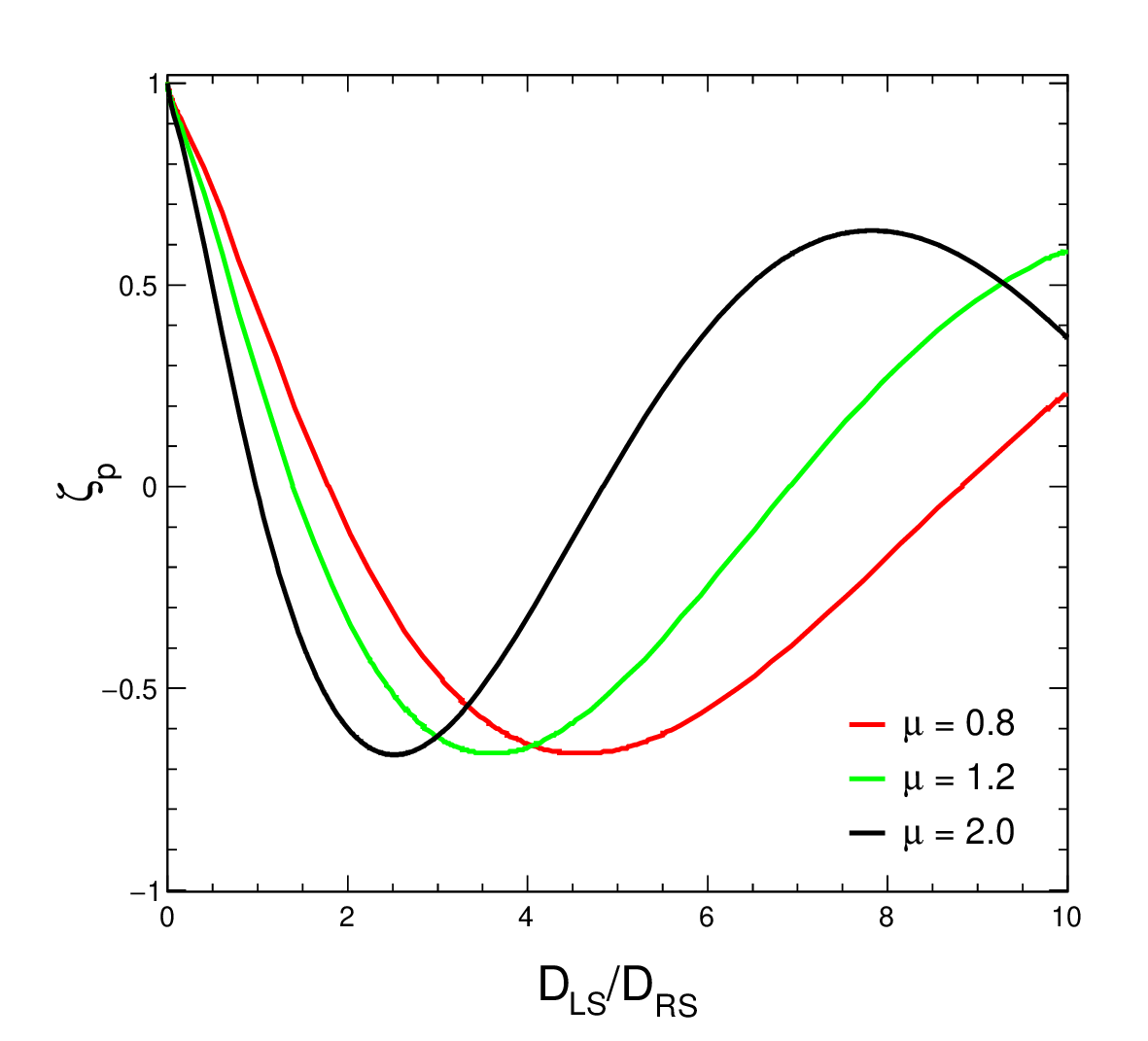}\\
\includegraphics[scale=0.29]{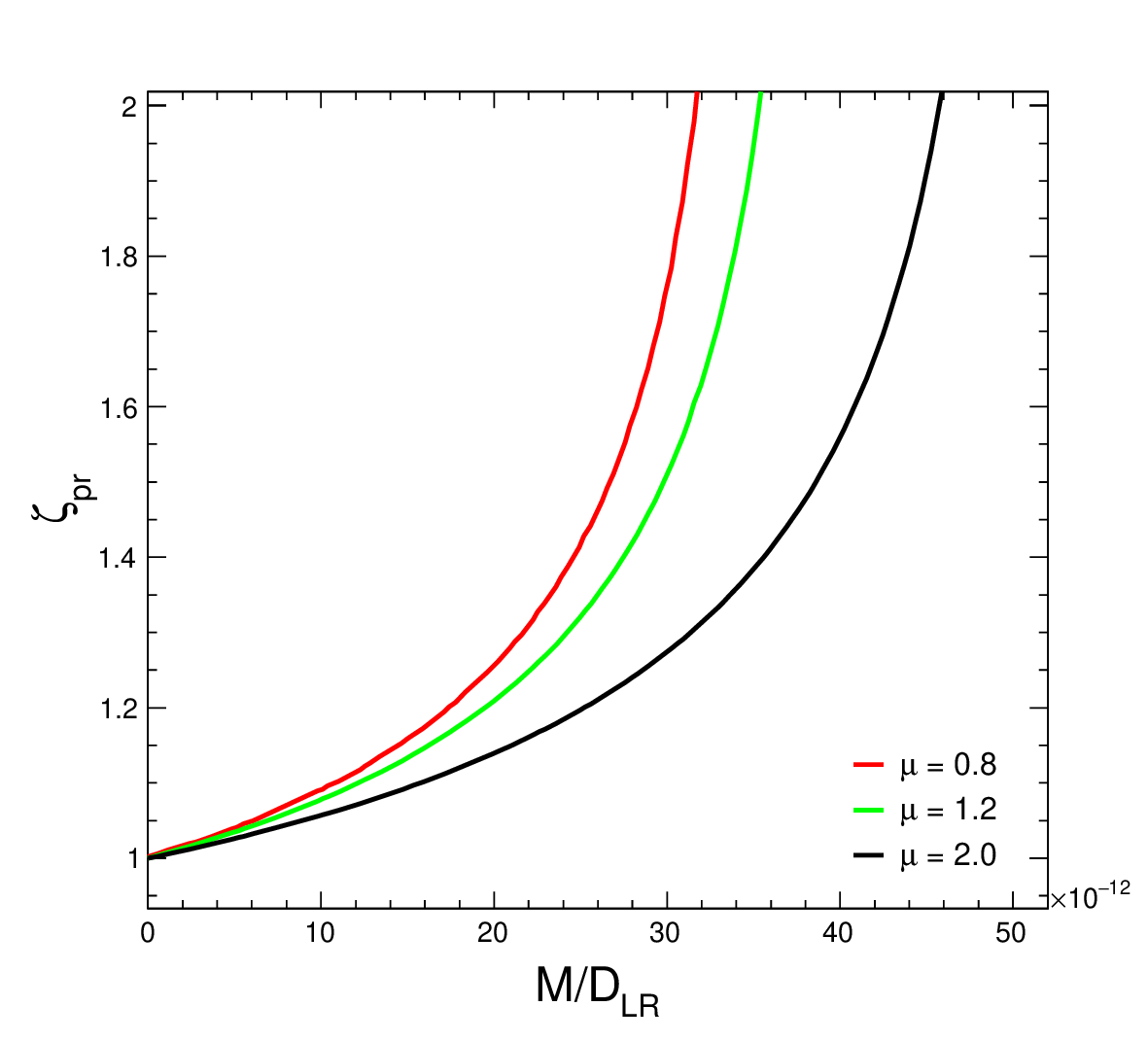}
\includegraphics[scale=0.29]{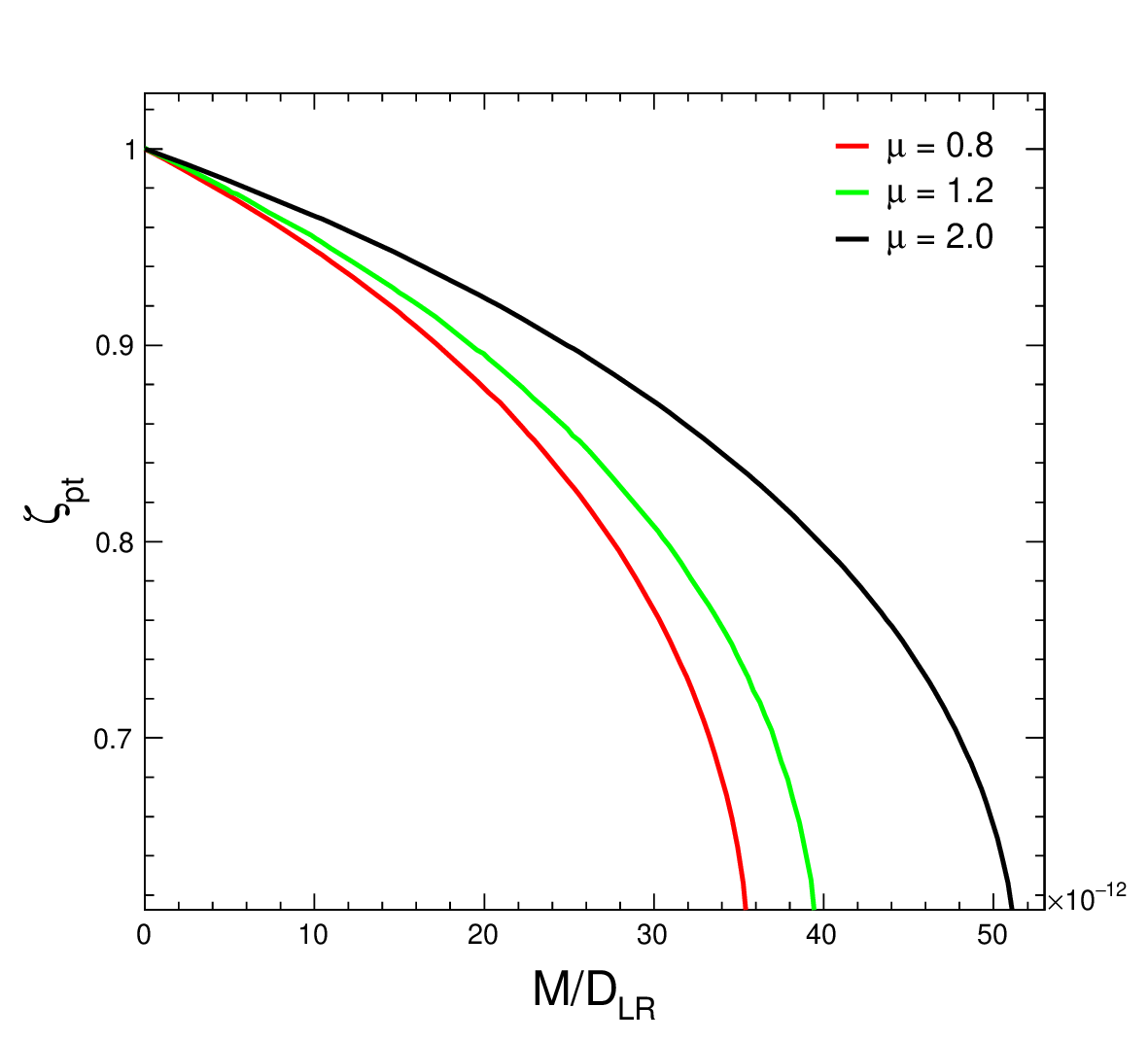}
\includegraphics[scale=0.29]{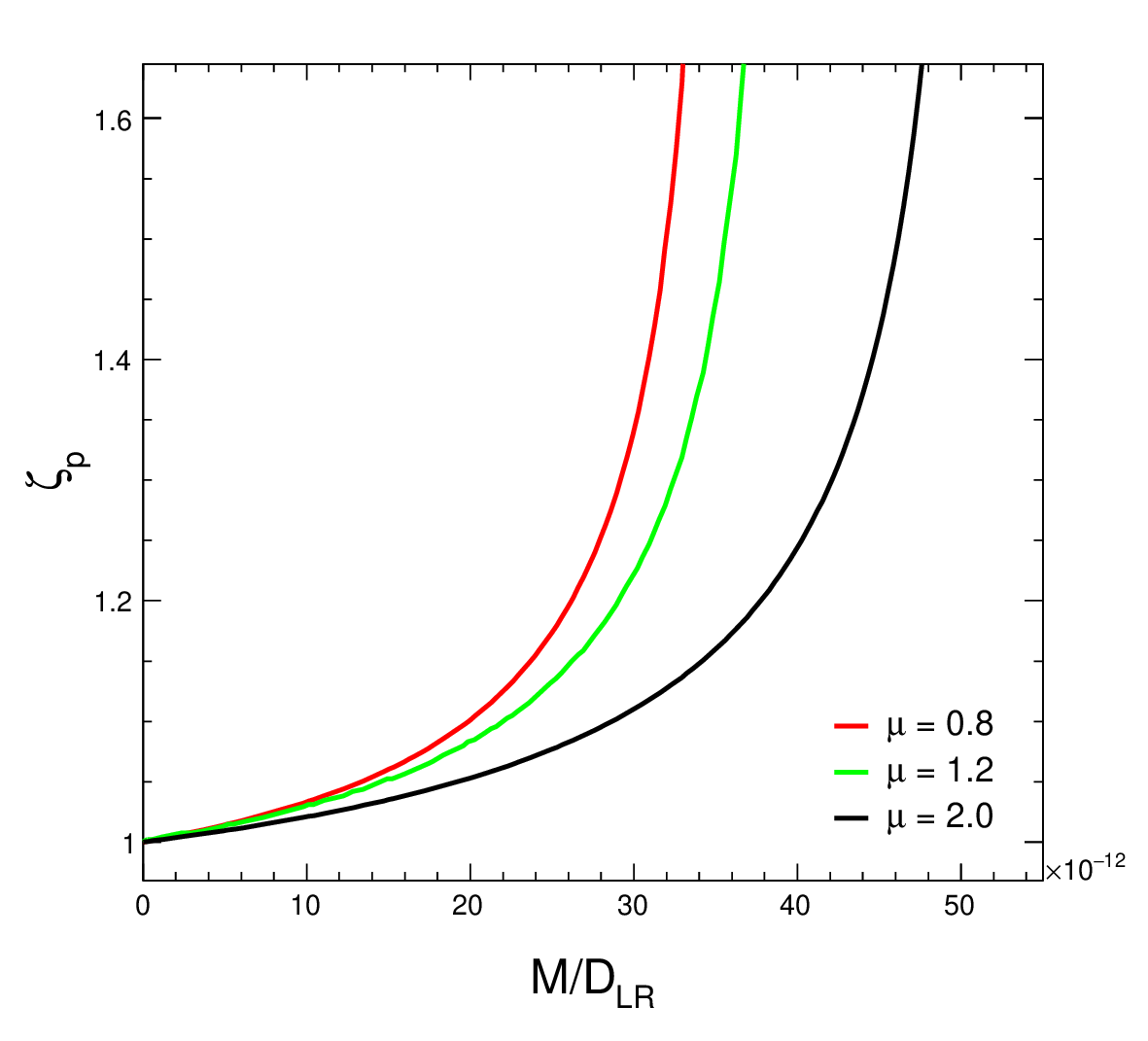}
\caption{The radial, tangential and total magnification of primary images are plotted a functions of $\beta$, $D_{LS}/D_{RS}$ and $M/D_{LR}$ for different values of $\mu$.}
 \label{fig.4}
\end{figure}

The distortion parameter reads \cite{virbhadra1}
\begin{equation}
\Delta = \frac{\zeta_t}{\zeta_r},
\label{eqn.38}
\end{equation}
which can be recast in a logarithmic form as
\begin{equation}
\delta = \log_{10} \left|\frac{\zeta_t}{\zeta_r}\right|.
\label{eqn.39}
\end{equation}

Now, following Refs.~\cite{schneider, schneider1} we can solve the lens equation~(\ref{eqn.32}) and arrive at the angular positions of the primary and secondary images respectively as
\begin{equation}
\begin{split}
\theta_p &= \frac{1}{2} \left(\beta + \sqrt{\beta^2 + 4\Theta_\mathrm{E}^2} \right) \\
\theta_s &= \frac{1}{2} \left(\beta - \sqrt{\beta^2 + 4\Theta_\mathrm{E}^2} \right) .
\end{split}
\label{eqn.40}
\end{equation}

\begin{figure}[h!]
\includegraphics[scale=0.29]{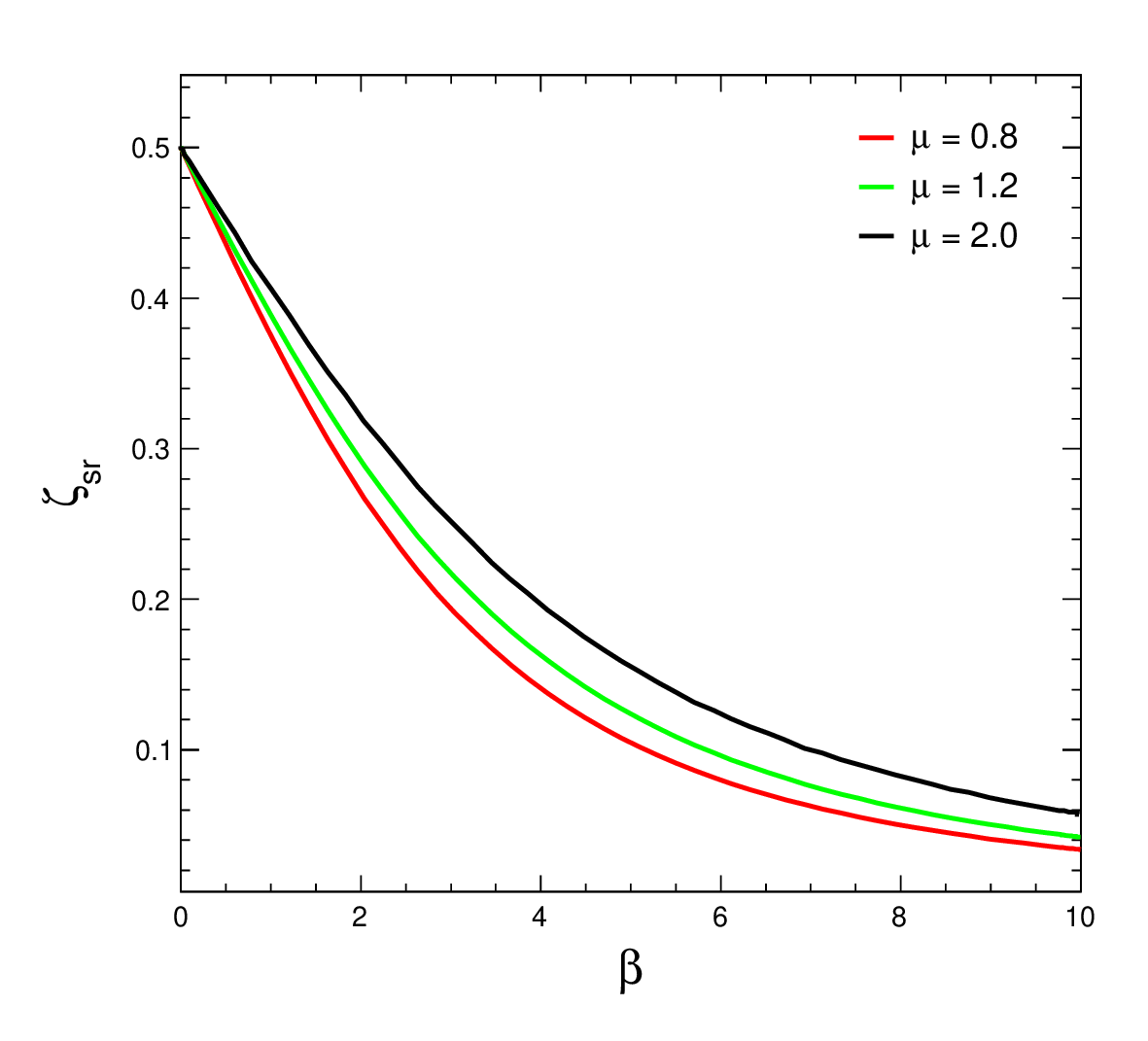}
\includegraphics[scale=0.29]{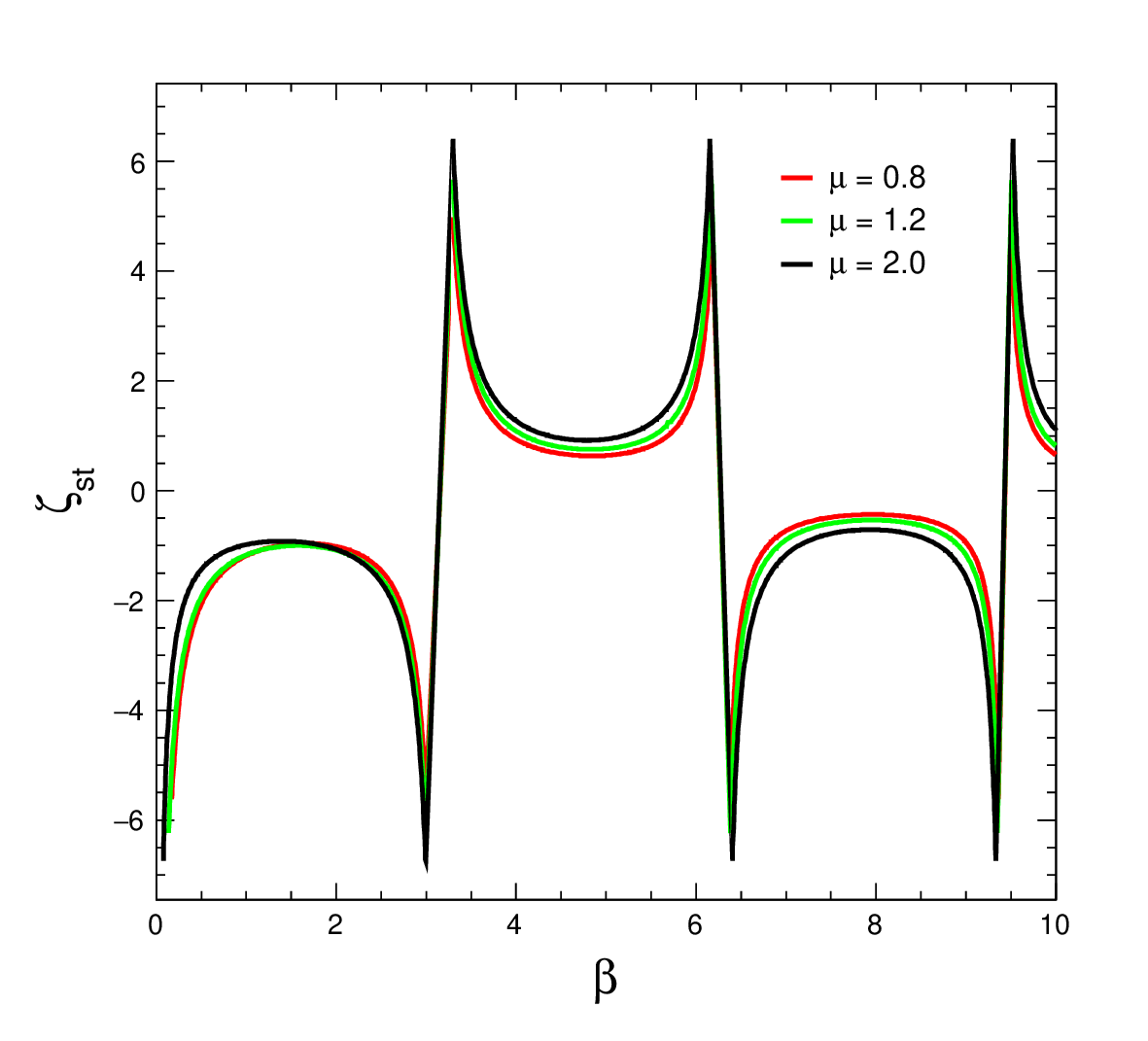}
\includegraphics[scale=0.29]{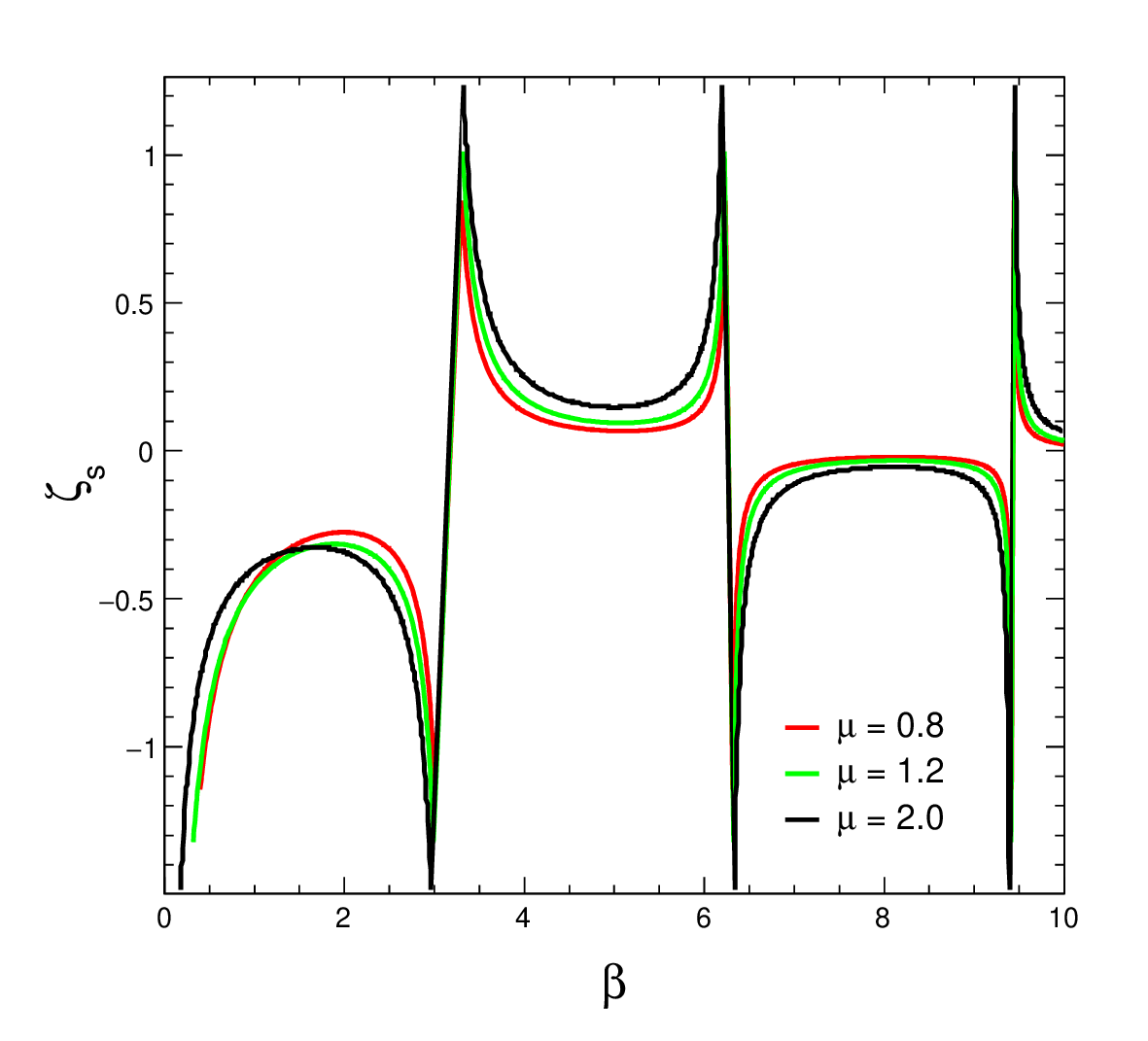}\\
\includegraphics[scale=0.29]{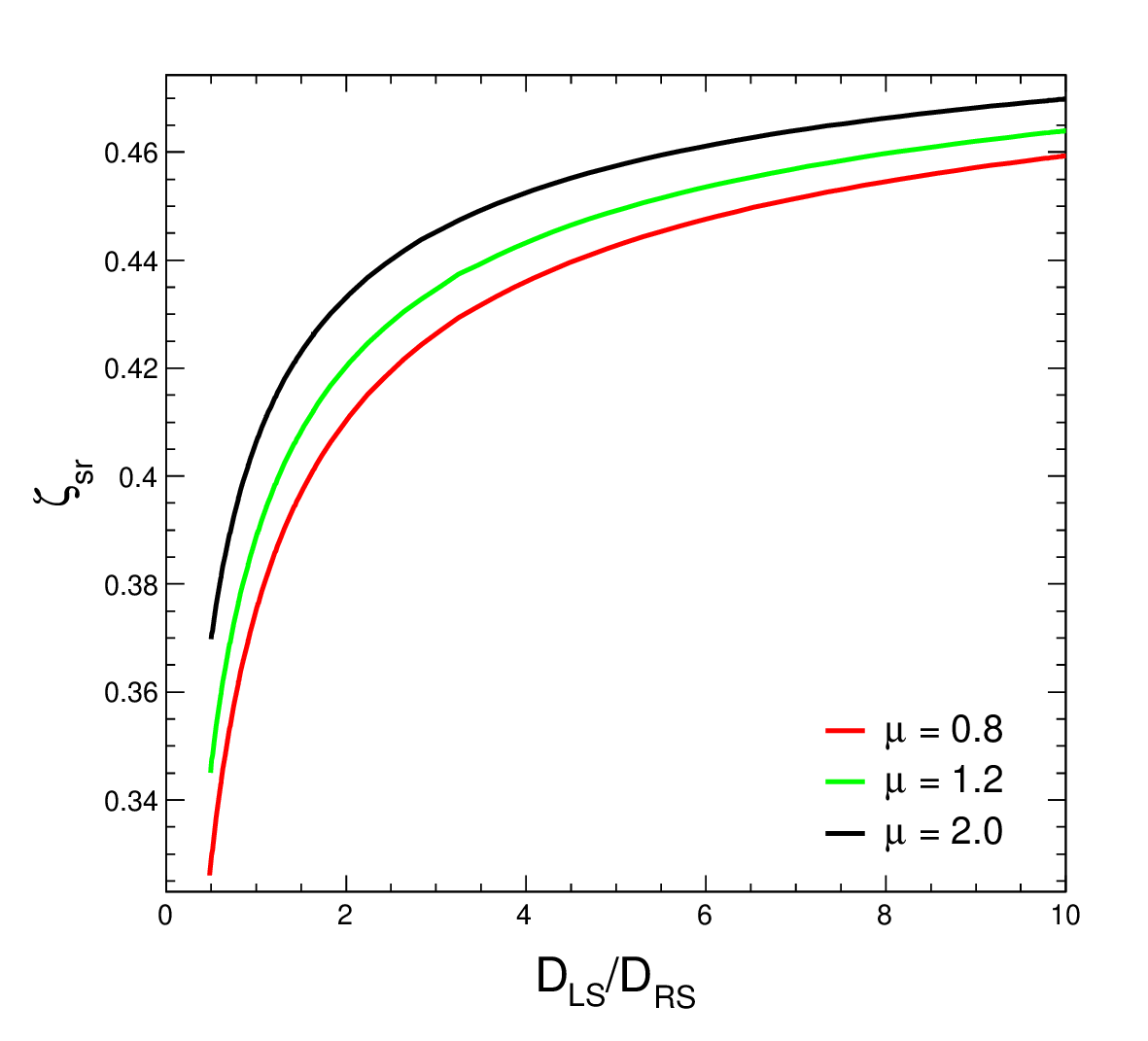}
\includegraphics[scale=0.29]{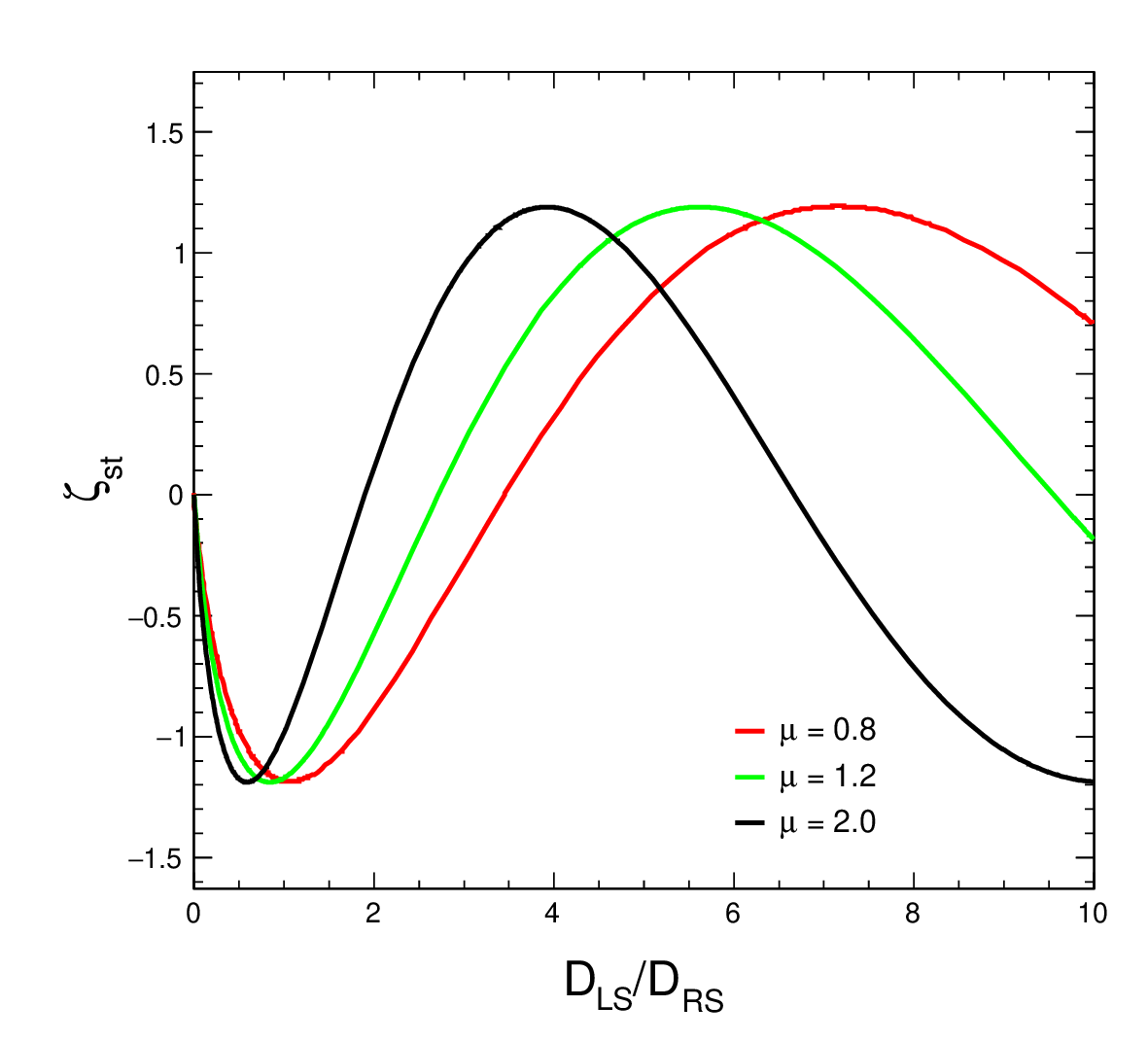}
\includegraphics[scale=0.29]{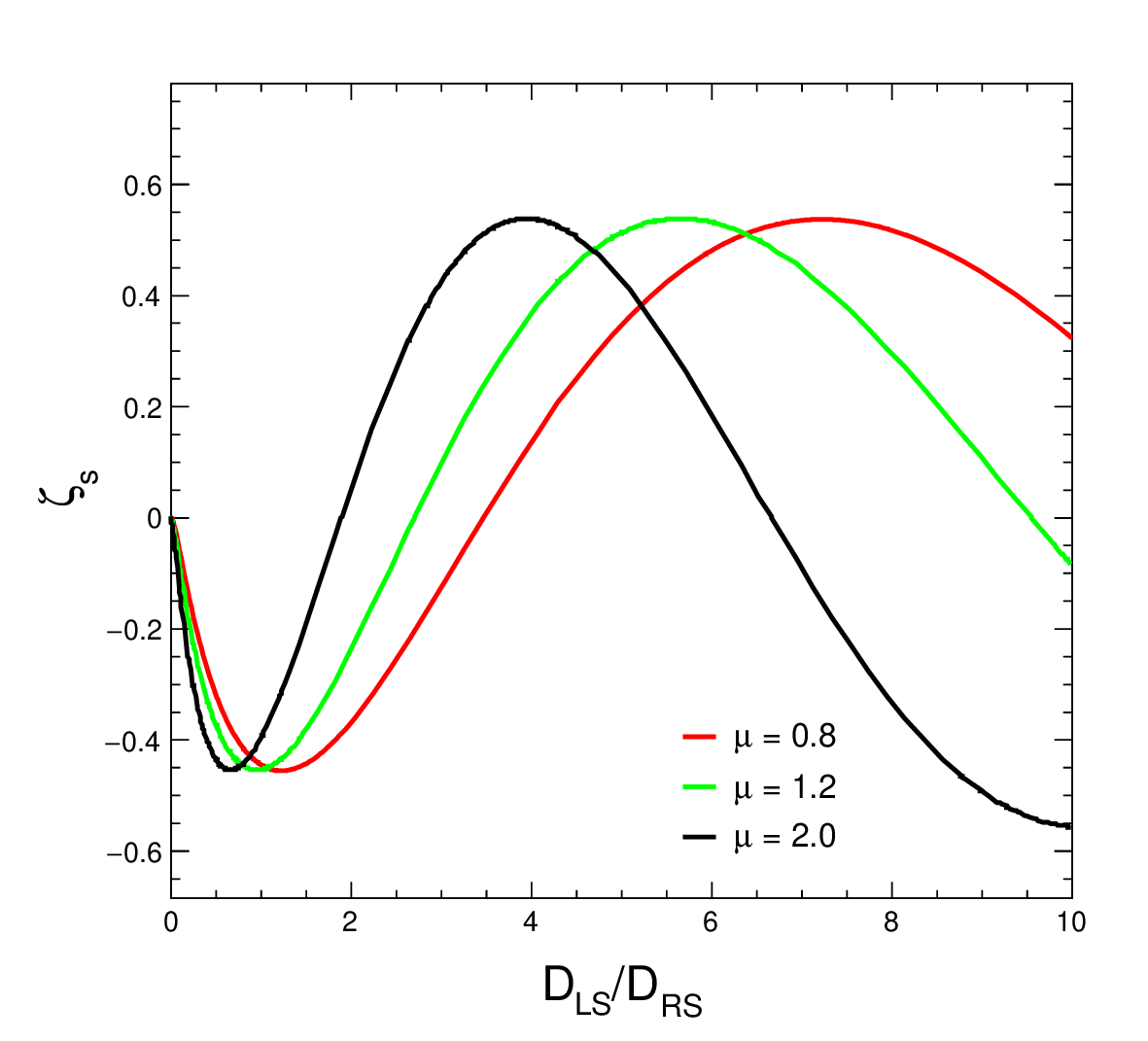}\\
\includegraphics[scale=0.29]{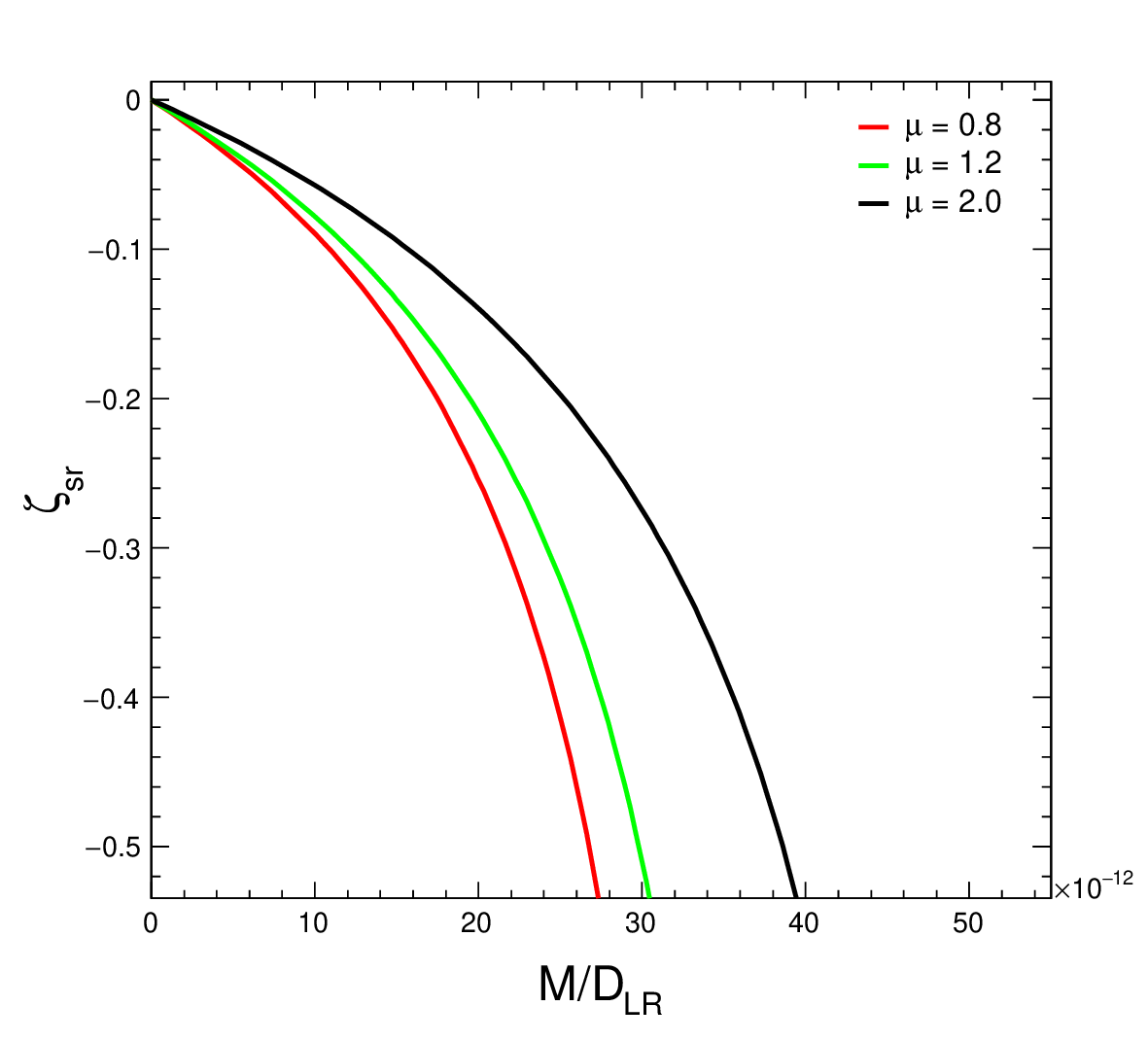}
\includegraphics[scale=0.29]{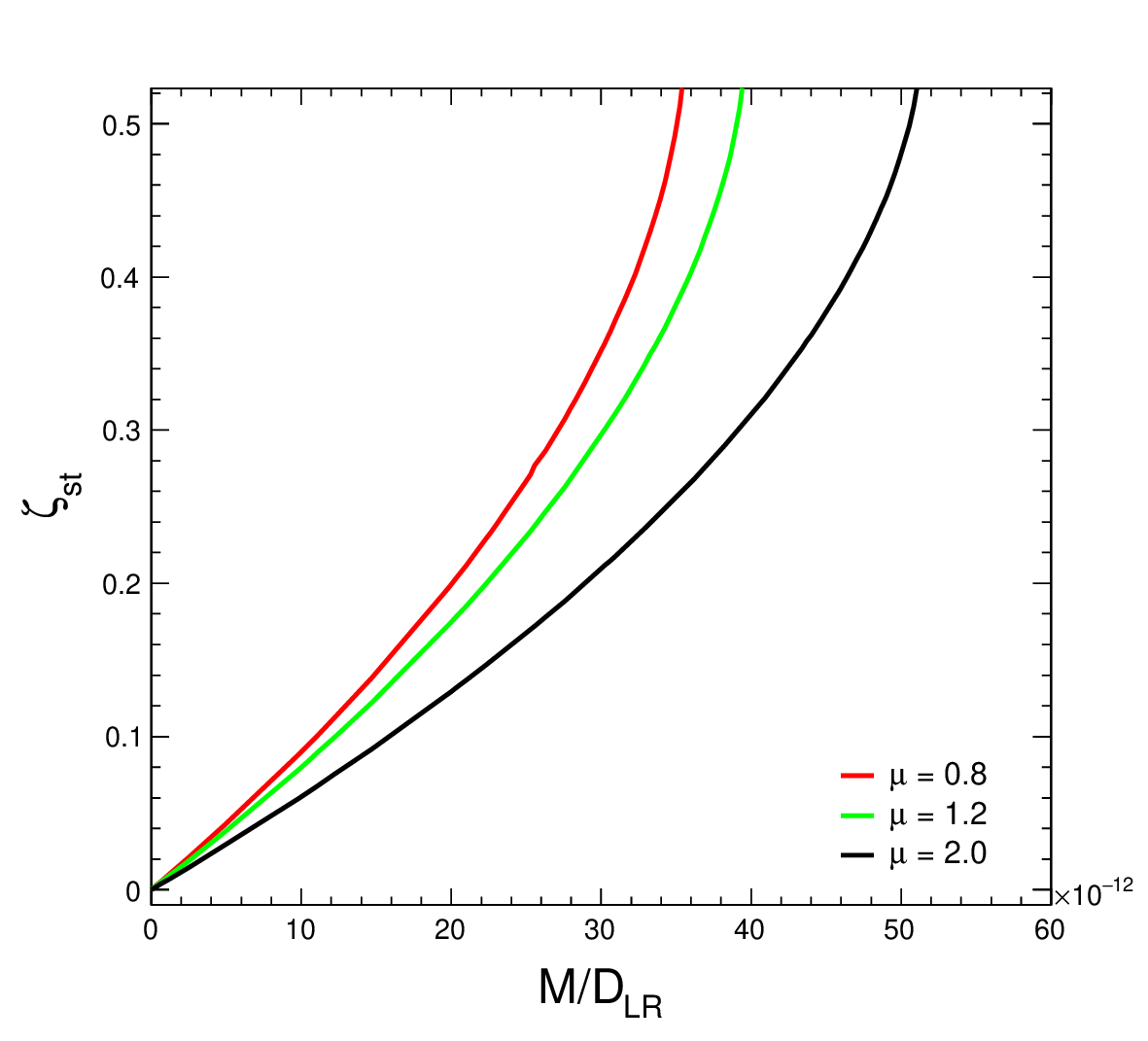}
\includegraphics[scale=0.29]{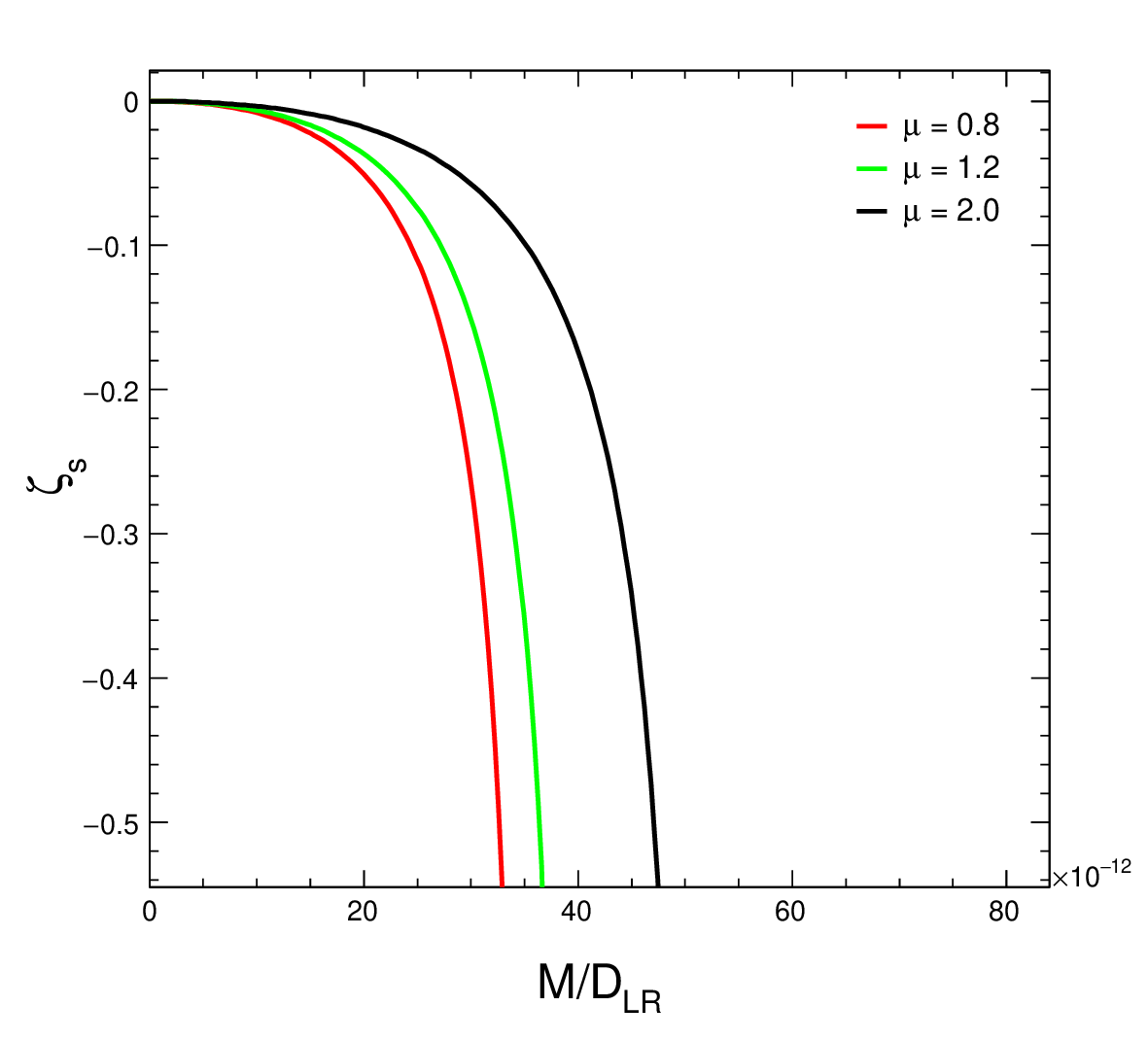}
\caption{The radial, tangential and total magnification of secondary images are plotted a functions of $\beta$, $D_{LS}/D_{RS}$ and $M/D_{LR}$ for different values of $\mu$.}
 \label{fig.5}
\end{figure}

Furthermore, the above relations can be used to deduce the expression for the distortion of the primary ($\Delta_p$) and secondary ($\Delta_s$) images. Substituting these equations in Eq.~(\ref{eqn.37}) and then using Eq.~(\ref{eqn.38}), we arrive at
\begin{equation}
\Delta_p = - \Delta_s = \left|\frac{\sqrt{\beta^2 + 16\frac{D_{LS}}{D_{RS}}\frac{M}{D_{LR}}}}{\beta}\right|.
\label{eqn.41}
\end{equation}
Accordingly, $\Delta = 0$ which is in agreement with the hypothesis put forward by Virbhadra in Ref.~\cite{virbhadra1}. It suggests that in weak field gravitational lensing, distortions in image positions cancel each other out, resulting in a vanishing net distortion, i.e. $\Delta = 0$.

\begin{figure}[h!]
\includegraphics[scale=0.29]{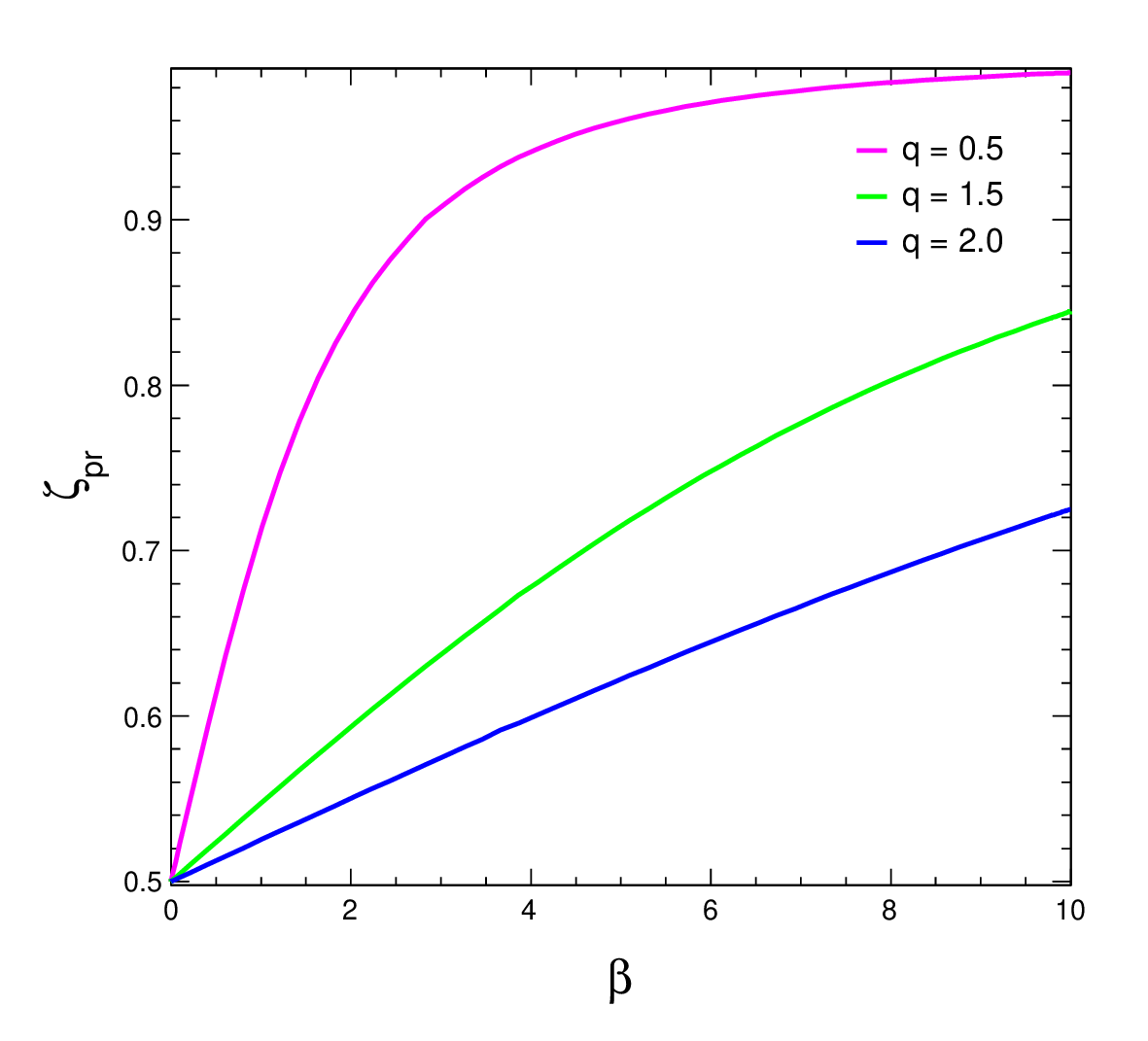}
\includegraphics[scale=0.29]{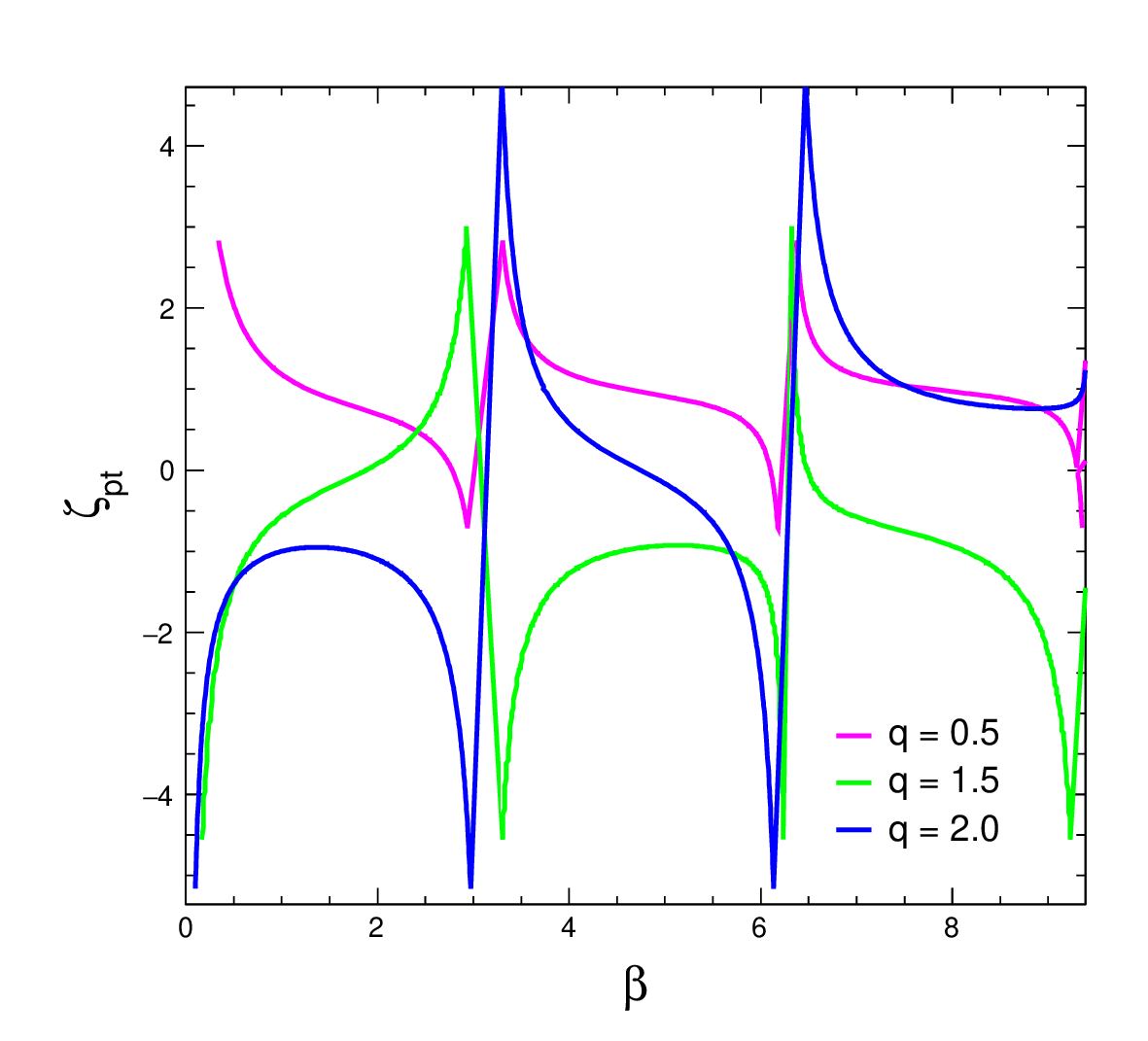}
\includegraphics[scale=0.29]{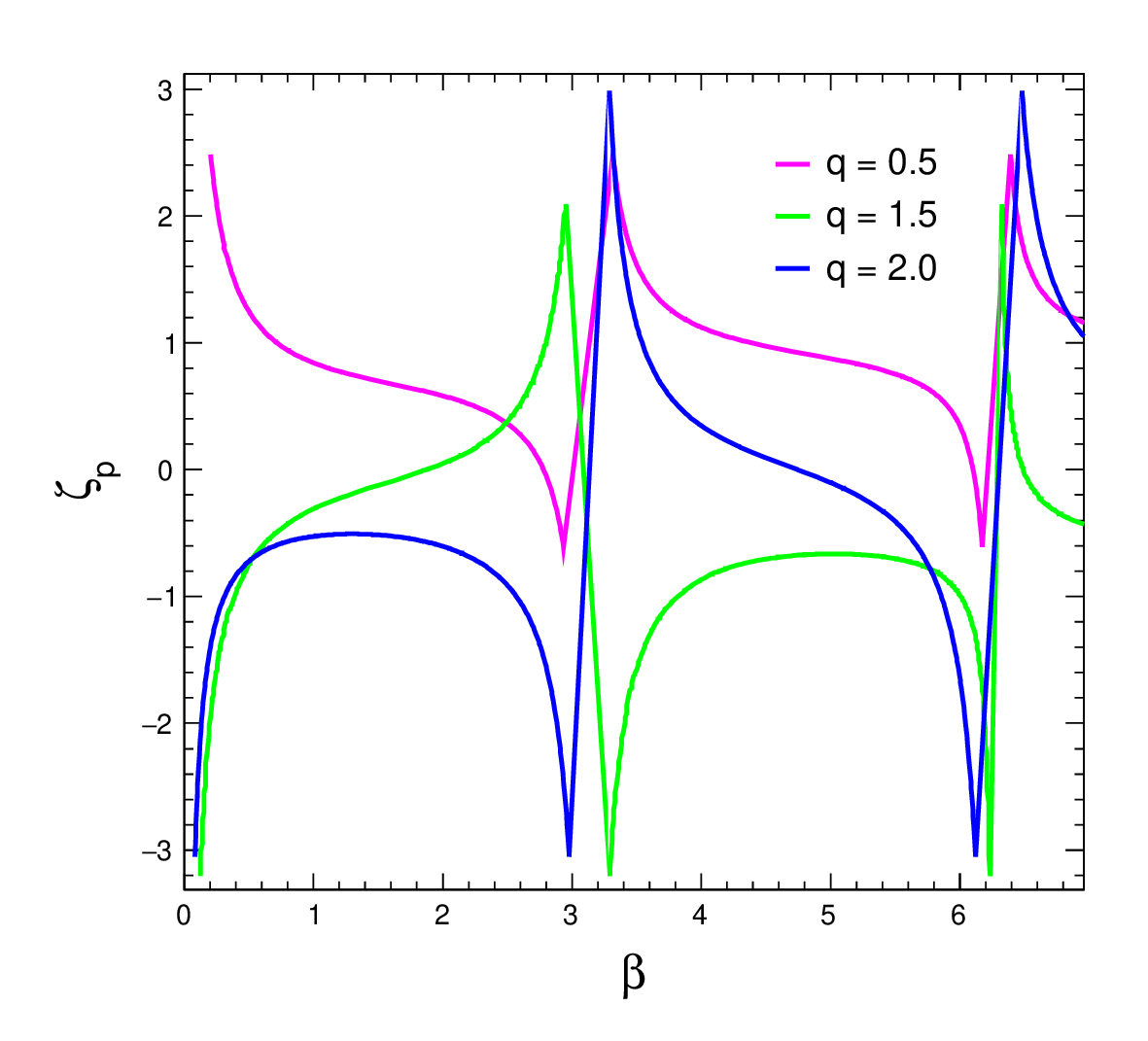}\\
\includegraphics[scale=0.29]{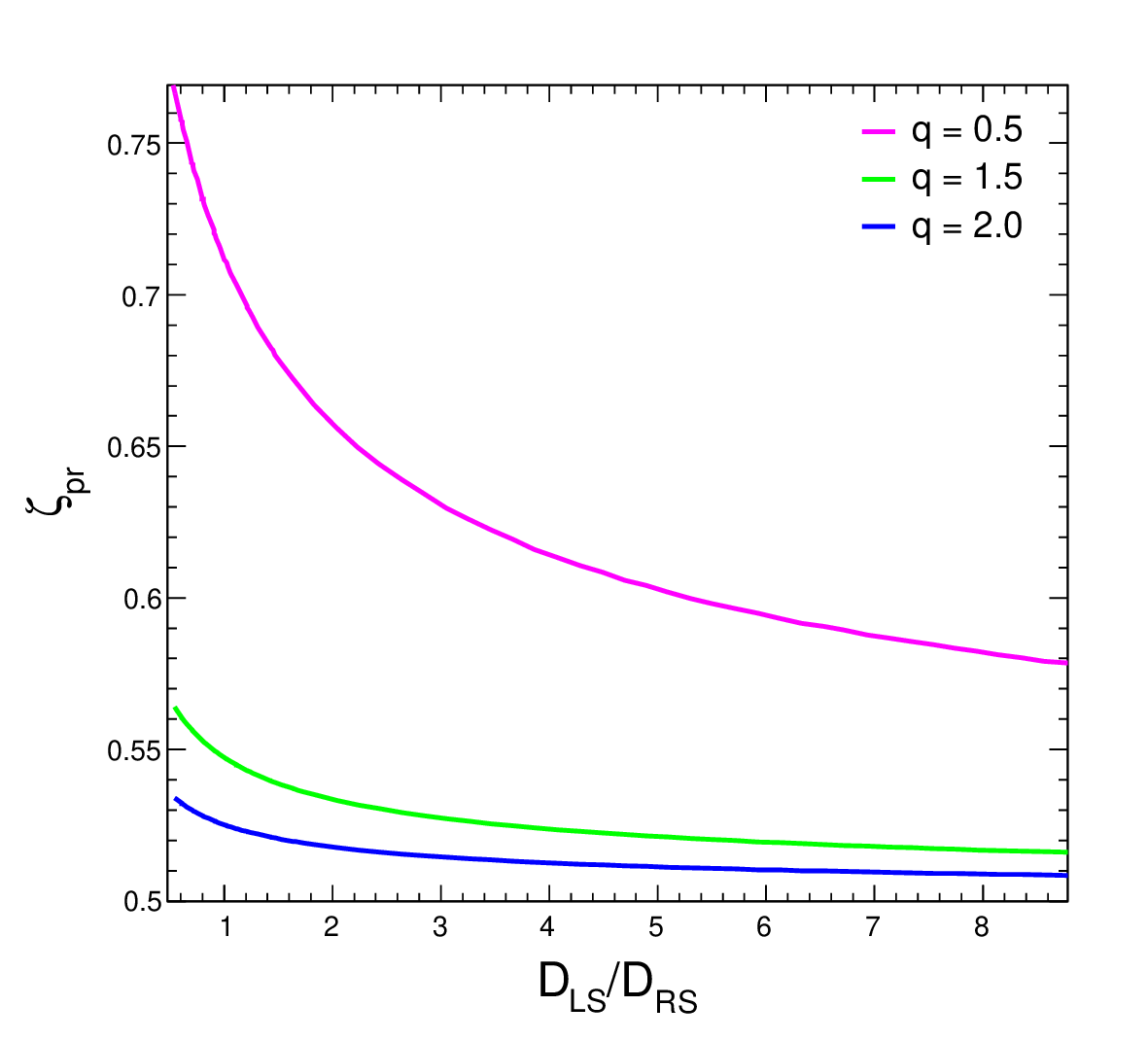}
\includegraphics[scale=0.29]{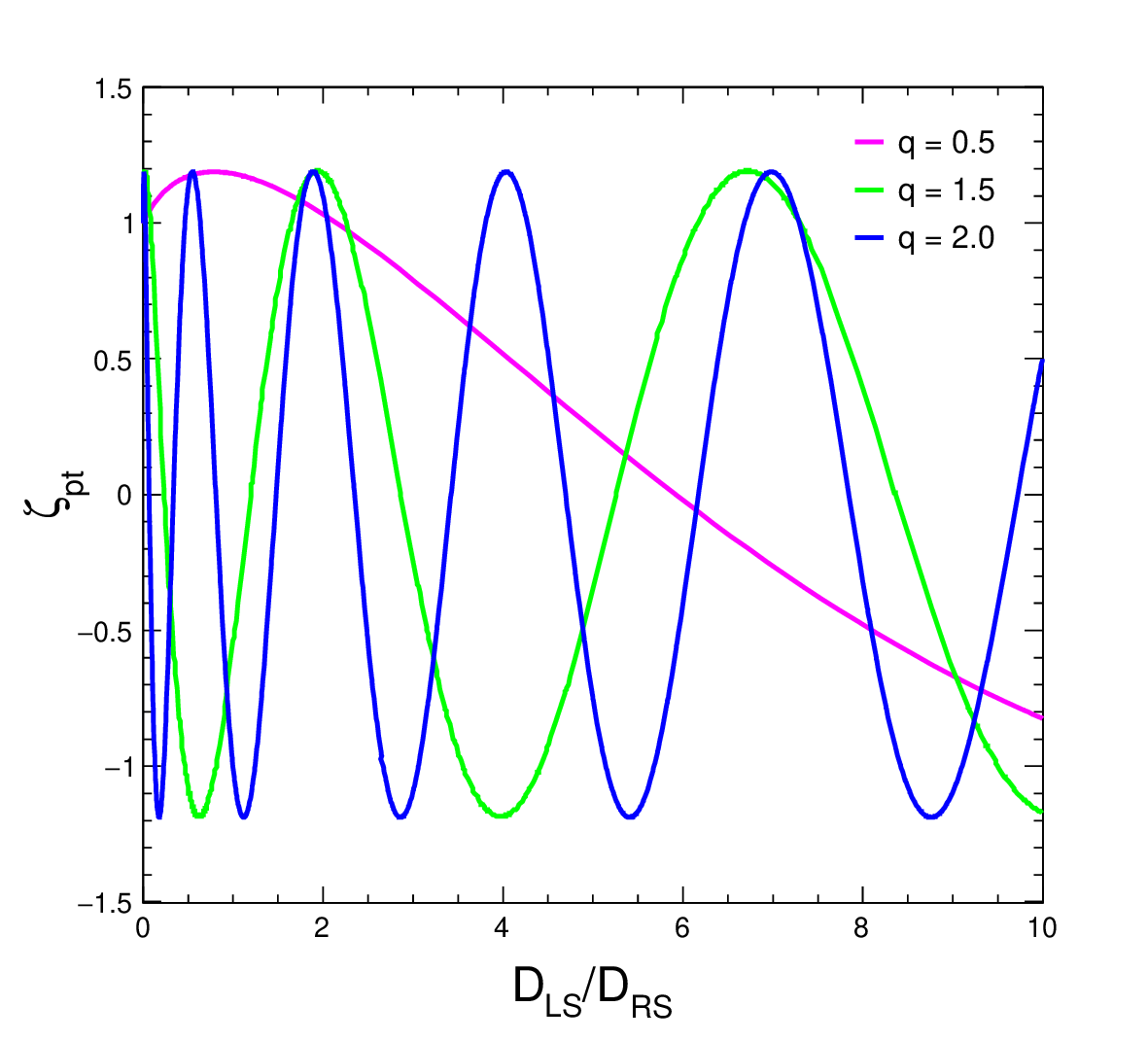}
\includegraphics[scale=0.29]{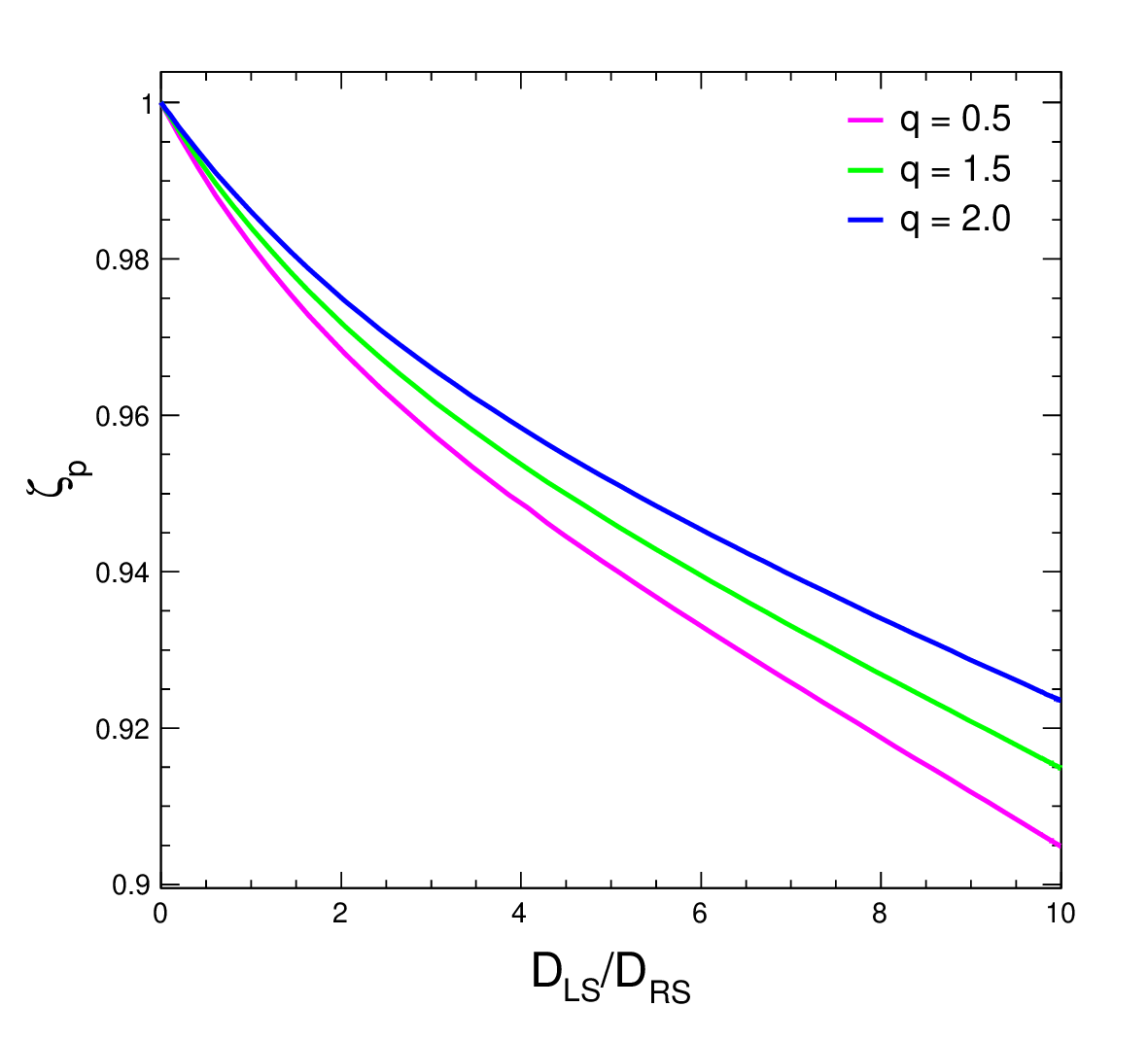}\\
\includegraphics[scale=0.29]{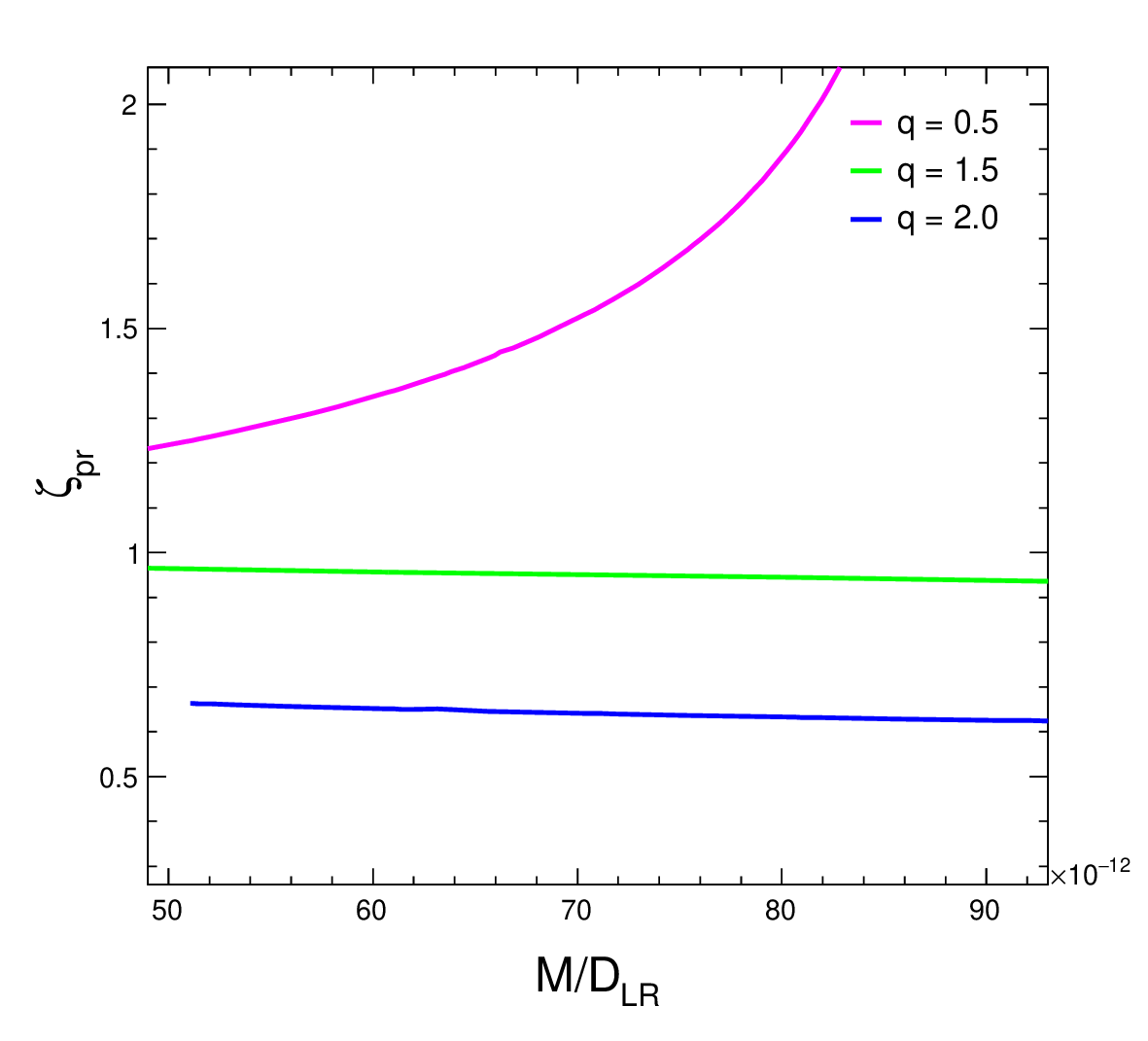}
\includegraphics[scale=0.29]{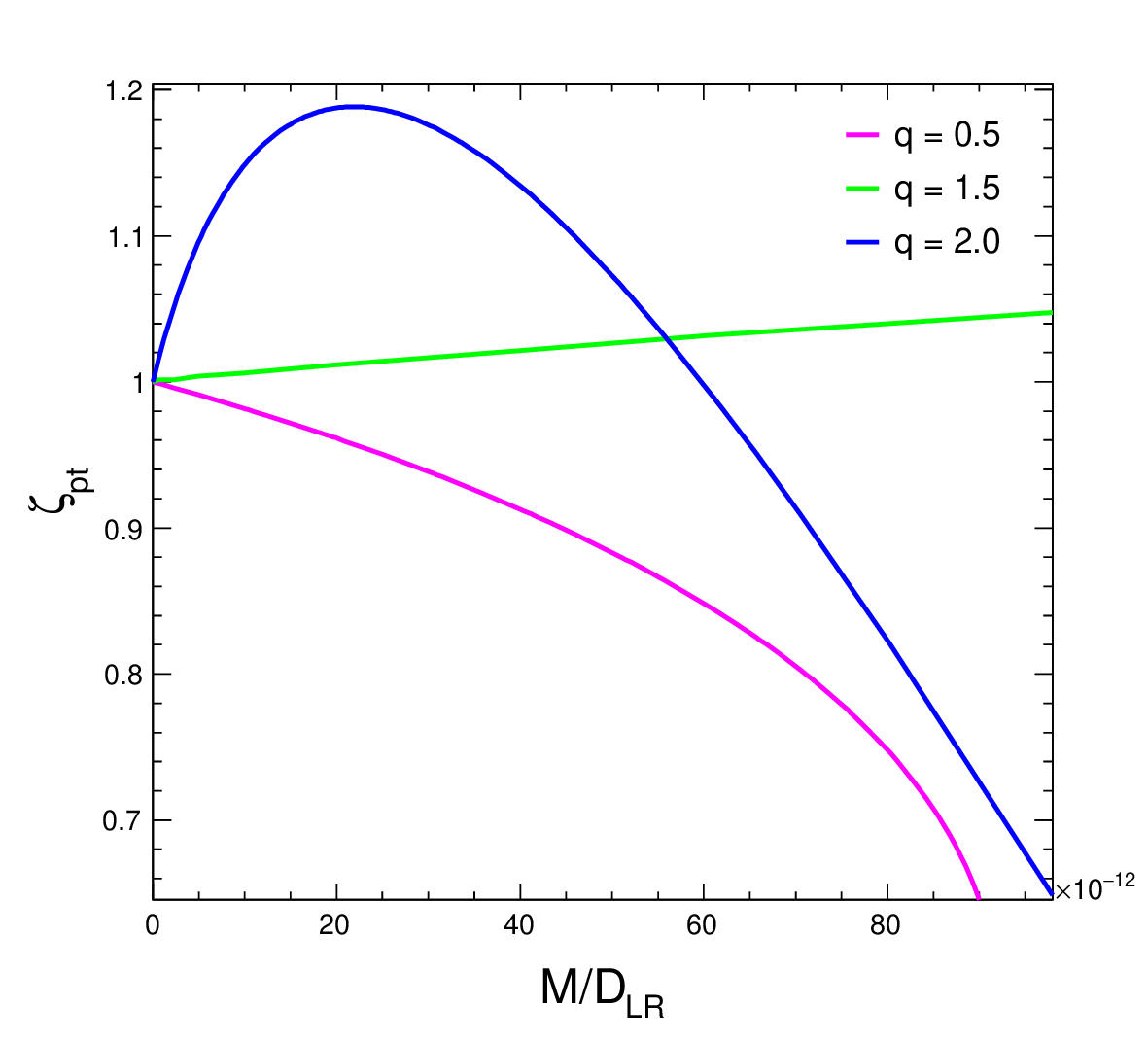}
\includegraphics[scale=0.29]{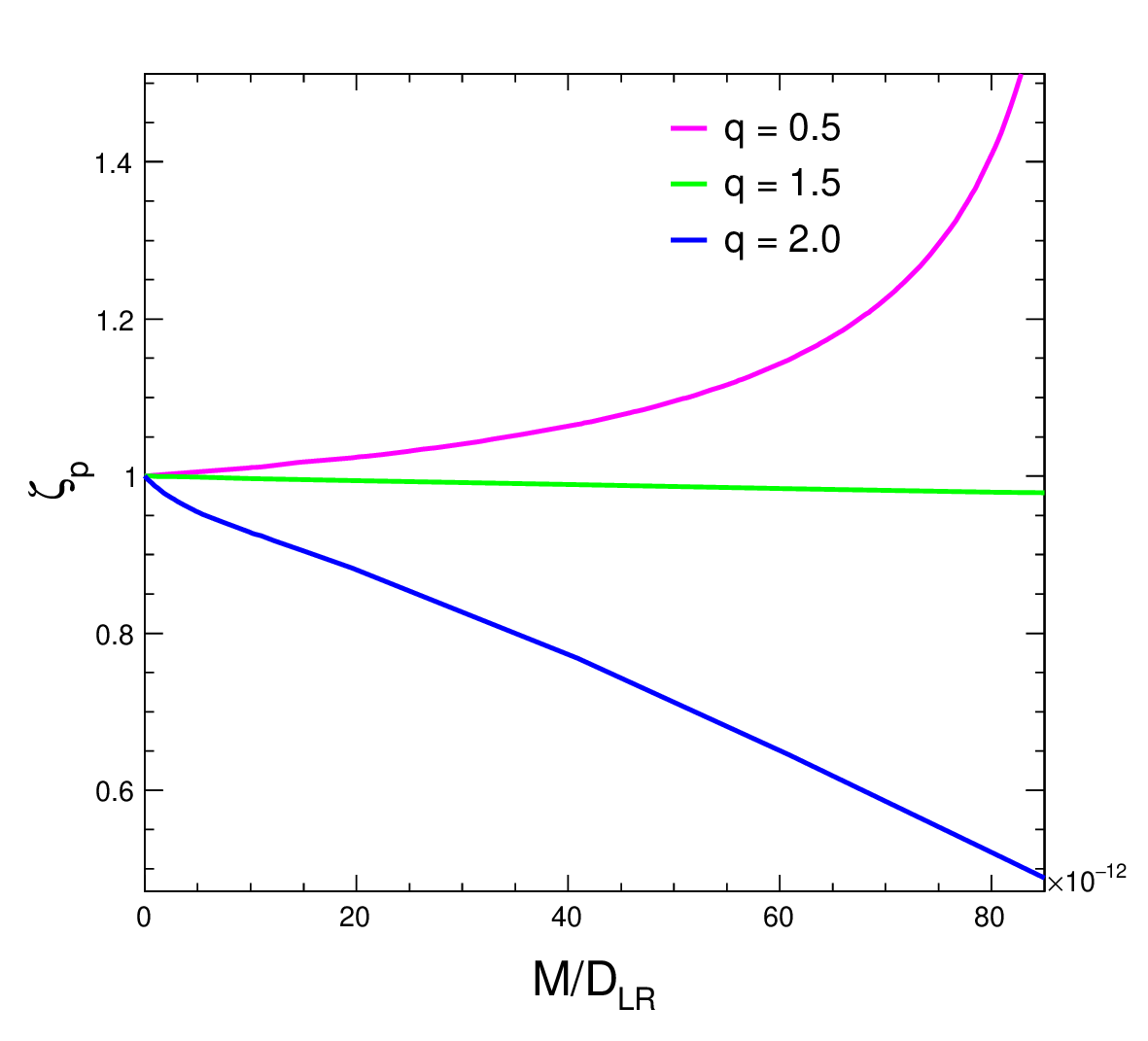}
\caption{The radial, tangential and total magnification of primary images are plotted a functions of $\beta$, $D_{LS}/D_{RS}$ and $M/D_{LR}$ for different values of $q$.}
 \label{fig.6}
\end{figure}

\begin{figure}[h!]
\includegraphics[scale=0.29]{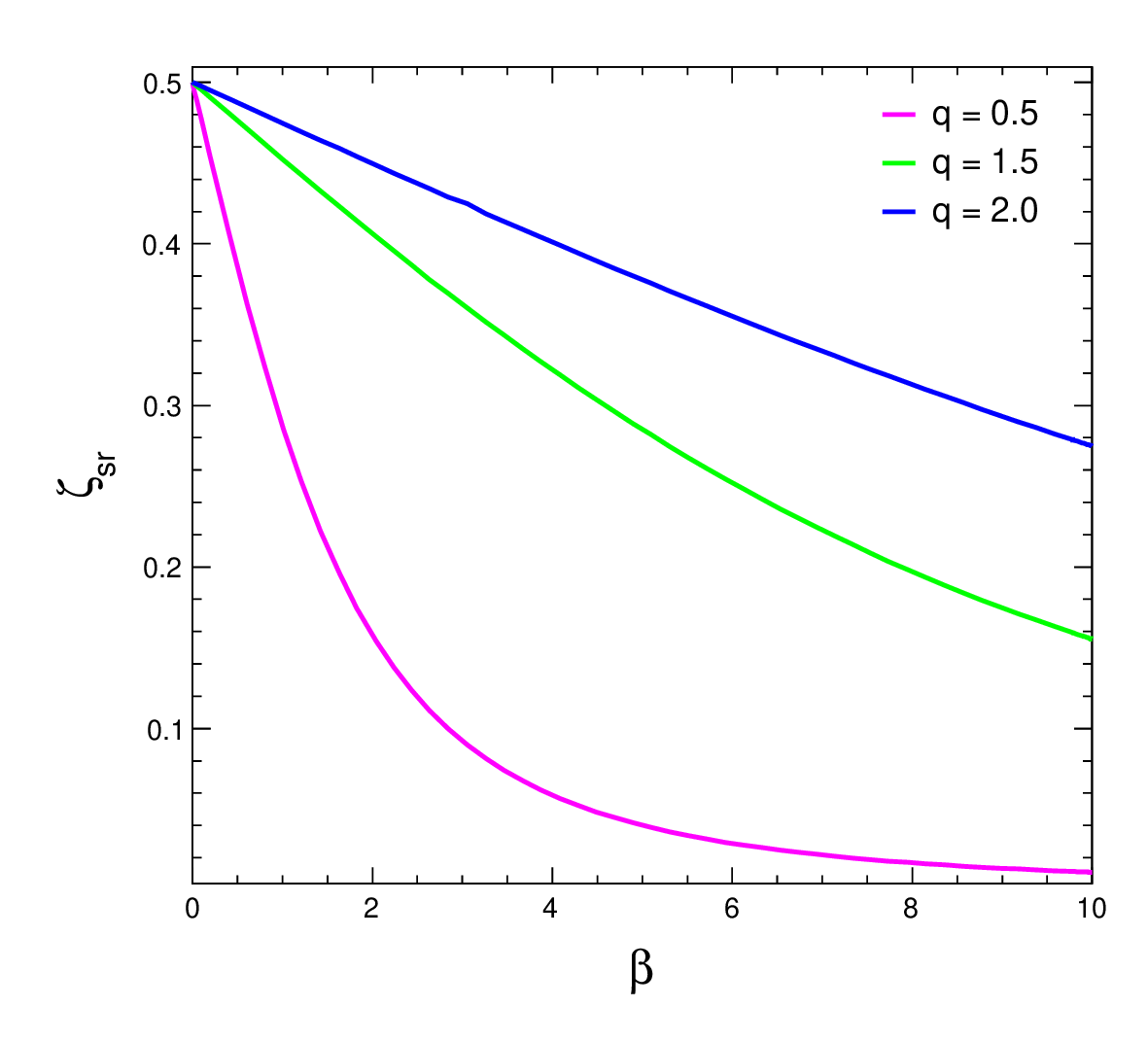}
\includegraphics[scale=0.29]{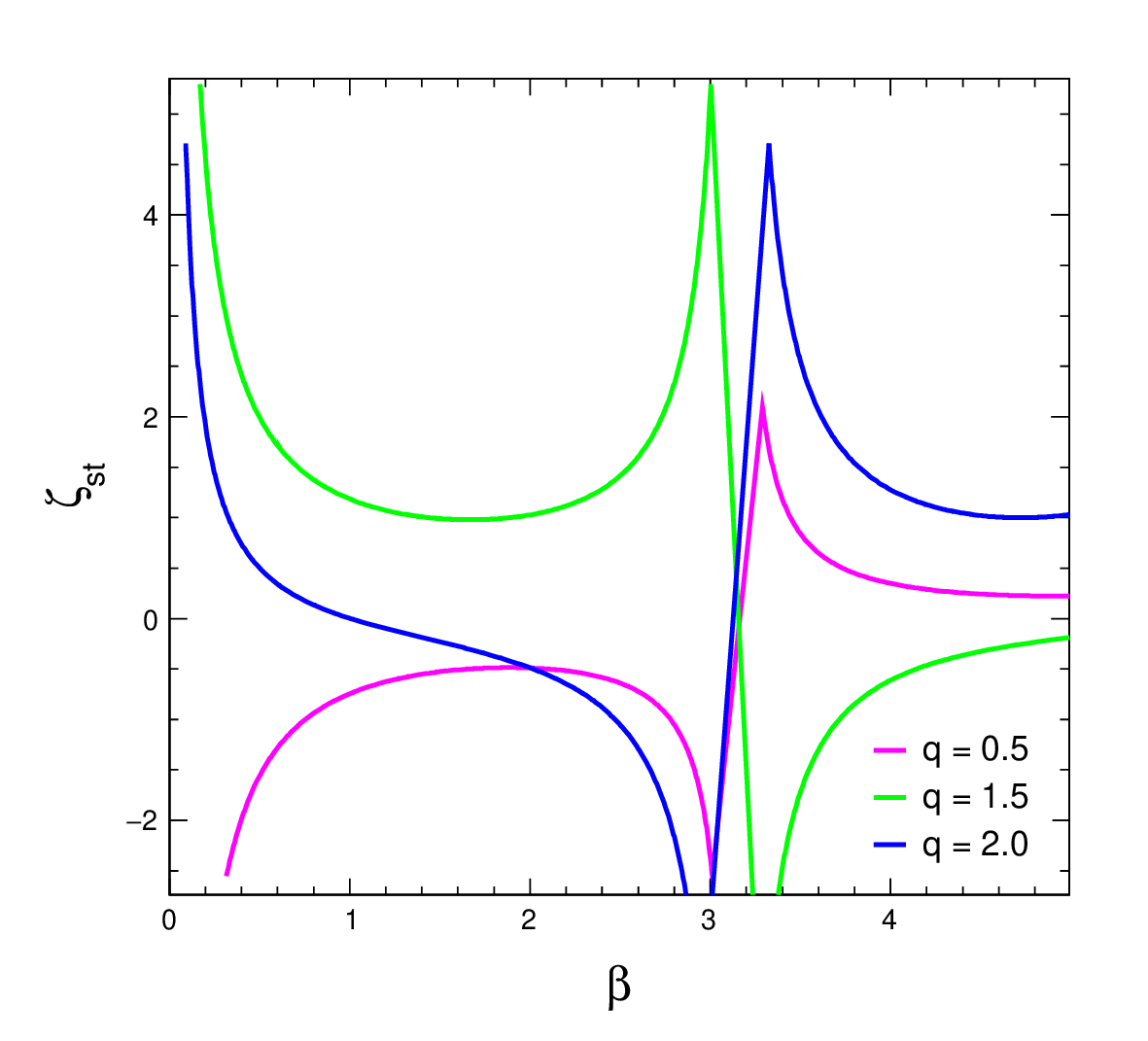}
\includegraphics[scale=0.29]{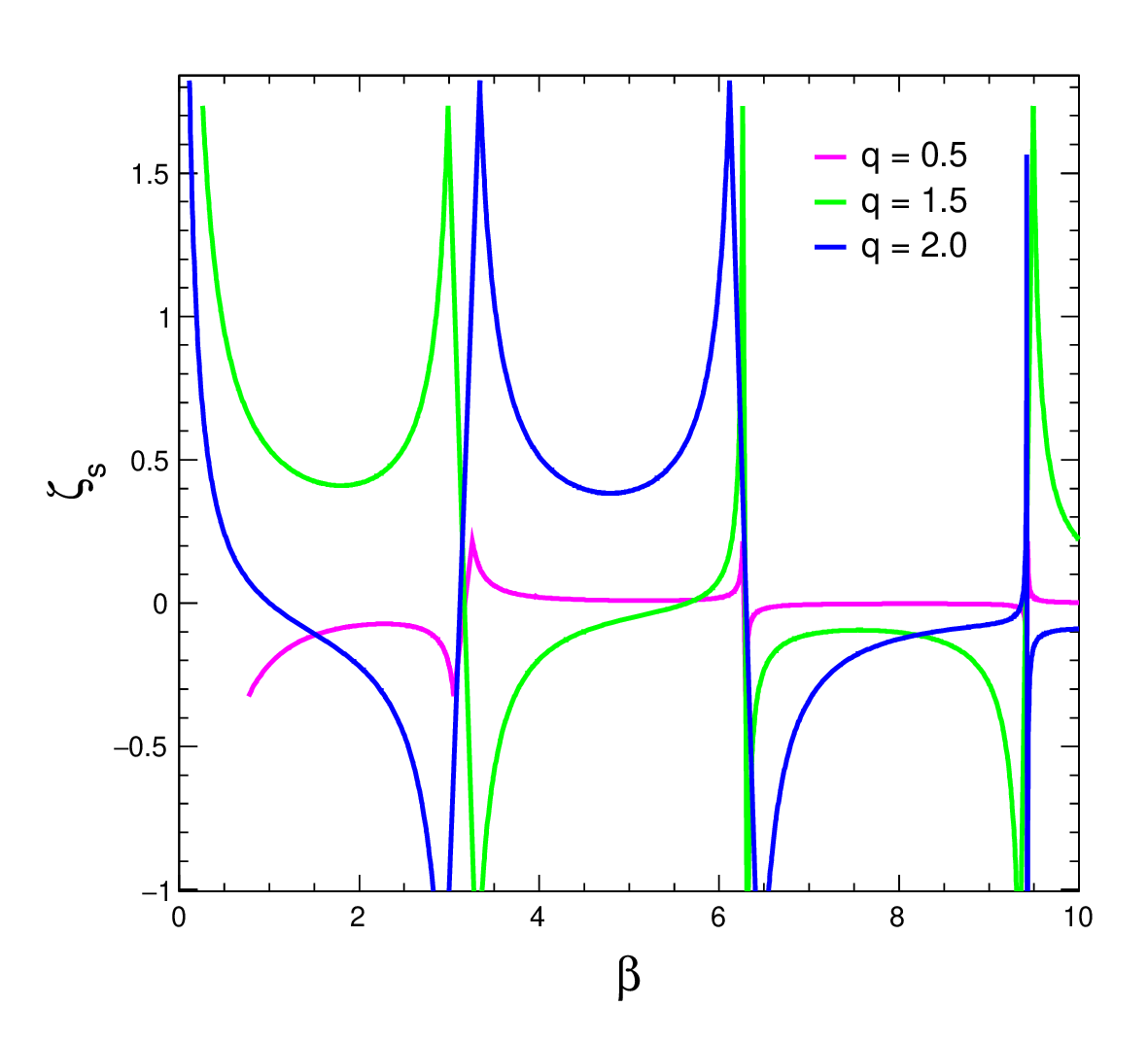}\\
\includegraphics[scale=0.29]{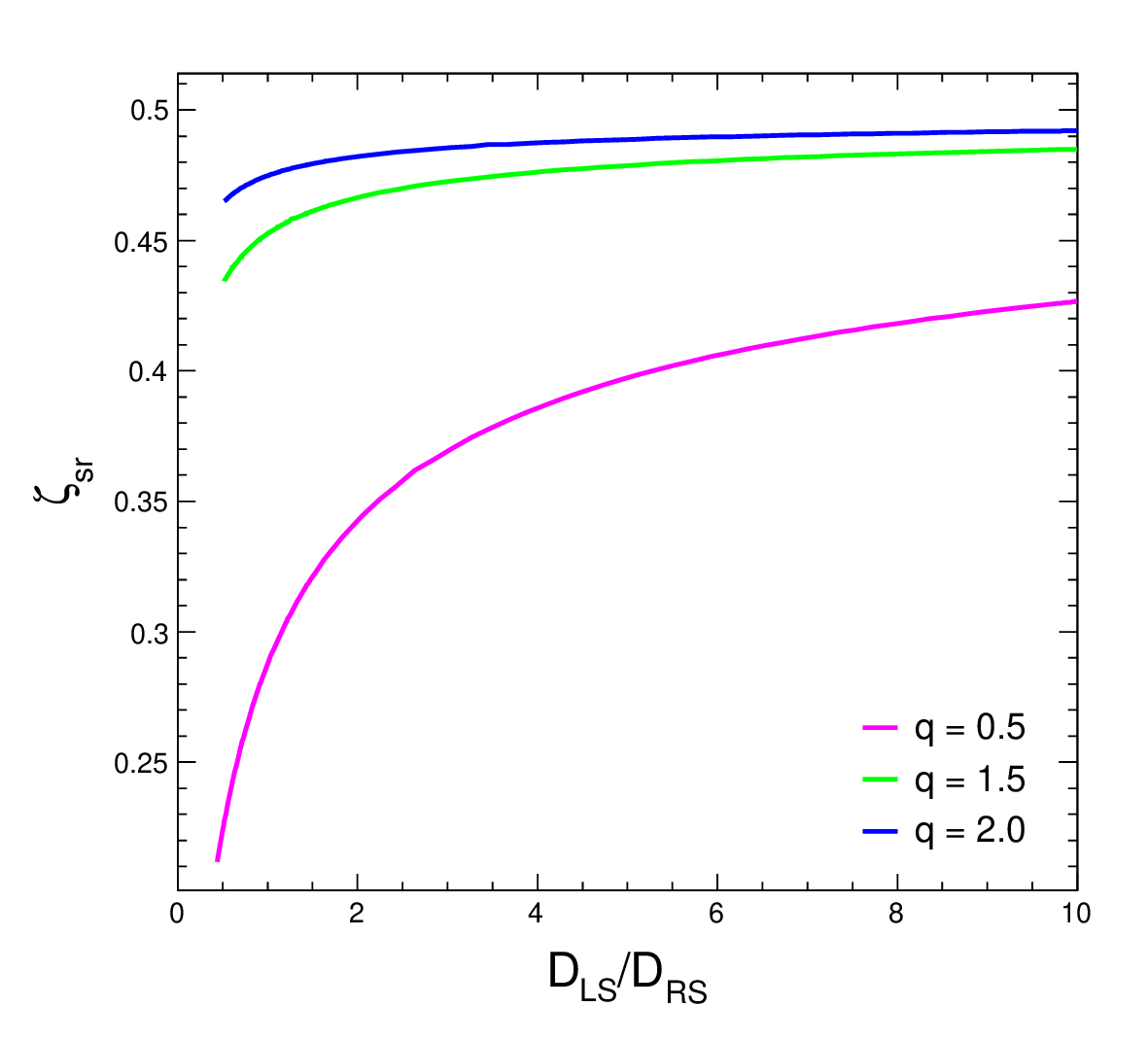}
\includegraphics[scale=0.29]{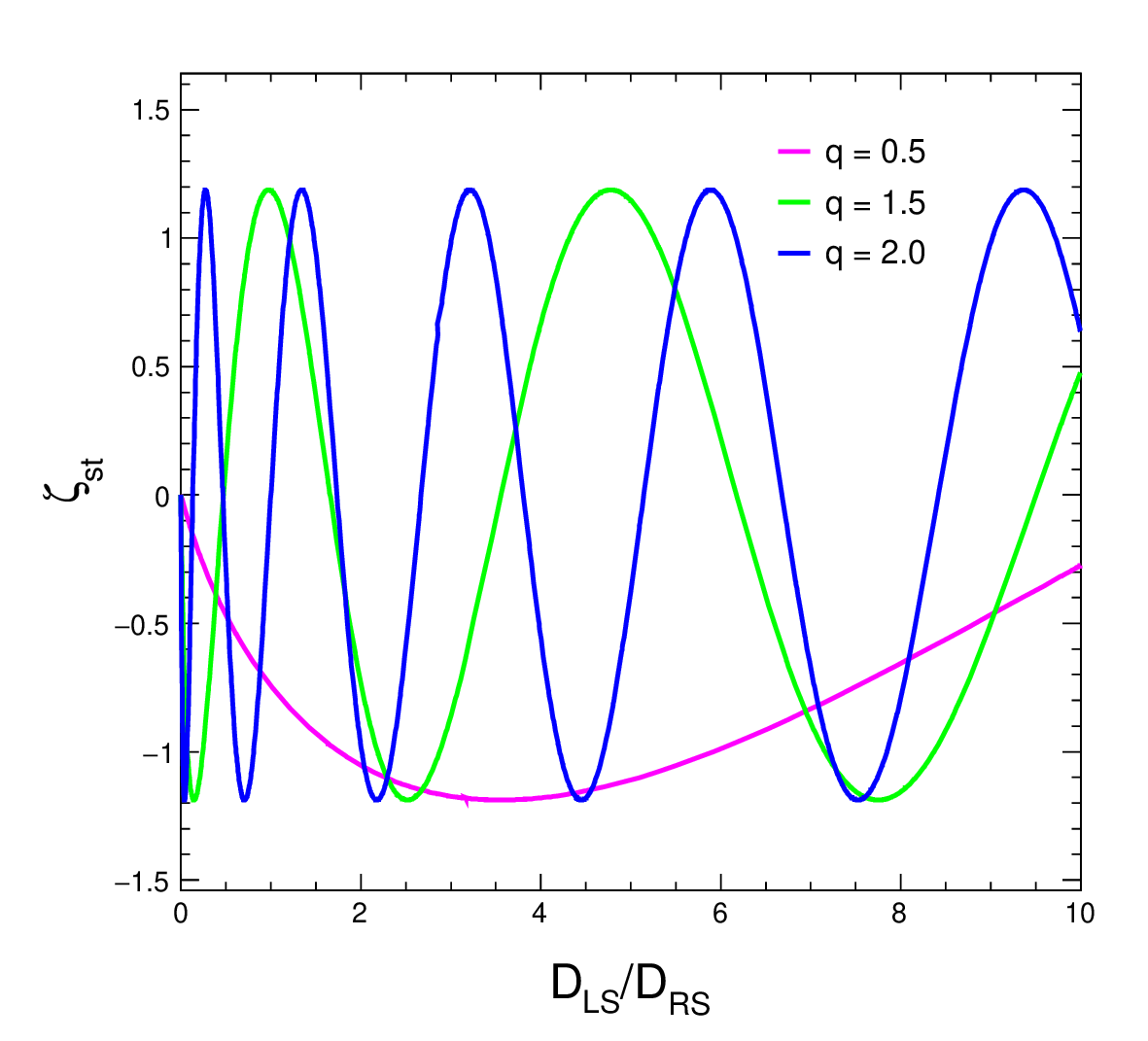}
\includegraphics[scale=0.29]{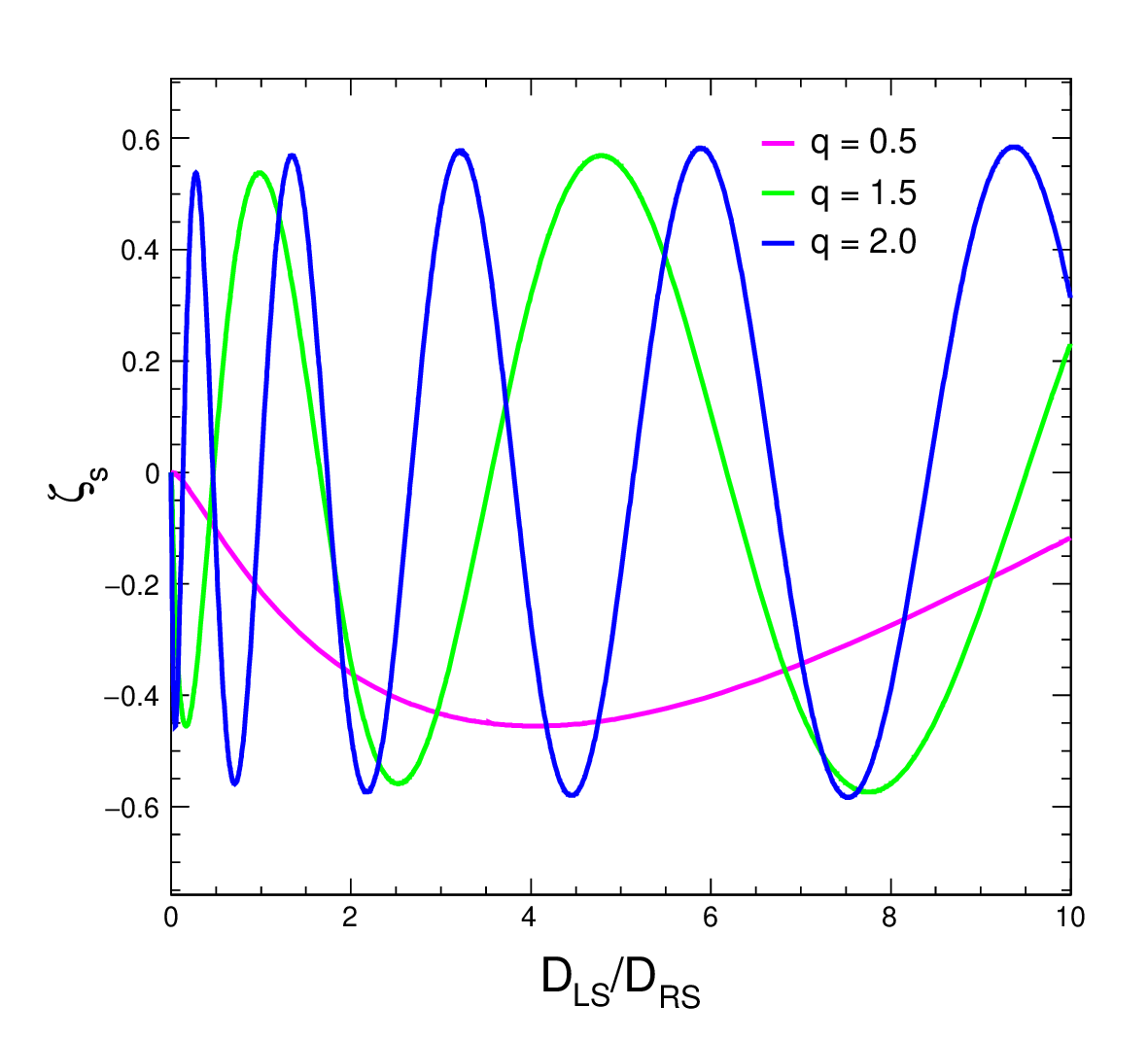}\\
\includegraphics[scale=0.29]{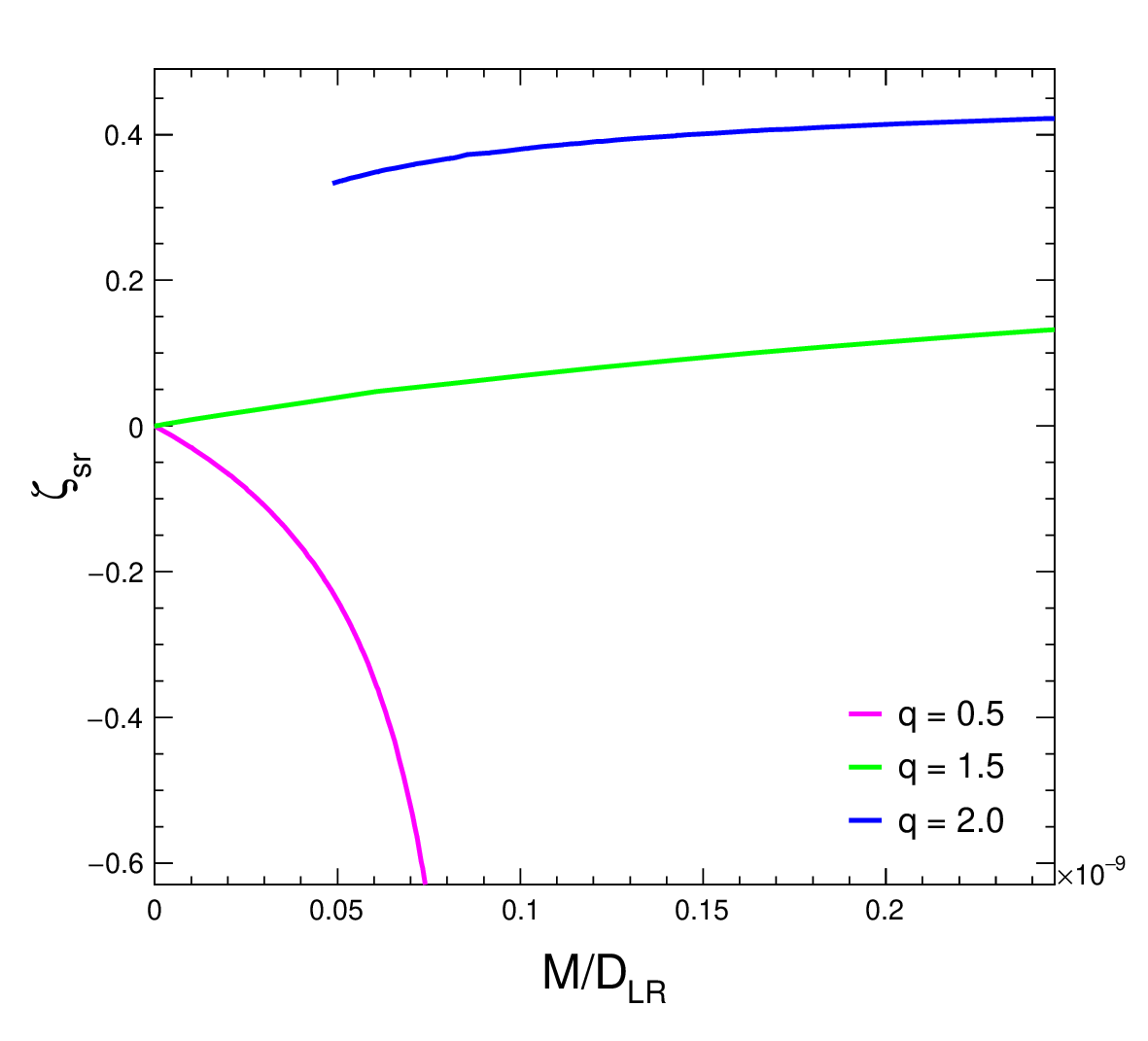}
\includegraphics[scale=0.29]{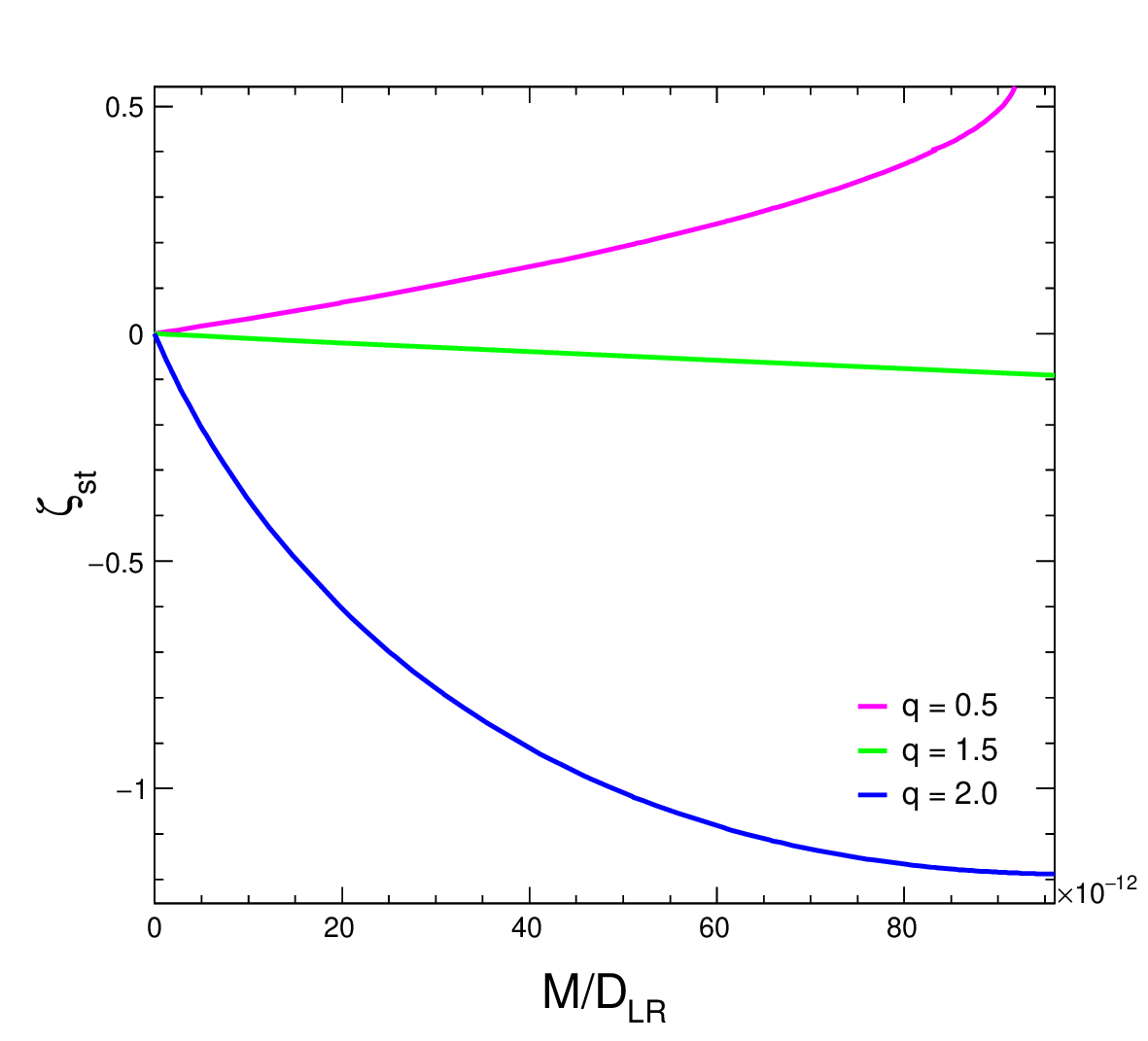}
\includegraphics[scale=0.29]{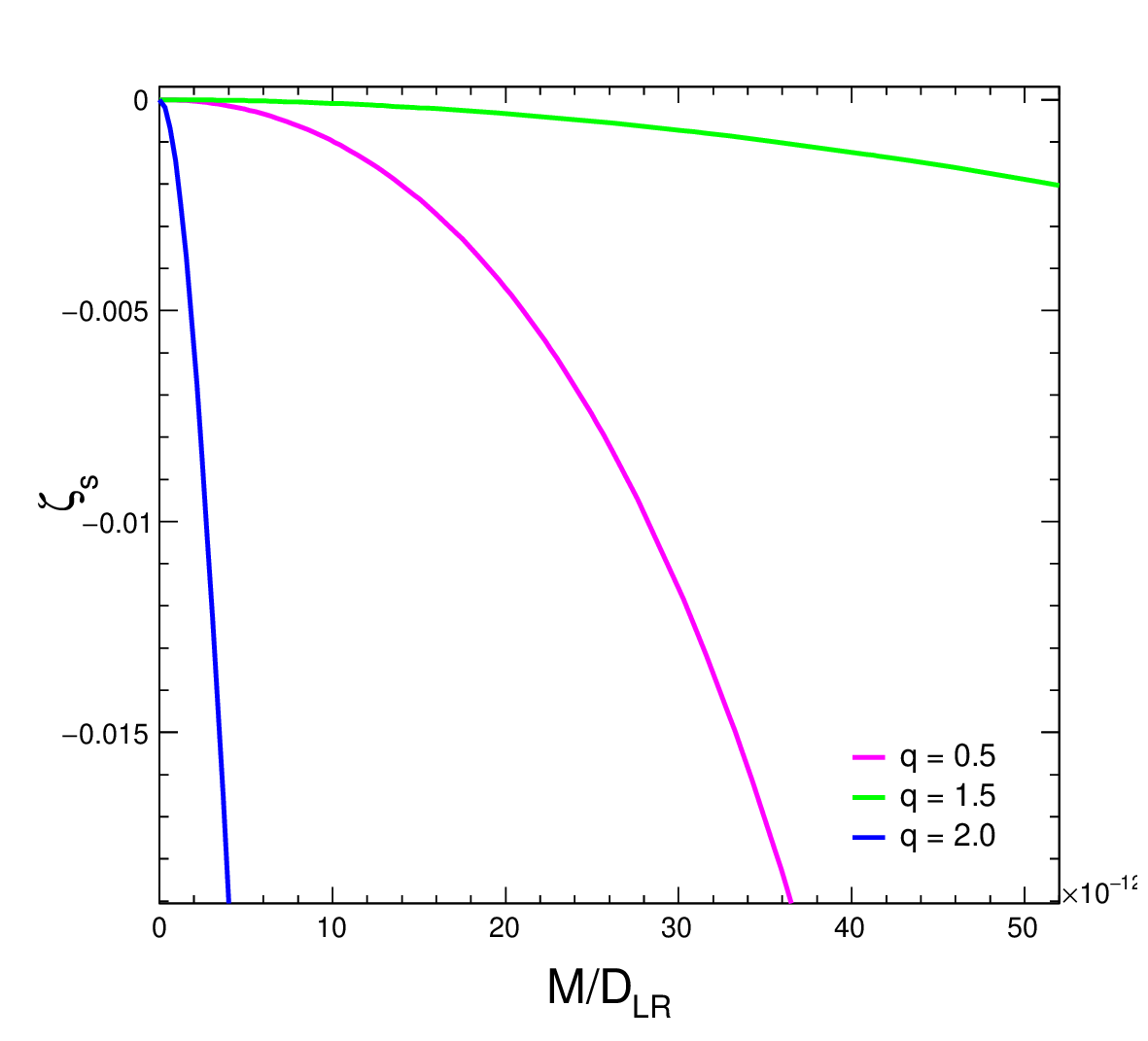}
\caption{The radial, tangential and total magnification of secondary images are plotted a functions of $\beta$, $D_{LS}/D_{RS}$ and $M/D_{LR}$ for different values of $q$.}
 \label{fig.7}
\end{figure}

\begin{figure}[h!]
\includegraphics[scale=0.29]{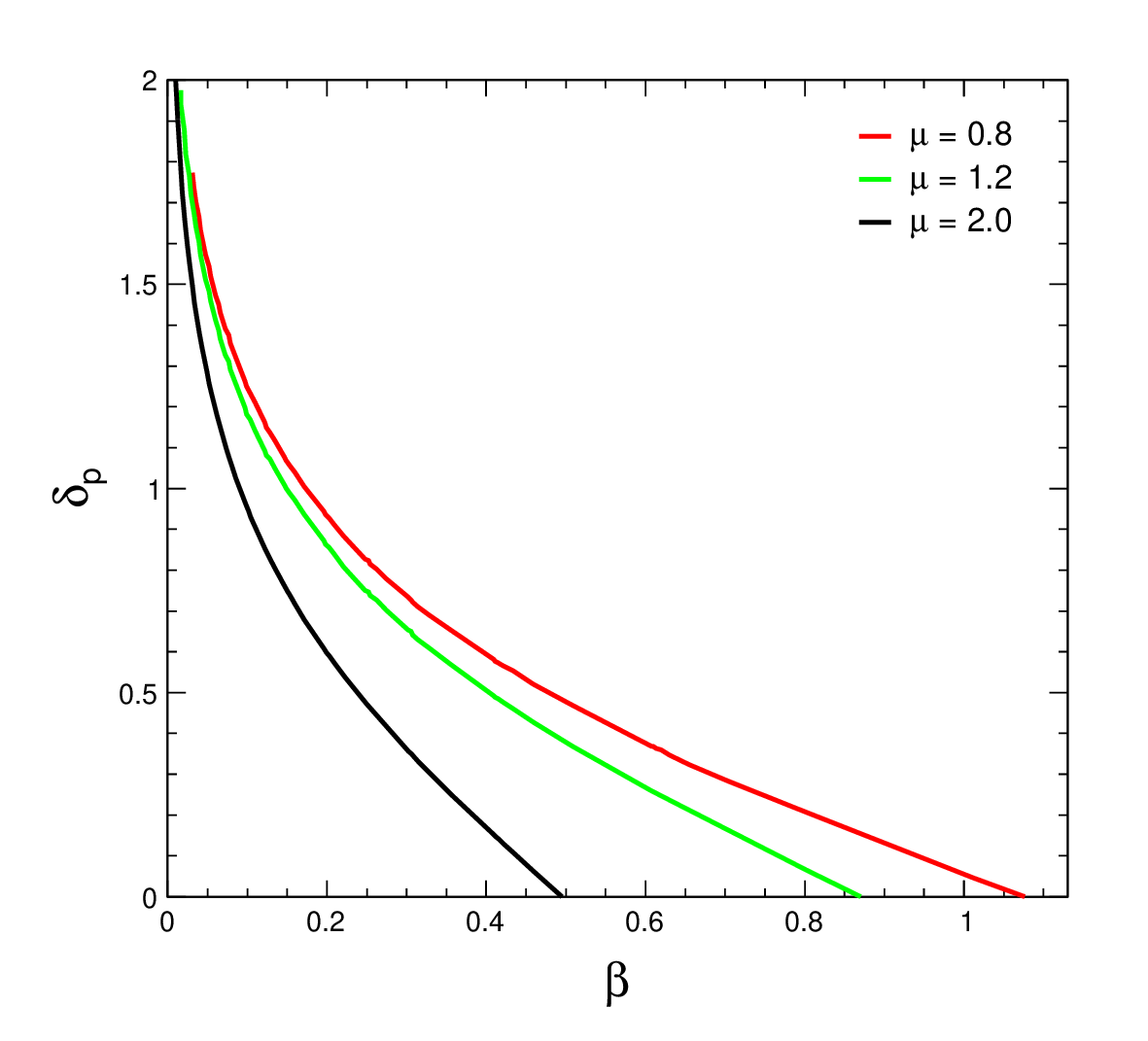}
\includegraphics[scale=0.29]{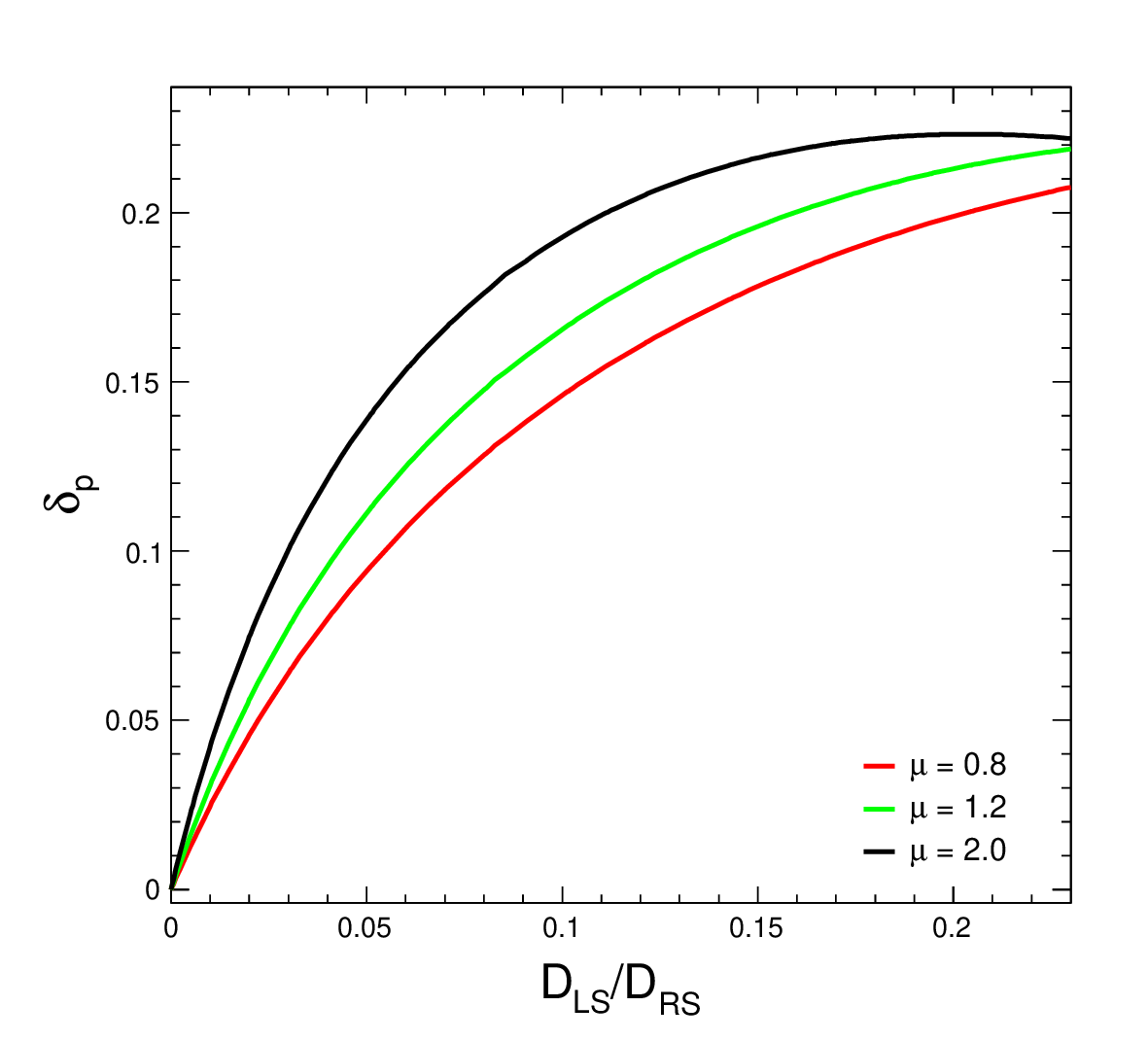}
\includegraphics[scale=0.29]{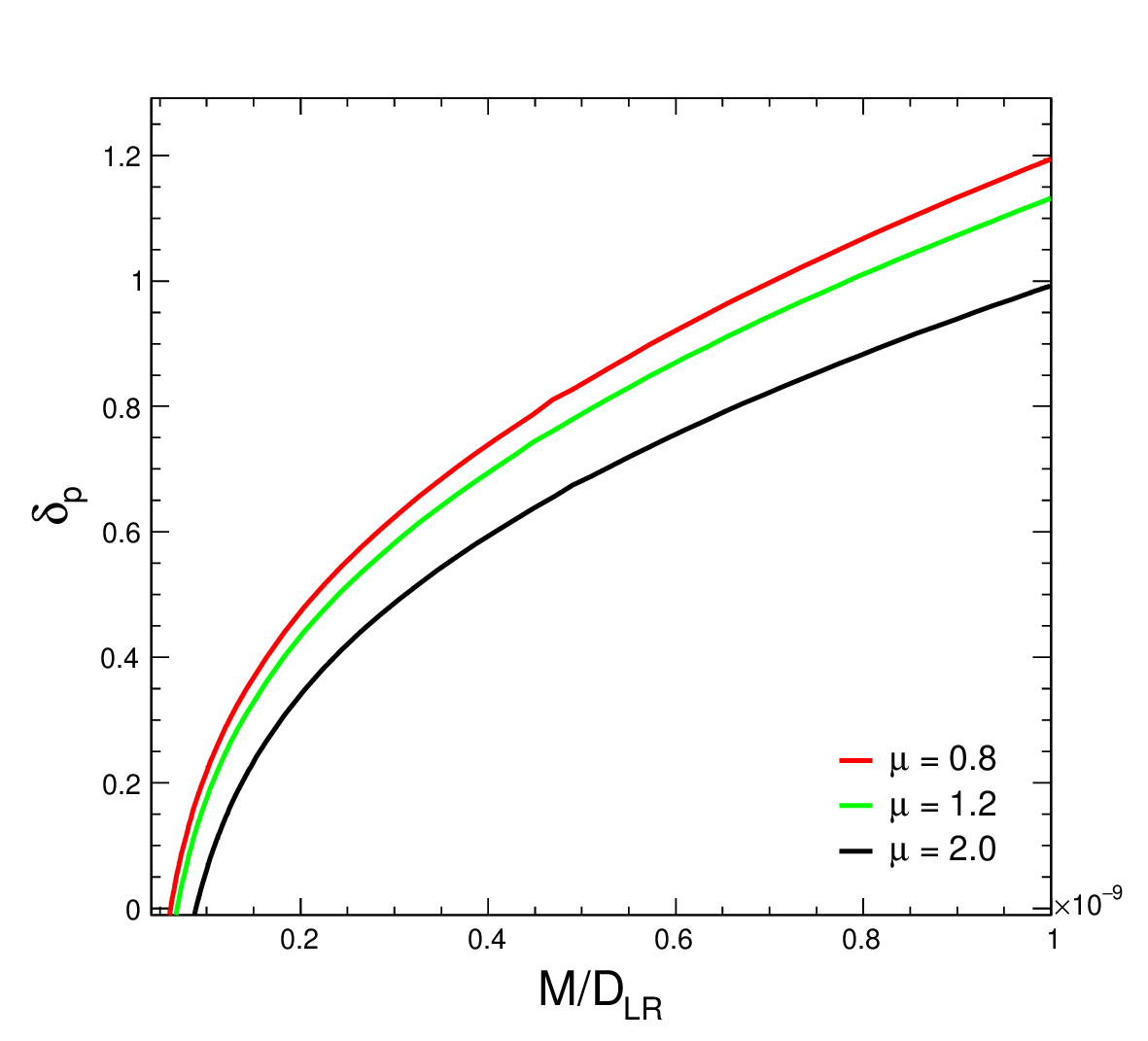}\\
\includegraphics[scale=0.29]{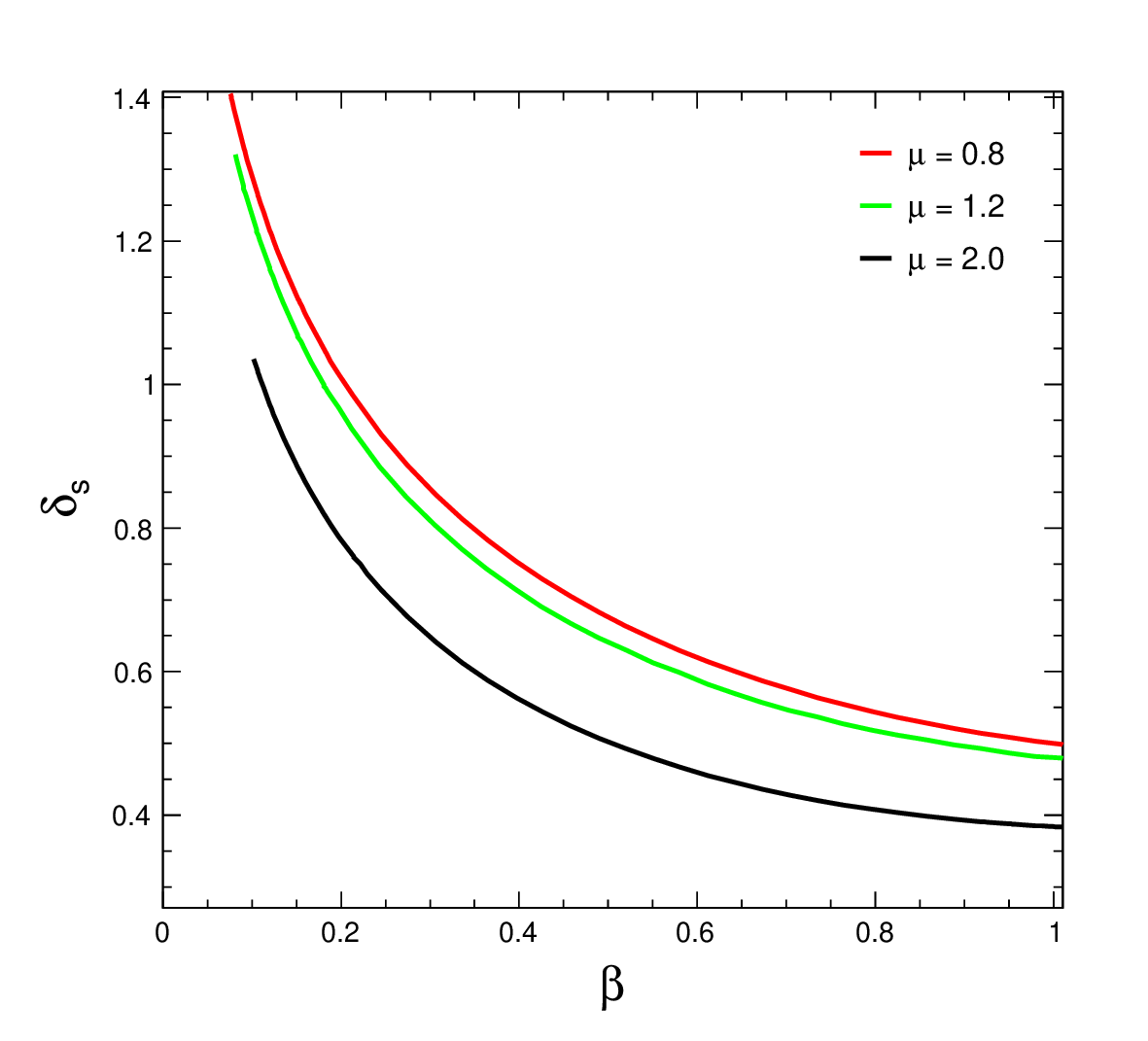}
\includegraphics[scale=0.29]{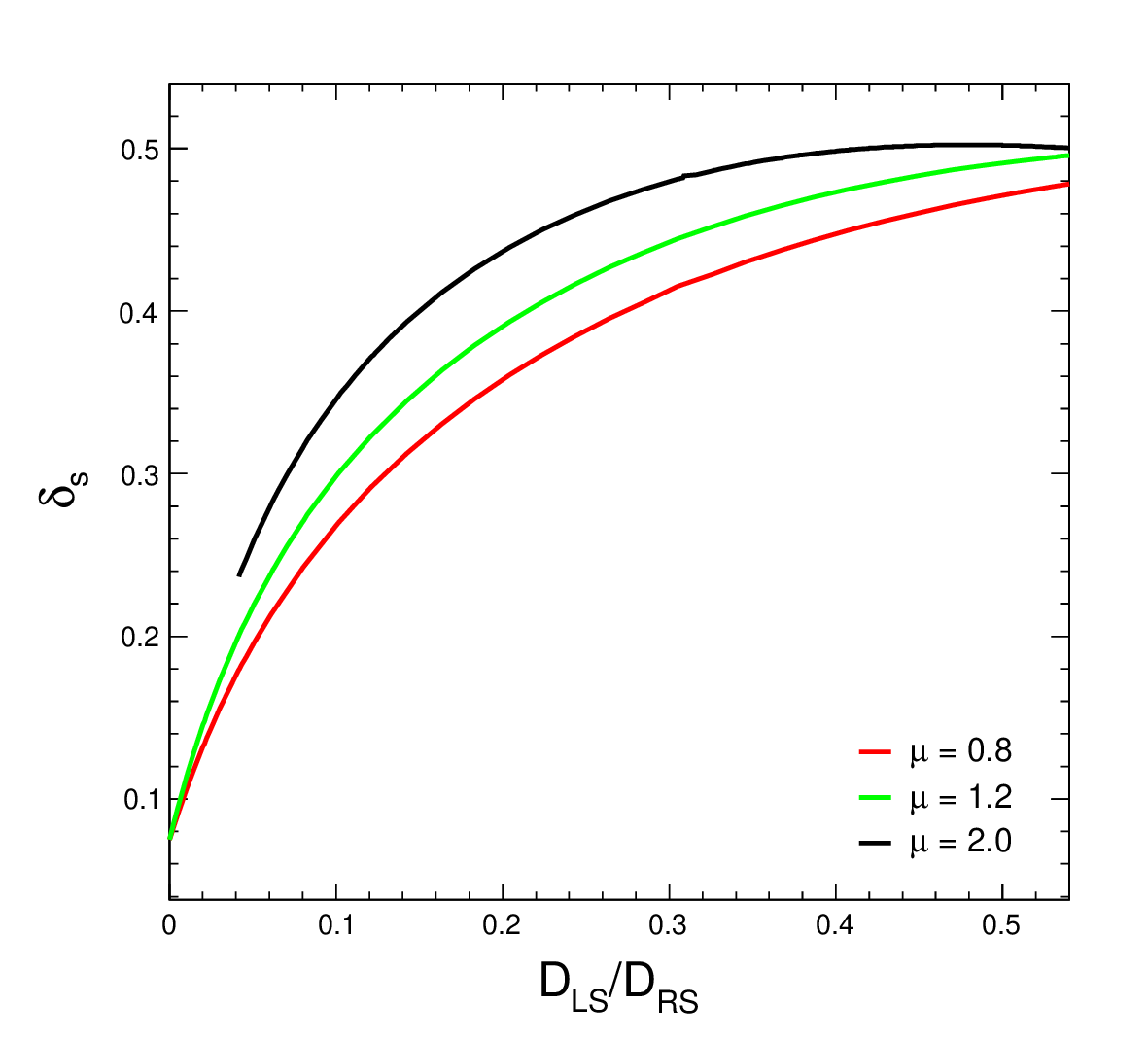}
\includegraphics[scale=0.29]{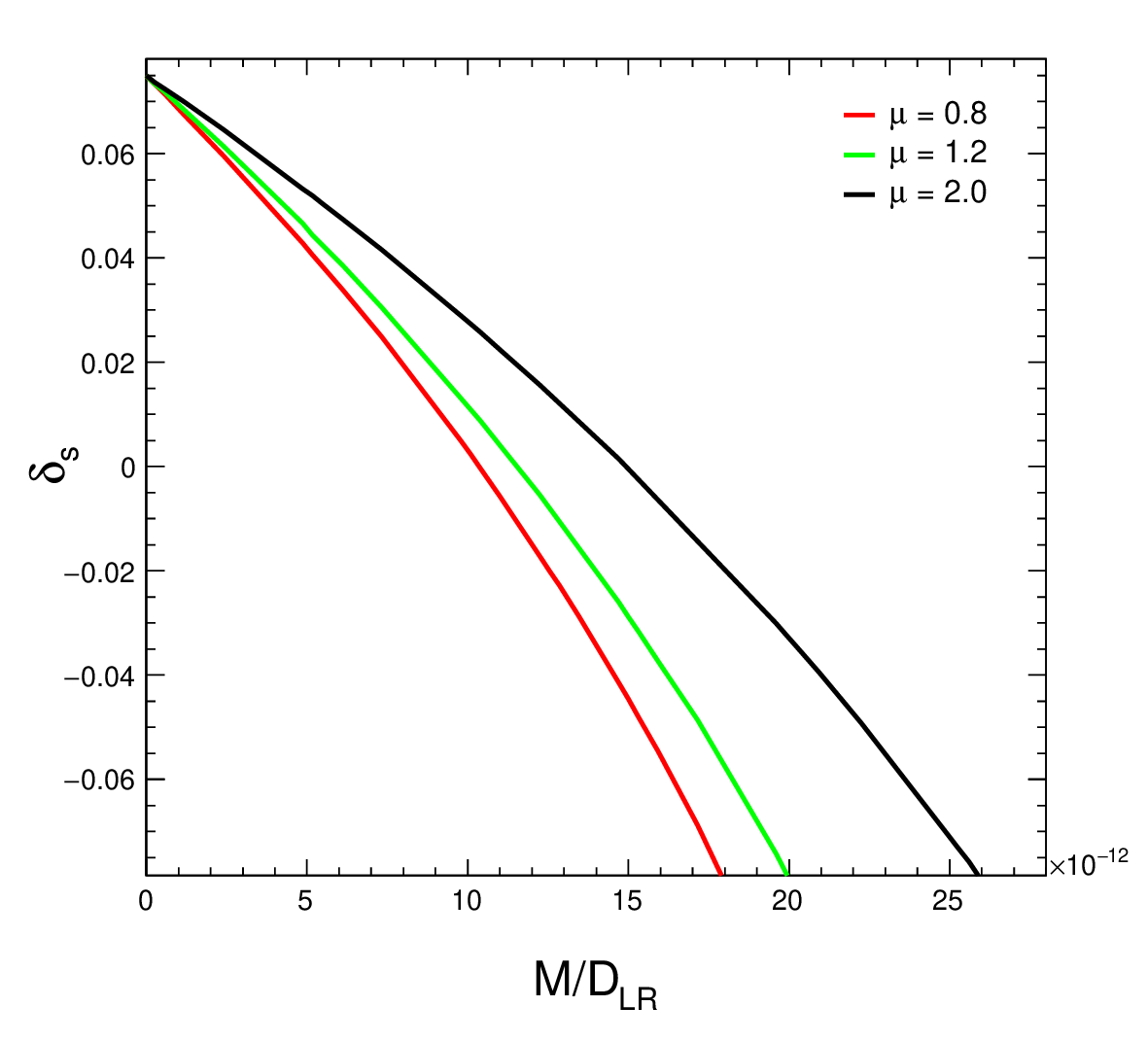}
\caption{The distortion parameter of primary and secondary images are plotted a functions of $\beta$, $D_{LS}/D_{RS}$ and $M/D_{LR}$ for different values of $\mu$.}
 \label{fig.8}
\end{figure}

\begin{figure}[h!]
\includegraphics[scale=0.29]{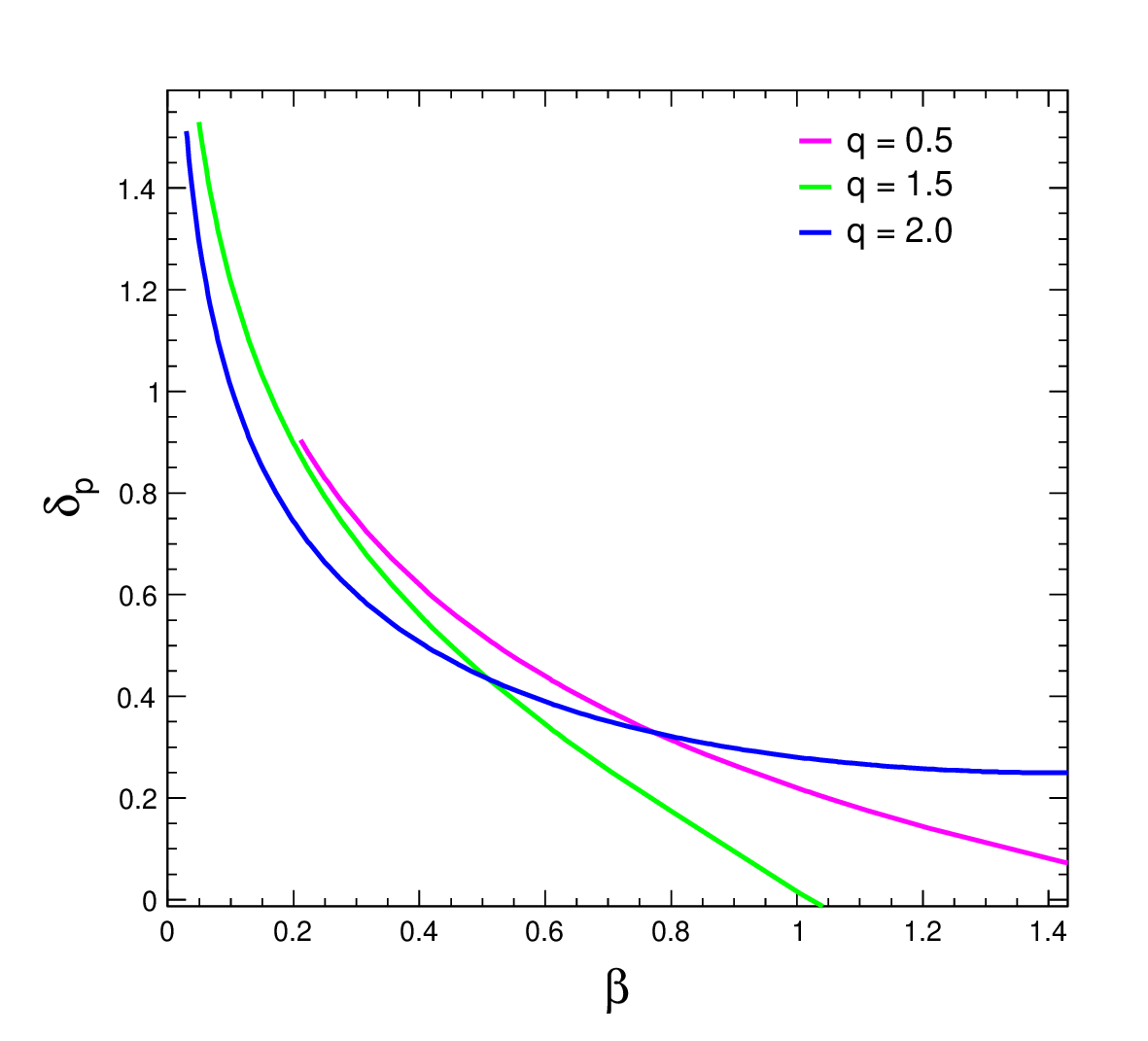}
\includegraphics[scale=0.29]{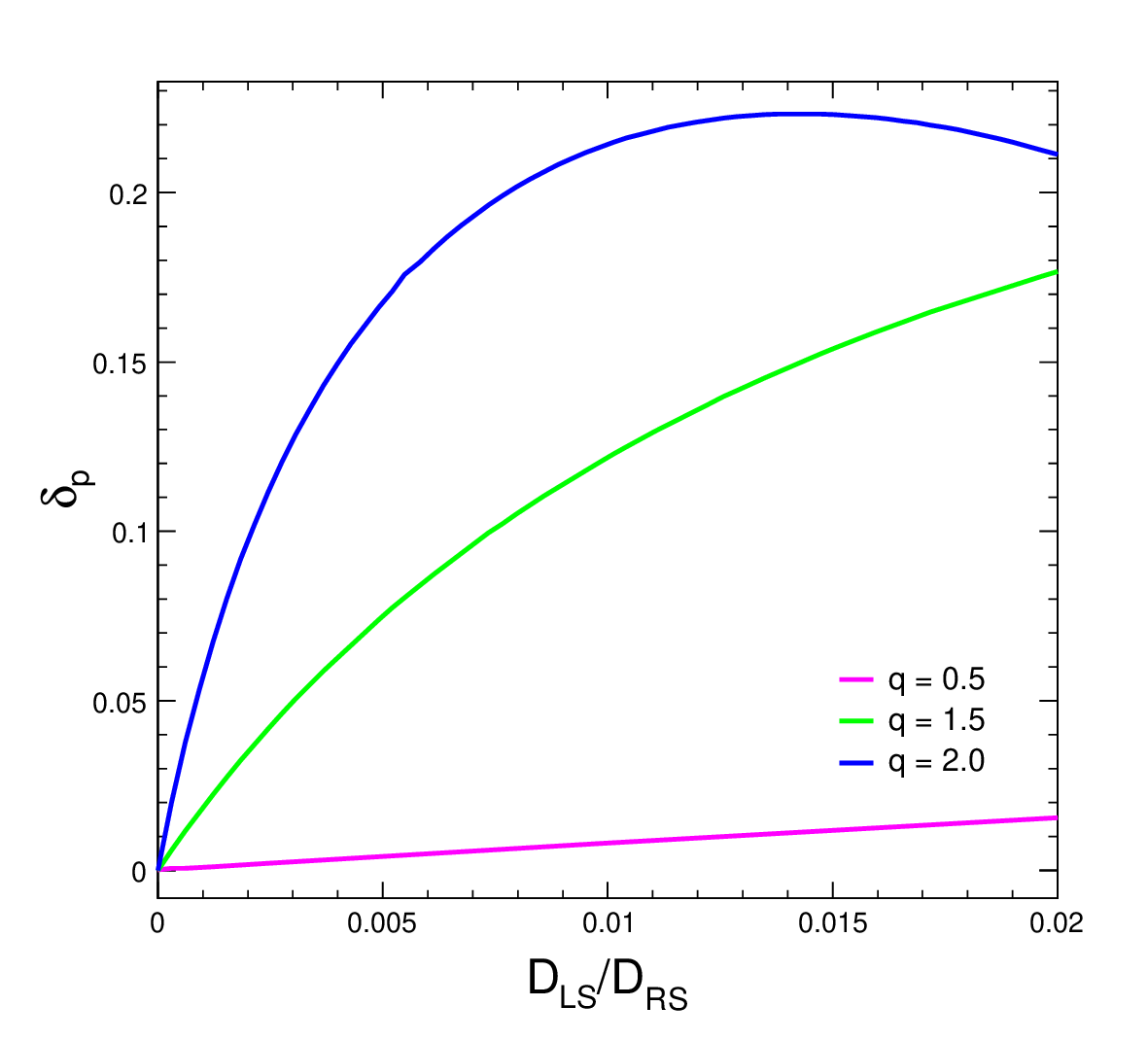}
\includegraphics[scale=0.29]{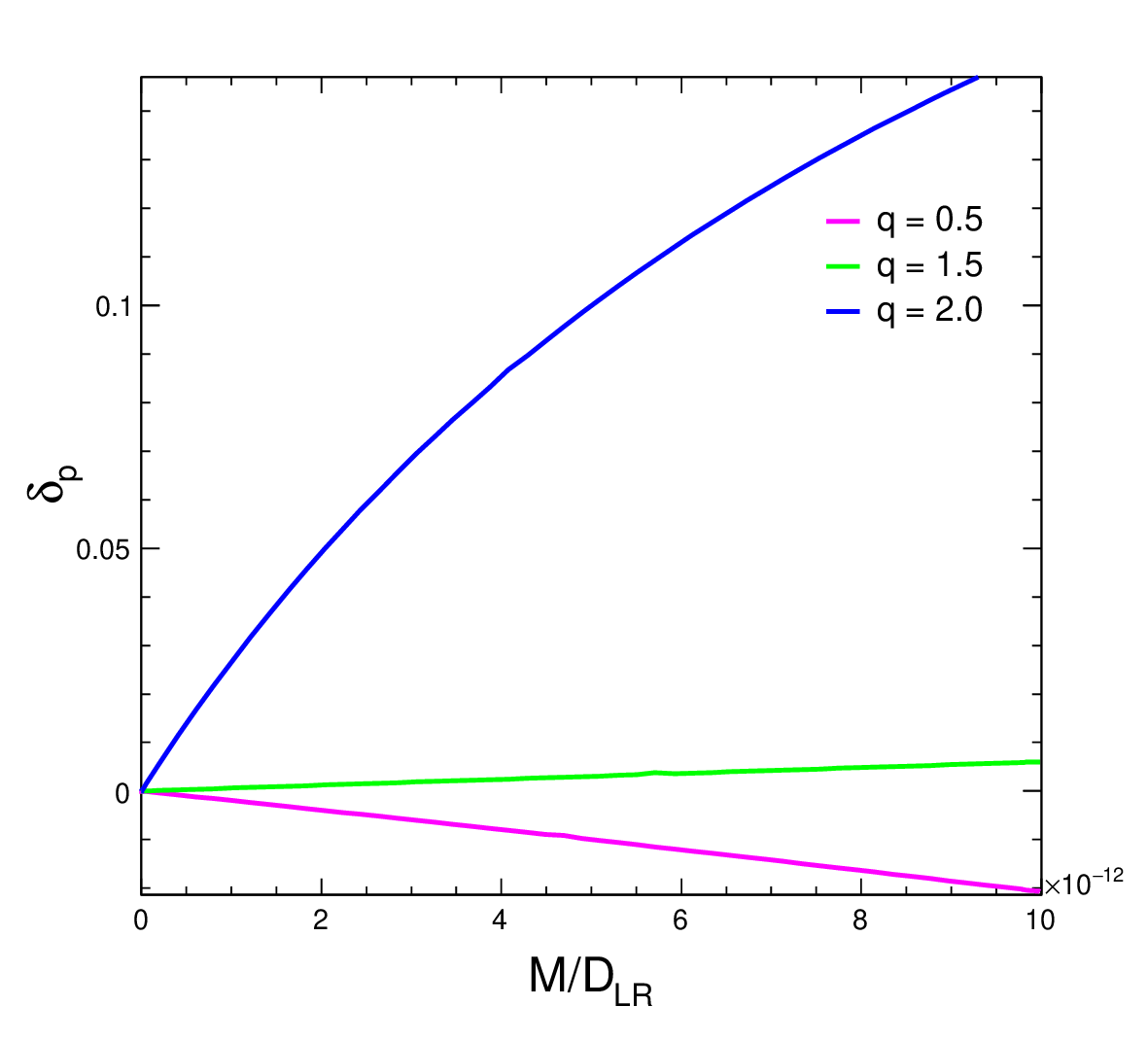}\\
\includegraphics[scale=0.29]{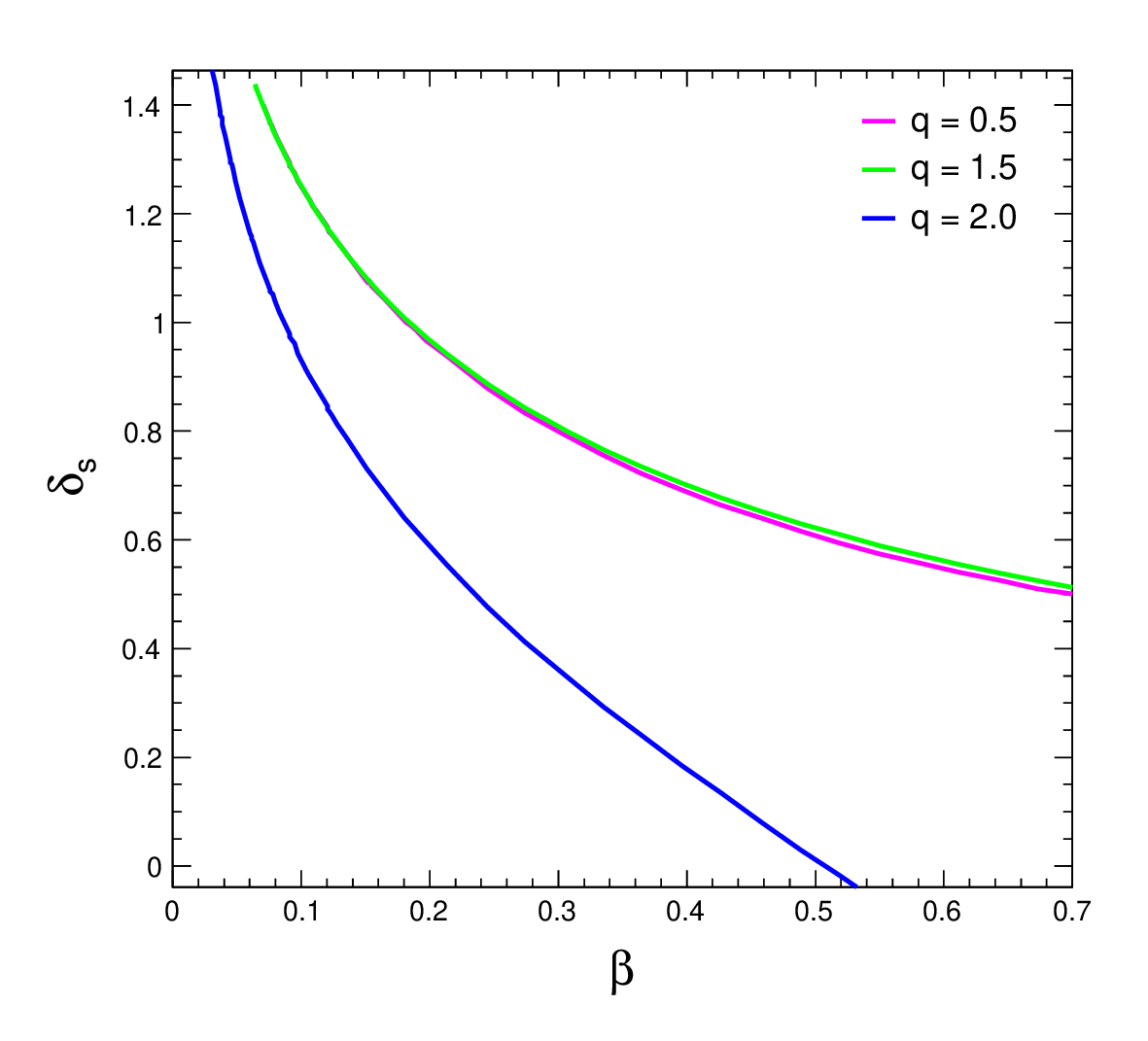}
\includegraphics[scale=0.29]{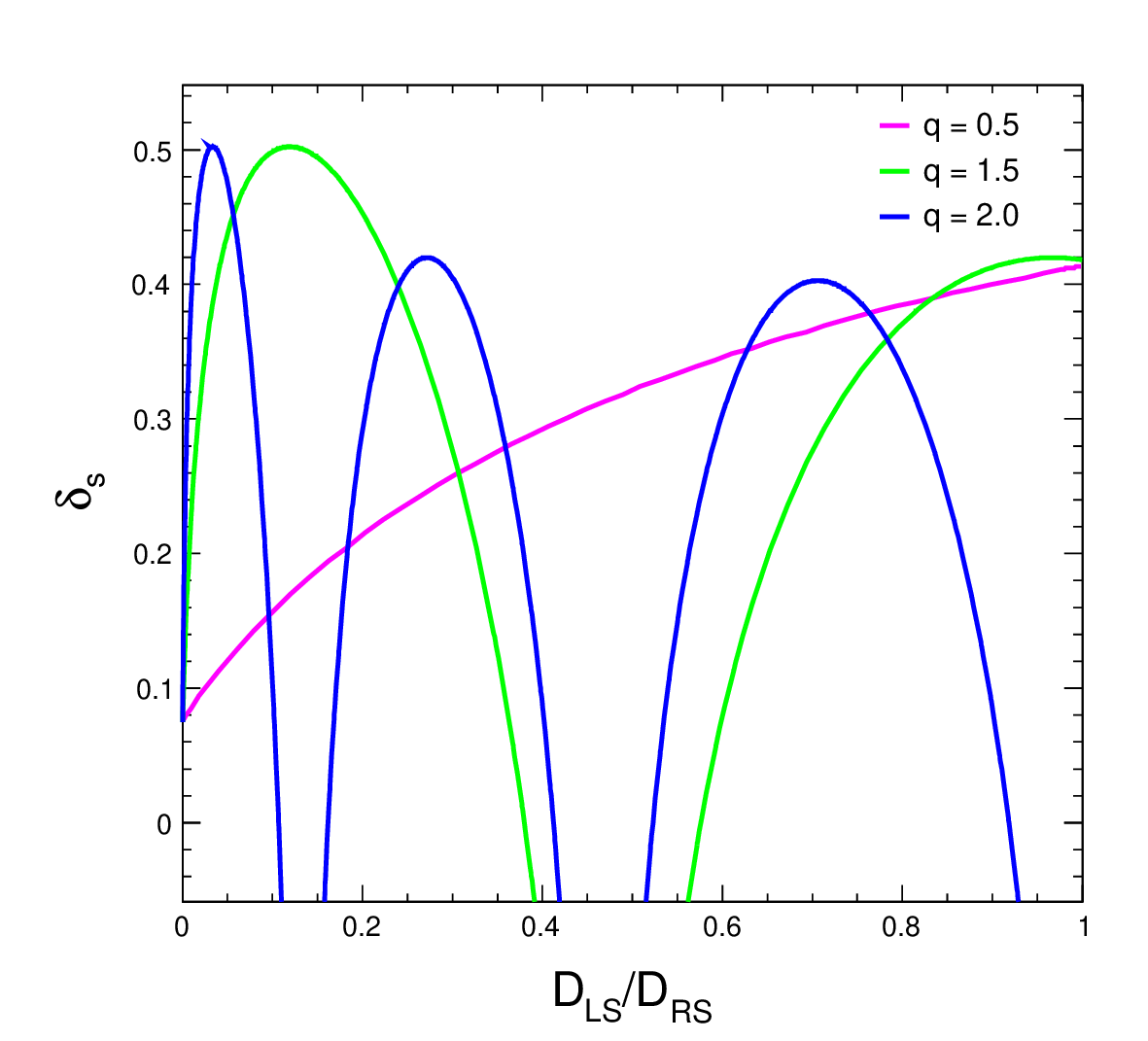}
\includegraphics[scale=0.29]{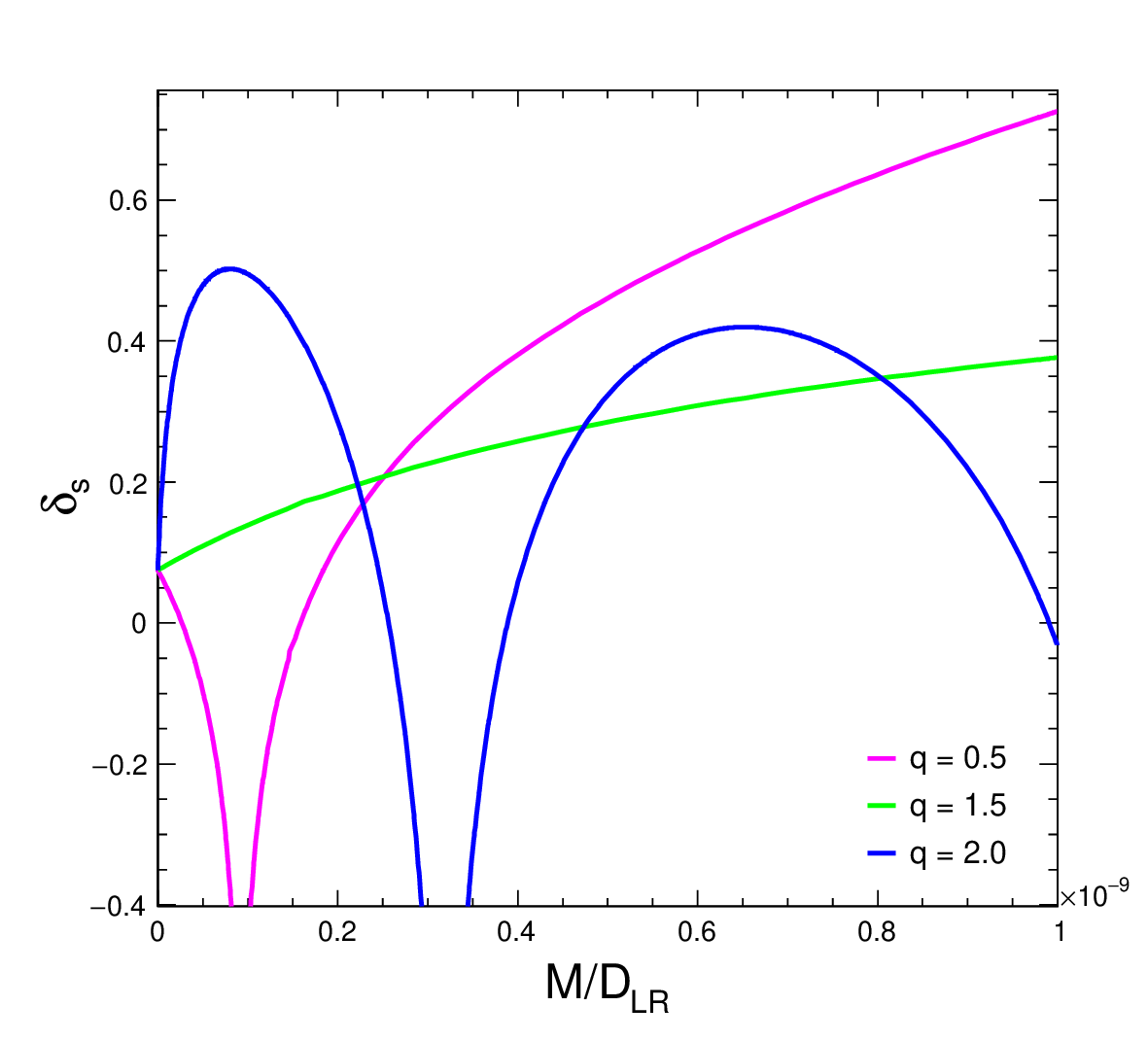}
\caption{The distortion parameter of primary and secondary images are plotted a functions of $\beta$, $D_{LS}/D_{RS}$ and $M/D_{LR}$ for different values of $q$.}
 \label{fig.9}
\end{figure}

In Figures~\ref{fig.4} and \ref{fig.5}, we have portrayed the radial, tangential and total magnification of primary and secondary images respectively as a function of $\beta$, $D_{LS}/D_{RS}$ and $M/D_{LR}$ for different values of the EH parameter $\mu$, while those for different values of $q$ are displayed in Figures~\ref{fig.6} and \ref{fig.7}. Following Eq.~(\ref{eqn.39}), the distortion parameter is also depicted as a function of $\beta$, $D_{LS}/D_{RS}$ and $M/D_{LR}$ for different $q$ and $\mu$ values in Figures~\ref{fig.8} and \ref{fig.9} respectively in order to analyze the impact of these parameters on the distortion of primary and secondary images. Here, we consider the SgrA$^*$ black hole (mass, $M =4\times 10^6 M_\odot$ and lens to receiver distance, $D_{LR} = 8.3$ kpc) as the lens in the MEH theory.

The effect of the EH parameter $\mu$ and the charge parameter $q$ on the magnification and distortion of primary and secondary images is evident from the respective figures. As illustrated in Figure~\ref{fig.4}, for varying $\mu$, the radial magnification of primary images increases with $\beta$ and $M/D_{LR}$ while it decreases with $D_{LS}/D_{RS}$. The tangential magnification of primary images, however, shows some fluctuations with $\beta$ and $D_{LS}/D_{RS}$ for different $\mu$ values, while it decreases with $M/D_{LR}$. Consequently, fluctuations are observed in the total magnification of primary images with $\beta$ and $D_{LS}/D_{RS}$, and an increase is seen with increasing $M/D_{LR}$. For secondary images, as shown in Figure~\ref{fig.5} for different $\mu$ values, the radial magnification decreases with $\beta$ and $M/D_{LR}$ while it increases with $D_{LS}/D_{RS}$. The tangential magnification also shows fluctuations for increasing $\beta$ and $D_{LS}/D_{RS}$ and depicts an increase with rising $M/D_{LR}$. The resulting total magnification shows fluctuations with increasing $\beta$ and $D_{LS}/D_{RS}$ whereas with increasing $M/D_{LR}$, it decreases. Figure~\ref{fig.6} portrays for different $q$ values, the radial magnification of primary images increases with $\beta$ and $M/D_{LR}$, while it falls for increasing $D_{LS}/D_{RS}$. The tangential magnifications depict fulctuations with rise in $\beta$ and $D_{LS}/D_{RS}$ while it decreases with increasing $M/D_{LR}$. As a result, the total magnification of primary images show fluctuations with rising $\beta$, a decrease with increasing $D_{LS}/D_{RS}$ and for rising $M/D_{LR}$, it increases for smaller values of $q$ and fall for higher values of $q$. Furthermore, for different $q$ values, the secondary images are depicted in Figure~\ref{fig.7}. Here, it can be observed that with increasing $\beta$, the radial magnification decreases while the tangential magnification shows a fluctuation and the total magnification consequently results in a fluctuating pattern. With the rise in $D_{LS}/D_{RS}$, the radial magnification also rises for increasing $q$ while the tangential and total magnifications present a fluctuating behaviour. The radial magnification as a function of $M/D_{LR}$ decreases for small $q$ value while it increases for higher $q$ value. The tangential magnification decreases with $M/D_{LR}$ for higher $q$ values. The total magnification decreases with rise in $M/D_{LR}$.

Illustrated in Figures~\ref{fig.8} and \ref{fig.9} are the distortion of primary and secondary images as a function of $\beta$, $D_{LS}/D_{RS}$ and $M/D_{LR}$ for different values of $\mu$ and $q$ respectively. With rise in $\beta$, the distortion in the primary and secondary images tend to fall for higher values of $\mu$ as well as for higher values of $q$. The distortion of primary images as a function of $D_{LS}/D_{RS}$ increases for higher values of $\mu$ and $q$. An increase is also observed in the distortion of secondary images as a function of $D_{LS}/D_{RS}$ for higher values of $\mu$ whereas an increasing fluctuation is observed for higher values of $q$. With respect to rising $M/D_{LR}$, the distortion of primary images show an increase with higher values of $\mu$ and $q$. A fall in the distortion of secondary images can be seen for higher values of $\mu$ and an increasing fluctuating pattern can be observed for higher values of $q$.

\section{Time Delay} 
\label{sec.5}

When light propagates from a source to a receiver via two distinct paths, the resulting difference in arrival times is known as the time delay. In this section, we investigate the influence of the charge parameter and the EH parameter on the time delay of light traveling through the neighborhood of a lensed black hole in MEH gravitational theory. To derive the time delay of light, it is important to deduce how the time coordinate $t$, as measured by a distant observer, varies along the photon trajectory. Hence, using Eq.s~(\ref{eqn.21}) and (\ref{eqn.22}), the time derivative of the radial co-ordinate can be given as
\begin{equation}
\frac{dr}{dt} = \frac{dr}{d\lambda}\cdot \frac{d\lambda}{dt} = \pm A(r) \sqrt{1 - A(r)\frac{\xi^2}{r^2}}.
\label{eqn.42}
\end{equation}

Upon emission from the source, the photons tend to travel inward towards the black hole causing the radial coordinate $r$ to initially decrease with time until it reaches the distance of closest approach, $r_0$, to the black hole. Beyond this turning point, the radial coordinate increases as time progresses. Hence, considering $r_S$ as the source position and $r_R$ as the receiver position and using the relation~(\ref{eqn.42}), the time delay associated with light propagation in gravitational lensing observation can be computed to be
\begin{equation}
\begin{split}
\Delta T & = T - T_0 = \int_{r_S}^{r_0} \frac{dt}{dr} dr + \int_{r_0}^{r_R} \frac{dt}{dr} dr - T_0 \\
& = - \int_{r_S}^{r_0} \frac{dr}{A(r) \sqrt{1 - A(r)\frac{\xi^2}{r^2}}} + \int_{r_0}^{r_R} \frac{dr}{A(r) \sqrt{1 - A(r)\frac{\xi^2}{r^2}}} - T_0 \\ 
& = \int_{r_0}^{r_S} \frac{dr}{A(r) \sqrt{1 - A(r)\frac{\xi^2}{r^2}}} + \int_{r_0}^{r_R} \frac{dr}{A(r) \sqrt{1 - A(r)\frac{\xi^2}{r^2}}} - \sqrt{r_S^2 - r_0^2} - \sqrt{r_R^2 - r_0^2}
\end{split}
\label{eqn.43}
\end{equation}
where, $T_0 = \sqrt{r_S^2 - r_0^2} + \sqrt{r_R^2 - r_0^2}$ is the time taken by light to travel from the source to the receiver in the absence of a lensing gravitational field. It is clear from the above expression that the time delay rises monotonically as $r_S$ and $r_R$ increases.

Evaluating Eq.~(\ref{eqn.43}) for the black hole metric~(\ref{eqn.10}), we can estimate the time delay of light due to the gravitational field of the black hole in MEH theory as
\begin{equation}
\begin{split}
\Delta T & =\frac{64 M \mu  r_0^3 q^4}{15 b^6 \left(b^2-r_0^2\right) \sqrt{r_0^2-b^2}} + \frac{21 \mu  \tan ^{-1}\left(\frac{b}{\sqrt{r_0^2-b^2}}\right) q^4}{80 b^5} + \frac{16 M \mu  r_R q^4}{5 b^4 \left(b^2-r_R^2\right) \sqrt{r_R^2-b^2}} \\
& + \frac{16 M \mu  r_s q^4}{5 b^4 \left(b^2-r_s^2\right) \sqrt{r_s^2-b^2}} -\frac{21 \mu  \tan ^{-1}\left(\frac{b}{\sqrt{r_R^2-b^2}}\right) q^4}{160 b^5} - \frac{21 \mu  \tan ^{-1}\left(\frac{b}{\sqrt{r_s^2-b^2}}\right) q^4}{160 b^5} \\ 
& - \frac{21 \mu  q^4}{80 b^4 \sqrt{r_0^2-b^2}} + \frac{7 \mu  q^4}{80 b^2 r_0^2 \sqrt{r_0^2-b^2}} + \frac{3 \mu  q^4}{40 r_0^4 \sqrt{r_0^2-b^2}} - \frac{32 M \mu  r_0 q^4}{5 b^4 \left(b^2-r_0^2\right) \sqrt{r_0^2-b^2}} \\
& + \frac{8 M \mu  q^4}{5 b^2 r_0 \left(b^2-r_0^2\right) \sqrt{r_0^2-b^2}} + \frac{2 M \mu  q^4}{15 r_0^3 \left(b^2-r_0^2\right) \sqrt{r_0^2-b^2}} + \frac{3 b^2 M \mu  q^4}{10 r_0^5 \left(b^2-r_0^2\right) \sqrt{r_0^2-b^2}} \\
& + \frac{21 \mu  q^4}{160 b^4 \sqrt{r_R^2-b^2}} - \frac{7 \mu  q^4}{160 b^2 r_R^2 \sqrt{r_R^2-b^2}} - \frac{3 \mu  q^4}{80 r_R^4 \sqrt{r_R^2-b^2}} - \frac{32 M \mu  r_R^3 q^4}{15 b^6 \left(b^2-r_R^2\right) \sqrt{r_R^2-b^2}} \\
& - \frac{4 M \mu  q^4}{5 b^2 r_R \left(b^2-r_R^2\right) \sqrt{r_R^2-b^2}} - \frac{M \mu  q^4}{15 r_R^3 \left(b^2-r_R^2\right) \sqrt{r_R^2-b^2}} - \frac{3 b^2 M \mu  q^4}{20 r_R^5 \left(b^2-r_R^2\right) \sqrt{r_R^2-b^2}} \\
& + \frac{21 \mu  q^4}{160 b^4 \sqrt{r_s^2-b^2}} - \frac{7 \mu  q^4}{160 b^2 r_s^2 \sqrt{r_s^2-b^2}} - \frac{3 \mu  q^4}{80 r_s^4 \sqrt{r_s^2-b^2}} - \frac{32 M \mu  r_s^3 q^4}{15 b^6 \left(b^2-r_s^2\right) \sqrt{r_s^2-b^2}} \\
& - \frac{4 M \mu  q^4}{5 b^2 r_s \left(b^2-r_s^2\right) \sqrt{r_s^2-b^2}} - \frac{M \mu  q^4}{15 r_s^3 \left(b^2-r_s^2\right) \sqrt{r_s^2-b^2}} - \frac{3 b^2 M \mu  q^4}{20 r_s^5 \left(b^2-r_s^2\right) \sqrt{r_s^2-b^2}} \\
& + \frac{r_R^2}{\sqrt{r_R^2-b^2}} + \frac{r_s^2}{\sqrt{r_s^2-b^2}} - 4 M \log \left(\sqrt{r_0^2-b^2}+r_0\right) + 2 M \log \left(\sqrt{r_R^2-b^2} + r_R\right) \\
& + 2 M \log \left(\sqrt{r_s^2-b^2}+r_s\right) + \frac{M r_R}{\sqrt{r_R^2-b^2}} + \frac{M r_s}{\sqrt{r_s^2-b^2}}+\frac{2 b^2}{\sqrt{r_0^2-b^2}}-\frac{2 r_0^2}{\sqrt{r_0^2-b^2}} \\
& -\frac{2 M r_0}{\sqrt{r_0^2-b^2}}-\frac{b^2}{\sqrt{r_R^2-b^2}}-\frac{b^2}{\sqrt{r_s^2-b^2}}.
\end{split}
\label{eqn.44}
\end{equation} 

The above expression shows that the time delay of light will be affected by the EH parameter as well as the charge parameter of the black hole under consideration.

\section{Conclusion and Outlook}
\label{sec.6}

Gravitational lensing is a crucial tool to probe the nature of black holes across various gravity theories. The framework of MEH theory combines mimetic gravity with the Euler-Heisenberg Lagrangian, which accounts for the non-linear quantum electrodynamics, to study modified spacetimes and black hole solutions. In this work, we have investigated the gravitational lensing signatures in the weak field limit of a black hole in the MEH theory. To this end, we assume a finite distance between the lens and the source as well as the receiver and derive the deflection angle of light. Next we derive the expression for the Einstein ring radius and the magnification and distortion parameters from the lens equation. Furthermore, we briefly discuss the time delay of light due lensing by the black hole. The effects of the EH parameter and the charge parameter of the black hole on each of these gravitational lensing features in the weak field limit are analyzed. The observables are studied by employing SgrA$^*$ black hole as the lens with $M = 4\times 10^6 M_\odot$ and $D_{LR} = 8.3$ kpc. Our numerical results presented in this work show the influences coming from the black hole charge parameter and the EH parameter on the gravitational lensing signatures.

To obtain the deflection angle of light around the black hole, we implement the GBT which
was extended by Ishihara {\it et al}. This method is independent of asymptotic flatness of the spacetime. The deflection angle is examined as a function of the impact parameter and the effects of the EH parameter $\mu$ and the charge parameter $q$ on the deflection angle are analyzed. For each case the computed results are compared with that of a black hole in GR theory. It is evident from the figures that the deflection angle decreases with increase in the impact parameter. Infact, it is seen that the deflection angle pattern is quite similar to that in GR except for a slight dip in the deflection angle for small impact parameter values. Consequently, by increasing the EH parameter and decreasing the charge parameter, it is possible to recover the Schwazschild behaviour for the black hole in MEH theory. Next, we compute the Einstein ring radius and study its behaviour for different values of $\mu$, $q$ and lens to source distance ($D_{LS}$). It is found from the analysis that the radii increases with rise in the values of $q$, $\mu$ and $D_{LS}$.

Furthermore, the magnification and distortion parameters are derived from the lens equation. For the black hole under study, the radial, tangetial and total magnifications for the primary and secondary images are portrayed as a function of $\beta$, $D_{LS}/D_{RS}$ and $M/D_{LR}$. For different values of $\mu$ and $q$, each case is analyzed and it is clear from this study that the EH parameter and the charge parameter have significant impact on the magnification of the primary and secondary images. Similarly, the distortion of primary and secondary images is also influenced by the EH parameter and the charge parameter of the black hole under consideration. Finally, we briefly discuss the time delay of light resulting from the gravitational lensing of light in MEH theory. The analytical expression for the time delay is computed and it is observed from the expression that the time delay rises monotonically as the coordinates of the source position and the receiver position increases. As a futuristic outlook, we plan to further explore more properties of black holes within the framework of MEH theory and search for possible observational evidence to validate these theoretical predictions. Also, we look to have observational constraints on the model parameters of the black hole by using EHT data as a future prospect of the study.

\bibliographystyle{apsrev}
\end{document}